\documentclass[structabstract]{aa}  
\usepackage{graphicx}
\usepackage{txfonts}
\usepackage{natbib}
\usepackage{multirow}
\usepackage{longtable}
\usepackage{pdflscape}
\usepackage{fixltx2e}
\usepackage[colorinlistoftodos]{todonotes}

\def\4he{$^4$He}

\def\kms{\mathrm{km\,s}^{-1}}
\def\e#1{\times 10^{#1}}

\def\msol{\mathrm{M}_\odot}

\def\up#1{$^{#1}$}
\def\down#1{$_{#1}$}

\def\h2{$\mathrm{H}_2$}
\def\so2{$\mathrm{SO}_2$}
\def\mic{\mathrm{ }\mu\mathrm{m}}
\def\spy{\;\msol~\mathrm{ yr}^{-1}}

\begin{document}
   \title{Sulphur molecules in the circumstellar envelopes of M-type AGB stars\thanks{Herschel is an ESA space observatory with science instruments provided by European-led Principal Investigator consortia and with important participation from NASA.}}

   \author{T. Danilovich
          \inst{1}
          \and
          E. De Beck\inst{1}
          \and
          J. H. Black\inst{1}
          \and
          H. Olofsson \inst{1}
          \and
          K. Justtanont\inst{1}
          }

   \institute{Onsala Space Observatory, Department of Earth and Space Sciences, Chalmers University of Technology, 439 92 Onsala, Sweden  \\
             \email{taissa@chalmers.se}
             }

   \date{Received 10 December 2015 / Accepted 25 January 2016}

 
  \abstract
   {}
   {The sulphur compounds SO and \so2 have not been widely studied in the circumstellar envelopes of asymptotic giant branch (AGB) stars. By presenting and modelling a large number of SO and \so2 lines in the low mass-loss rate M-type AGB star R Dor, and modelling the available lines of those molecules in a further four M-type AGB stars, we aim to determine their circumstellar abundances and distributions.}
   {We use a detailed radiative transfer analysis based on the accelerated lambda iteration method to model circumstellar SO and \so2 line emission. We use molecular data files for both SO and \so2 that are more extensive than those previously available. }
   {Using 17 SO lines and 98 \so2 lines to constrain our models for R~Dor, we find an SO abundance of $(6.7\pm 0.9)\e{-6}$ and an \so2 abundance of $5\e{-6}$ with both species having high abundances close to the star. We also modelled \up{34}SO and found an abundance of $(3.1\pm0.8)\e{-7}$, giving an \up{32}SO/\up{34}SO ratio of $21.6\pm8.5$. We derive similar results for the circumstellar SO and \so2 abundances and their distributions for the low mass-loss rate object W Hya. For the higher mass-loss rate stars, we find shell-like SO distributions with peak abundances that decrease and peak abundance radii that increase with increasing mass-loss rate. The positions of the peak SO abundance agree very well with the photodissociation radii of \h2O.  We also modelled \so2 in two higher mass-loss rate stars but our models for these were less conclusive.}
  {We conclude that for the low mass-loss rate stars, the circumstellar SO and \so2 abundances are much higher than predicted by chemical models of the extended stellar atmosphere. These two species may also account for all the available sulphur. For the higher mass-loss rate stars we find evidence that SO is most efficiently formed in the circumstellar envelope, most likely through the photodissociation of \h2O and the subsequent reaction between S and OH. The S-bearing parent molecule does not appear to be \h2S. The \so2 models for the higher mass-loss rate stars are less conclusive, but suggest an origin close to the star for this species. This is not consistent with current chemical models. The combined circumstellar SO and \so2 abundances are significantly lower than that of sulphur for these higher mass-loss rate objects.}
   

   \maketitle
%

\section{Introduction}

Low- to intermediate-mass stars eventually evolve from the main sequence to the asymptotic giant branch (AGB). AGB stars lose mass rapidly, producing a circumstellar envelope (CSE) of atomic and molecular matter and dust, rich in chemical diversity. The variety and abundances of molecules that can be found in the circumstellar envelopes of AGB stars depend on the chemistry of the individual star. For example, carbon stars, which have carbon-to-oxygen ratio C/O $> 1$, are most likely to have a variety of C-bearing molecules in their CSEs \citep[e.g.][]{Gong2015}, while oxygen-rich M-type stars, with C/O $< 1$, are more likely to contain a variety of O-bearing molecules \citep[e.g. ][]{Justtanont2012}.

SO and \so2 are two such O-bearing molecules. They are thought to exist in shells in the CSE, having been formed through the photodissociation of the parent molecule \h2S and subsequent reactions with O and OH \citep{Cherchneff2006,Willacy1997}. Observations of the red supergiant VY CMa by \cite{Adande2013} contradict this view, however; the modelling results show small, concentrated envelopes of SO and \so2 and indications that \so2 may itself be formed directly, rather than being a photodissociation product of \h2S or another molecule and are found in a hollow shell around the star. Similarly, when \cite{Decin2010} modelled \so2 emission around the AGB star IK~Tau, they had difficulty reconciling their observations with a shell model.

The \citet{Yamamura1999} ISO/SWS detections of the 7.4~$\mic$ $\nu_3$ \so2 band in a few AGB stars suggest that \so2 is formed in the warmest regions of the CSE. Analysis of these data by \cite{Yamamura1999} and \citet{Cami1999} indicates that the \so2 is mostly likely formed within a few stellar radii of the star at a temperature of $\sim$ 600 K. \citet{Cami1999} also find that the excitation of \so2 to the $\nu_3$ band varies with pulsation period.
Their simple models put the outer radius of \so2 within $\sim5R_*$.


In this paper we present new observations of circumstellar SO and \so2 from an APEX spectral survey of the M-type AGB star R~Dor. We combine these results with SO and \so2 detections from \textit{Herschel}/HIFI, developing comprehensive models of the SO and \so2 distributions around R~Dor using 17 SO lines and 98 \so2 lines, all spectrally resolved. 

We also model the sparse detections of SO and \so2 emission towards 
the other M-type AGB stars observed with \textit{Herschel}/HIFI, supplemented with archival data where available.


\section{Sample and observations}\label{obs}

The stars included in this study come from the sample of M-type AGB stars observed as part of the HIFISTARS Guaranteed Time Key Programme \citep[][and see Sect. \ref{hifi} for details]{Justtanont2012}. The OH/IR stars are excluded, as is Mira, which has a complicated and asymmetric CSE induced by a white dwarf companion \citep[see][]{Ramstedt2014a}. That leaves a sample of five M-stars, four of which had SO and \so2 lines detected by HIFI. The remaining star, TX~Cam, has previously been detected in SO at lower frequencies.

Some basic information about the five sources is given in Table \ref{radecs}.

\begin{table}[t]
\caption{Basic information about our five sources.}\label{radecs}
\begin{center}
\begin{tabular}{ccccc}
\hline\hline
Star & RA & Dec & Variability &Spec type\\
\hline
IK~Tau & 03 53 28.87 & $+$11 24 21.7& M &M9\\
R~Dor & 04 36 45.59 & $-$62 04 37.8 & SRB&M8e\\
TX~Cam & 05 00 50.39 &$+$56 10 52.6& M&  	M8.5\\
W~Hya & 13 49 02.00 & $-$28 22 03.5& M & M7.5-9e\\
R~Cas & 23 58 24.87 &$+$51 23 19.7& M &	M6.5-9e\\
\hline
\end{tabular}
\end{center}
\tablefoot{RA and Dec are given in J2000 co-ordinates. The variability types are M = Mira variable, SRB = semi-regular variable type B.}
\end{table}%

\subsection{APEX data}\label{apexobs}

We performed a spectral survey of R~Dor in the ranges $213-321.5$\,GHz and $338.5-368.5$\,GHz ($\lambda=0.8-1.4$\,mm) using the Swedish Heterodyne Facility Instrument \citep[SHeFI;][]{Vassilev2008} on the Atacama Pathfinder Experiment telescope (APEX). The data were observed over several observing seasons between May 2011 and June 2015. The observations were carried out using beam switching with a standard beam throw of 3\arcmin. A detailed description of this survey will be presented by De Beck et al., (\textit{in prep.}).

Data reduction was carried out using the \textsc{Gildas/Class}\footnote{\texttt{http://www.iram.fr/IRAMFR/GILDAS/}} package. Scans with very unstable baselines were ignored and bad channels were blanked. After masking the regions with line emission, polynomial baselines of typically first degree were subtracted from the averaged spectra to obtain a 0\,K baseline. Rms noise levels throughout the survey are around $2-10$\,mK at a velocity resolution of 1\,km\,s$^{-1}$. The spectra were then converted to main beam temperatures using efficiency correction factors of $\eta_{mb} = 0.75$ for $\nu< 270$~GHz, $\eta_{mb} = 0.74$ for $270 < \nu< 320$~GHz, and $\eta_{mb} = 0.73$ for $\nu> 320$~GHz. The half-power beam-widths were calculated using the general formula
\begin{equation}
\theta =7.8 \left(\frac{800}{\nu}\right)
\end{equation}
where $\nu$ is in GHz and $\theta$ is in arcseconds. The beam-widths across our frequency range are between 17--29$\arcsec$.

The detections of SO using APEX are listed in Table \ref{apexso} and the \so2 detections are listed in Table \ref{apexso2}. There were also some detections of SO and \so2 isotopologues: \up{34}SO, and tentative detections of \up{34}\so2 and SO\up{18}O. We model \up{34}SO, but are unable to perform a full radiative transfer analysis for the other isotopologues. See Table \ref{isotopologues} for a list of isotopologue detections and for the full discussion, see Sect. \ref{isoresults}.

In terms of other S-bearing molecules, there were no conclusive detections of either CS (out of three possible transitions in the range $J_\mathrm{up}= 5$ to $J_\mathrm{up}= 7$) or SiS (out of nine possible transitions in the range $J_\mathrm{up}= 12$ to $J_\mathrm{up}= 20$). There was a tentative detection of CS ($6\to5$) but it is blended with \up{29}SiO($7\to6$, $v=3$) line and hence allows no reliable conclusion on the detection of CS. (We note that there are several other detections of \up{29}SiO in the survey, but none of CS.) No other S-bearing molecules were detected in this survey.

\begin{table}[tp]
\caption{SO observations towards R~Dor using APEX, listed in order of descending energy of the upper level.}\label{apexso}
\begin{center}
\begin{tabular}{crrrc}
\hline\hline
		Transition		&	$\nu\;\;\;$	&	$E_\mathrm{up}$	& $\theta\;$&  $I_{mb}$ \\
&[GHz]&[K]&[\arcsec]&[K $\kms$]\\
\hline	
$	8_8\to7_7	$	&	344.311	&	88	&	18	&	5.04	\\
$	8_7\to7_6	$	&	340.714	&	81	&	18	&	4.59	\\
$	8_9\to7_8	$	&$\dagger$ 	346.528	&	79	&	18	&	4.54	\\
$	7_7\to6_6	$	&	301.286	&	71	&	21	&	4.74	\\
$	7_6\to6_5	$	&	296.550	&	65	&	21	&	4.12	\\
$	7_8\to6_7	$	&	304.078	&	62	&	21	&	6.40	\\
$	6_6\to5_5	$	&	258.256	&	57	&	24	&	3.49	\\
$	6_5\to5_4	$	&	251.826	&	51	&	25	&	3.13	\\
$	6_7\to5_6	$	&	261.844	&	48	&	24	&	5.20	\\
$	5_5\to4_4	$	&	215.221	&	44	&	29	&	2.25	\\
$	5_6\to4_5	$	&	219.949	&	35	&	28	&	4.21	\\
$	3_3\to2_3	$	&	339.341	&	26	&	18	&	\phantom{0}0.125	\\
$	2_2\to1_2	$	&	309.502	&	19	&	20	&	\phantom{0}0.100	\\
\hline
\end{tabular}
\end{center}
\tablefoot{$\dagger$ indicates a line overlap with SO$_2$.}
\end{table}%

\begin{table*}[tp]
\caption{SO\down{2} observations towards R~Dor using APEX, listed in order of descending energy of the upper level.}\label{apexso2}
\begin{center}
\begin{tabular}{crrrc|crrrc}
\hline\hline
		Transition		&	$\nu\;\;\;$	&	$E_\mathrm{up}$	& $\theta\;$&  $I_{mb}$&		Transition		&	$\nu\;\;\;$	&	$E_\mathrm{up}$	& $\theta\;$&  $I_{mb}$\\
&[GHz]&[K]&[\arcsec]&[K $\kms$]&&[GHz]&[K]&[\arcsec]&[K $\kms$]\\
\hline
$	25_{4,22}\to26_{1,25	}	$	&*	279.497	&	1085	&	22	&	0.373	&	$	16_{3,13}\to16_{2,14	}	$	&	214.689	&	148	&	29	&	0.415	\\
$	40_{4,36}\to40_{3,37	}	$	&	341.403	&	808	&	18	&	0.217	&	$	17_{1,17}\to16_{0,16	}	$	&	313.661	&	136	&	20	&	2.881	\\
$	36_{5,31}\to36_{4,32	}	$	&	341.674	&	679	&	18	&	0.148	&	$	14_{4,10}\to14_{3,11	}	$	&	351.874	&	136	&	18	&	0.860	\\
$	36_{4,32}\to36_{3,33	}	$	&	281.689	&	662	&	22	&	0.257	&	$	15_{3,13}\to15_{2,14	}	$	&	275.240	&	133	&	23	&	0.745	\\
$	34_{5,29}\to34_{4,30	}	$	&	360.290	&	612	&	17	&	0.250	&	$	16_{1,15}\to15_{2,14	}	$	&	236.217	&	131	&	26	&	0.595	\\
$	34_{3,31}\to34_{2,32	}	$	&	342.762	&	582	&	18	&	0.255	&	$	13_{4,10}\to13_{3,11	}	$	&	357.165	&	123	&	17	&	0.919	\\
$	32_{4,28}\to32_{3,29	}	$	&	258.389	&	531	&	24	&	0.368	&	$	16_{0,16}\to15_{1,15	}	$	&	283.292	&	121	&	22	&	0.959	\\
$	32_{3,29}\to32_{2,30	}	$	&	300.273	&	519	&	21	&	0.453	&	$	15_{2,14}\to15_{1,15	}	$	&	248.057	&	119	&	25	&	0.423	\\
$	30_{4,26}\to30_{3,27	}	$	&	259.599	&	472	&	24	&	0.212	&	$	15_{2,14}\to14_{1,13	}	$	&	366.214	&	119	&	17	&	0.840	\\
$	30_{3,27}\to30_{2,28	}	$	&	263.544	&	459	&	24	&	0.231	&	$	14_{3,11}\to14_{2,12	}	$	&	226.300	&	119	&	28	&	0.596	\\
$	28_{4,24}\to28_{3,25	}	$	&	267.720	&	416	&	23	&	0.412	&	$	12_{4,8}\to12_{3,9	}	$	&	355.046	&	111	&	18	&	0.498	\\
$	28_{3,25}\to28_{2,26	}	$	&	234.187	&	403	&	27	&	0.289	&	$	13_{3,11}\to13_{2,12	}	$	&	267.537	&	106	&	23	&	0.600	\\
$	28_{3,25}\to27_{4,24	}	$	&$\dagger$ 	313.412	&	403	&	20	&	0.129	&	$	15_{1,15}\to14_{0,14	}	$	&	281.763	&	107	&	22	&	2.007	\\
$	28_{2,26}\to28_{1,27	}	$	&	340.316	&	392	&	18	&	0.365	&	$	11_{4,8}\to11_{3,9	}	$	&	357.388	&	100	&	17	&	0.980	\\
$	26_{4,22}\to26_{3,23	}	$	&	280.807	&	364	&	22	&	0.382	&	$	12_{3,9}\to12_{2,10	}	$	&	237.069	&	94	&	26	&	0.616	\\
$	26_{3,23}\to26_{2,24	}	$	&	213.068	&	351	&	29	&	0.410	&	$	14_{0,14}\to13_{1,13	}	$	&	244.254	&	94	&	26	&	1.702	\\
$	26_{3,23}\to25_{4,22	}	$	&	245.339	&	351	&	25	&	0.216	&	$	13_{2,12}\to13_{1,13	}	$	&	225.154	&	93	&	28	&	0.652	\\
$	26_{2,24}\to26_{1,25	}	$	&	296.169	&	341	&	21	&	0.306	&	$	13_{2,12}\to12_{1,11	}	$	&$\dagger$ 	345.339	&	93	&	18	&	2.283	\\
$	25_{3,23}\to25_{2,24	}	$	&	359.151	&	321	&	17	&	0.601	&	$	10_{4,6}\to10_{3,7	}	$	&	356.755	&	90	&	17	&	0.804	\\
$	24_{4,20}\to24_{3,21	}	$	&$\dagger$ 	296.535	&	317	&	21	&	0.520	&	$	11_{3,9}\to11_{2,10	}	$	&	262.257	&	83	&	24	&	0.639	\\
$	24_{1,23}\to24_{0,24	}	$	&	363.891	&	281	&	17	&	0.396	&	$	13_{1,13}\to12_{0,12	}	$	&$\dagger$ 	251.200	&	82	&	25	&	1.275	\\
$	22_{4,18}\to22_{3,19	}	$	&	312.543	&	273	&	20	&	0.424	&	$	9_{4,6}\to9_{3,7	}	$	&	357.672	&	81	&	17	&	0.497	\\
$	23_{2,22}\to23_{1,23	}	$	&	363.926	&	260	&	17	&	0.741	&	$	10_{3,7}\to10_{2,8	}	$	&	245.563	&	73	&	25	&	0.587	\\
$	21_{4,18}\to21_{3,19	}	$	&	363.159	&	252	&	17	&	0.815	&	$	8_{4,4}\to8_{3,5	}	$	&	357.581	&	72	&	17	&	0.686	\\
$	22_{2,20}\to22_{1,21	}	$	&	216.643	&	248	&	29	&	0.373	&	$	7_{4,4}\to7_{3,5	}	$	&	357.892	&	65	&	17	&	0.685	\\
$	22_{2,20}\to21_{3,19	}	$	&	286.416	&	248	&	22	&	0.168	&	$	9_{3,7}\to9_{2,8	}	$	&	258.942	&	64	&	24	&	0.561	\\
$	21_{3,19}\to21_{2,20	}	$	&	316.099	&	235	&	20	&	0.878	&	$	11_{1,11}\to10_{0,10	}	$	&	221.965	&	60	&	28	&	0.953	\\
$	19_{4,16}\to19_{3,17	}	$	&	359.771	&	214	&	17	&	0.430	&	$	6_{4,2}\to6_{3,3	}	$	&	357.926	&	59	&	17	&	0.527	\\
$	20_{1,19}\to20_{0,20	}	$	&	282.293	&	199	&	22	&	0.751	&	$	8_{3,5}\to8_{2,6	}	$	&$\dagger$ 	251.211	&	55	&	25	&	0.995	\\
$	20_{1,19}\to19_{2,18	}	$	&	338.612	&	199	&	18	&	1.255	&	$	5_{4,2}\to5_{3,3	}	$	&	358.013	&	53	&	17	&	0.398	\\
$	19_{3,17}\to19_{2,18	}	$	&	299.317	&	197	&	21	&	0.496	&	$	9_{2,8}\to8_{1,7	}	$	&	298.576	&	51	&	21	&	0.496	\\
$	20_{0,20}\to19_{1,19	}	$	&$\dagger$ 	358.216	&	185	&	17	&	2.373	&	$	4_{4,0}\to4_{3,1	}	$	&	358.038	&	49	&	17	&	0.234	\\
$	19_{2,18}\to19_{1,19	}	$	&	301.897	&	183	&	21	&	0.561	&	$	7_{3,5}\to7_{2,6	}	$	&	257.100	&	48	&	24	&	0.448	\\
$	17_{4,14}\to17_{3,15	}	$	&	357.963	&	180	&	17	&	0.917	&	$	6_{3,3}\to6_{2,4	}	$	&$\dagger$ 	254.281	&	41	&	25	&	0.776	\\
$	19_{1,19}\to18_{0,18	}	$	&	346.652	&	168	&	18	&	1.390	&	$	5_{3,3}\to5_{2,4	}	$	&	256.247	&	36	&	24	&	0.350	\\
$	16_{4,12}\to16_{3,13	}	$	&$\dagger$ 	346.524	&	165	&	18	&	3.864	&	$	5_{3,3}\to4_{2,2	}	$	&	351.257	&	36	&	18	&	0.705	\\
$	18_{1,17}\to18_{0,18	}	$	&	240.943	&	163	&	26	&	0.506	&	$	7_{2,6}\to6_{1,5	}	$	&	271.529	&	36	&	23	&	0.426	\\
$	18_{1,17}\to17_{2,16	}	$	&	288.520	&	163	&	22	&	0.852	&	$	4_{3,1}\to4_{2,2	}	$	&	255.553	&	31	&	24	&	0.215	\\
$	17_{3,15}\to17_{2,16	}	$	&	285.744	&	163	&	22	&	0.382	&	$	6_{2,4}\to5_{1,5	}	$	&	282.037	&	29	&	22	&	0.349	\\
$	18_{0,18}\to17_{1,17	}	$	&	321.330	&	152	&	19	&	2.091	&	$	3_{3,1}\to2_{2,0	}	$	&	313.280	&	28	&	20	&	0.381	\\
$	15_{4,12}\to15_{3,13	}	$	&	357.241	&	150	&	17	&	0.893	&	$	5_{2,4}\to4_{1,3	}	$	&	241.616	&	24	&	26	&	0.398	\\
$	17_{2,16}\to17_{1,17	}	$	&	273.753	&	149	&	23	&	0.378	&	$	4_{2,2}\to3_{1,3	}	$	&	235.152	&	19	&	27	&	0.211	\\
\hline
\end{tabular}
\end{center}
\tablefoot{* indicates a $\nu_2 = 1$ transition; $\dagger$ indicates a line overlap and hence an approximate line intensity.}
\end{table*}%

\begin{table}[tp]
\caption{SO and \so2 isotopologue observations towards R~Dor using APEX.}\label{isotopologues}
\begin{center}
\begin{tabular}{lcrrrc}
\hline\hline
 &		Transition		&	$\nu\;\;\;$	&	$E_\mathrm{up}$	& $\theta\;$&  $I_{mb}$ \\
&&[GHz]&[K]&[\arcsec]&[K $\kms$]\\
\hline	
$^{34}$SO	&	$	8_9\to7_8	$	&	339.857	&	77.3	&	18	&	0.26	\\
	&	$	7_7\to6_6	$	&	295.396	&	69.9	&	21	&	0.20	\\
	&	$	7_8\to6_7	$	&	298.258	&	61.1	&	21	&	0.42	\\
	&	$	6_6\to5_5	$	&	253.207	&	55.7	&	25	&	0.26	\\
	&	$	6_7\to5_6	$	&	256.878	&	46.7	&	24	&	0.38	\\
	&	$	5_6\to4_5	$	&	215.840	&	34.4	&	29	&	0.18	\\
$^{34}$SO$_2$	&	$	20_{0,20}\to19_{1,19}	$	&	357.102	&	184.6	&	17	&	0.25	\\
	&	$	17_{3,15}\to17_{2,16}	$	&	279.075	&	161.9	&	22	&	0.11	\\
	&	$	6_{3,3}\to5_{2,4}	$	&	362.158	&	40.6	&	17	&	0.24	\\
SO$^{18}$O	&	$	35_{10,26}\to36_{9,27}	$	&	288.482	&	786.3	&	22	&	0.19	\\
	&	$	19_{3,17}\to19_{2,18}	$	&	288.270	&	186.8	&	22	&	0.24	\\
	&	$	18_{0,18}\to17_{1,17}	$	&	303.476	&	143.4	&	21	&	0.18	\\
	&	$	17_{2,16}\to16_{2,15}	$	&	303.155	&	141.3	&	21	&	0.28	\\
	&	$	14_{4,10}\to14_{3,11}	$	&	344.874	&	129.6	&	18	&	0.24	\\
\hline
\end{tabular}
\end{center}
\end{table}%

\subsection{HIFI data}\label{hifi}

R~Dor, IK~Tau, R~Cas, TX~Cam and W~Hya were observed as part of the HIFISTARS Guaranteed Time Key Programme, using the \textit{Herschel}/HIFI instrument \citep{de-Graauw2010} to observe emission lines with high spectral resolution. The full results are presented in detail in \citet{Justtanont2012}. Since those data were published, there have been updates to the main beam efficiencies (Mueller et al., 2014\footnote{{\tt http://herschel.esac.esa.int/twiki/pub/Public/Hifi CalibrationWeb/HifiBeamReleaseNote\_Sep2014.pdf}}) and for this work we have re-reduced the HIFI data to take this into account \citep[using HIPE\footnote{http://www.cosmos.esa.int/web/herschel/data-processing-overview} version 12.1,][]{Ott2010}. We have also identified three additional \so2 lines that were not included in \citet{Justtanont2012}. The detected SO and \so2 HIFI lines are listed in Table \ref{hifiobs}. We note that no SO or \so2 lines were detected with HIFI in TX~Cam.

\begin{table*}[tp]
\caption{SO\down{2} and SO observations using HIFI, listed by molecule in order of descending energy of the upper level.}\label{hifiobs}
\begin{center}
\begin{tabular}{ccrrrccccc}
\hline\hline
Molecule	&		Transition		&	$\nu\;\;\;$	&	$E_\mathrm{up}$	& $\theta\;$&  IK Tau	&	R Dor	&	TX Cam	&	W Hya	&	R Cas\\
&&[GHz]&[K]&[\arcsec]&[K $\kms$]&[K $\kms$]&[K $\kms$]&[K $\kms$]&[K $\kms$]\\
\hline
SO	&	$	23_{24}\to22_{23}	$	&	988.618		&	575	&	21	&	-	&	-	&	-	&	0.76	&	-	\\
	&	$	15_{16}\to 14_{15}	$	&	645.875		&	253	&	33	&	-	&	1.5	&	$<$0.1	&	0.43	&	0.13	\\
	&	$	13_{13}\to 12_{12}	$	&	559.319		&	201	&	37	&	0.16	&	1.3	&	$<$0.08	&	0.55	&	0.16	\\
	&	$	13_{12}\to 12_{11}	$	&	558.087		&	195	&	37	&	0.44	&	1.4	&	$<$0.08	&	0.42	&	0.14	\\
	&	$	13_{14}\to 12_{13}	$	&	560.178		&	193	&	37	&	0.23	&	1.5	&	$<$0.08	&	0.58	&	0.26	\\
	\hline
SO$_2$	&	$	36_{2,34}\to 35_{3,33}	$	&*	661.510	&	630	&	32	&	-	&	0.37	&	-	&	-	&	-	\\
	&	$	37_{1,37}\to 36_{0,36}	$	&	659.421	&	609	&	32	&	-	&	0.68	&	-	&	0.12	&	-	\\
	&	$	36_{1,35}\to 35_{2,34}	$	&	658.632	&	606	&	32	&	0.16	&	0.61	&	-	&	0.35	&	0.26	\\
	&	$	32_{2,30}\to 31_{3,29}	$	&	571.532	&	505	&	36	&	-	&	\phantom{1} 0.20 $\dagger$	&	-	&	-	&	-	\\
	&	$	27_{7,21}\to 27_{6,22}	$	&*	657.885	&	468	&	32	&	-	&	0.21	&	-	&	-	&	-	\\
	&	$	32_{0,32}\to 31_{1,31}	$	&	571.553	&	459	&	36	&	-	&	\phantom{1} 0.39 $\dagger$	&	-	&	-	&	-	\\
	&	$	25_{7,19}\to 25_{6,20}	$	&*	659.338	&	419	&	32	&	-	&	0.25	&	-	&	-	&	-	\\
	&	$	24_{7,17}\to 24_{6,18}	$	&	659.898	&	396	&	32	&	-	&	\phantom{1} 0.23 $\dagger$	&	-	&	-	&	-	\\
	&	$	23_{7,17}\to 22_{6,16}	$	&	1102.115	&	374	&	19	&	-	&	0.93	&	-	&	-	&	-	\\
	&	$	22_{6,16}\to 22_{5,17}	$	&	557.283	&	321	&	37	&	-	&	0.29	&	-	&	0.14	&	-	\\
	&	$	15_{9,7}\to 14_{8,6}	$	&	1151.852	&	309	&	19	&	-	&	1.5\phantom{0}	&	-	&	-	&	-	\\
	&	$	21_{6,16}\to 21_{5,17}	$	&	558.391	&	301	&	37	&	-	&	0.22	&	-	&	0.20	&	-	\\
	&	$	13_{9,5}\to 12_{8,4}	$	&	1113.506	&	282	&	19	&	-	&	0.76	&	-	&	-	&	-	\\
	&	$	20_{6,14}\to 20_{5,15}	$	&	558.812	&	282	&	37	&	-	&	0.14	&	-	&	-	&	-	\\
	&	$	18_{6,12}\to 18_{5,13}	$	&	559.882	&	246	&	37	&	-	&	0.19	&	-	&	0.12	&	-	\\
	&	$	16_{6,10}\to 16_{5,11}	$	&	560.613	&	213	&	37	&	-	&	0.22	&	-	&	-	&	-	\\
\hline																				
\end{tabular}
\end{center}
\tablefoot{* indicates a line not listed in \citet{Justtanont2012}; $\dagger$ indicates a line blend and hence an approximate integrated intensity.}
\end{table*}%

\subsection{Archival data}

To supplement the HIFI data for IK~Tau, R~Cas, W~Hya, and TX~Cam, we have used observations found in the literature. These are listed in Table \ref{oldobs}. As the older data generally covers lower-energy transitions than those observed by HIFI, we are better able to constrain our models over a larger energy range. This is particularly important for R~Cas, IK~Tau and TX~Cam where the HIFI lines (or non-detections in the case of TX~Cam) are clustered close together energetically.

\begin{table*}[tp]
\caption{Archival observations of SO and \so2 towards IK~Tau, TX~Cam, R~Cas, and W~Hya.}\label{oldobs}
\begin{center}
\begin{tabular}{llccccccl}
\hline\hline
Source& Molecule		& Transition		&	$\nu$	&	$E_\mathrm{up}$	&Telescope& $\theta$&  $I_{mb}$ & Reference\\
&&&[GHz]&[K]&&[\arcsec]&[K $\kms$]\\
\hline	
IK~Tau&SO	&	$	8_8 \to 7_7	$	&	344.310	&	88	&	APEX	&	18	&	2.7	&	\cite{Kim2010}	\\		
&	&	$	7_7 \to 6_6	$	&	301.286	&	71	&	APEX	&	20	&	0.89	&	\cite{Kim2010}	\\		
&	&	$	5_6\to4_5	$	&	219.949	&	35	&	NRAO	&	30	&	6.5	&	\cite{Sahai1992}	\\
&	&	$	2_2 \to 1_1	$	&	86.094	&	19	&	IRAM	&	27	&	0.65	&	\cite{Omont1993}	\\
&	&	$	3_4\to 2_3	$	&	138.179	&	16	&	IRAM	&	17	&	13.5	&	\cite{Sahai1992}	\\
&	&	$	2_3 \to 1_2	$	&	99.300	&	9	&	OSO	&	37.5	&	4.2	&	\cite{Sahai1992}	\\
&	&	$	2_3 \to 1_2	$	&	99.300	&	9	&	OSO	&	37.5	&	3.6	&	\cite{Olofsson1998}	\\
&SO$_2$	&	$	17_{1,17} \to 16_{0,16}	$	&	313.660	&	136	&	APEX	&	20	&	11.3	&	\cite{Kim2010}	\\
&	&	$	14_{4,10} \to 14_{3,11}	$	&	351.873	&	136	&	APEX	&	18	&	0.55	&	\cite{Kim2010}	\\
&	&	$	14_{3,11} \to 14_{2,12}	$	&	226.300	&	119	&	IRAM	&	10.5	&	1.0	&	\cite{Decin2010}	\\
&	&	$	13_{2,12} \to 12_{1,11}	$	&	345.338	&	93	&	APEX	&	18	&	6.3	&	\cite{Kim2010}	\\
&	&	$	10_{1,9} \to 10_{0,10}	$	&	104.239	&	55	&	IRAM	&	24	&	1.83	&	\cite{Omont1993}	\\
&	&	$	10_{0,10} \to 9_{1,9}	$	&	160.828	&	50	&	IRAM	&	15	&	8.6	&	\cite{Omont1993}	\\
&	&	$	5_{3,3} \to 5_{2,4}	$	&	256.247	&	36	&	IRAM	&	9.5	&	3.4	&	\cite{Decin2010}	\\
&	&	$	5_{3,3} \to 4_{2,2}	$	&	351.257	&	36	&	APEX	&	18	&	1.4	&	\cite{Kim2010}	\\
&	&	$	4_{3,1} \to 4_{2,2}	$	&	255.553	&	31	&	IRAM	&	9.5	&	3.2	&	\cite{Decin2010}	\\
&	&	$	4_{3,1} \to 3_{2,2}	$	&	332.505	&	31	&	APEX	&	19	&	1.4	&	\cite{Kim2010}	\\
&	&	$	3_{3,1} \to 3_{2,2}	$	&	255.958	&	28	&	IRAM	&	9.5	&	2.2	&	\cite{Decin2010}	\\
&	&	$	3_{3,1} \to 2_{2,0}	$	&	313.279	&	28	&	APEX	&	20	&	2.2	&	\cite{Kim2010}	\\
&	&	$	3_{1,3} \to 2_{0,2}	$	&	104.029	&	8	&	IRAM	&	24	&	1.78	&	\cite{Omont1993}	\\
TX~Cam &SO	&	$	5_6\to4_5	$	&	219.949	&	35	&	IRAM	&	13	&	7.0	&	\cite{Bujarrabal1994}	\\
	&&	$	5_6\to4_5	$	&	219.949	&	35	&	NRAO	&	30	&	1.8	&	\cite{Sahai1992}	\\
&	&	$	2_3\to1_2	$	&	99.300	&	9	&	OSO	&	37.5	&	2.9	&	\cite{Sahai1992}	\\
&	&	$	2_3 \to 1_2	$	&	99.300	&	9	&	OSO	&	37.5	&	2.1	&	\cite{Olofsson1998}	\\
W~Hya &SO &	$	5_5\to4_4	$	&	215.221	&	31	&	SMA	&	1.5	&	\phantom{*}8.2*	&	\cite{Vlemmings2011}	\\	
	&&	$	2_3\to1_2	$	&	99.300	&	9	&	SEST	&	51	&	0.2	&	\cite{Olofsson1998}	\\
R~Cas &SO	
&	$	5_6\to4_5	$	&	219.949	&	35	&	NRAO	&	30	&	2.0	&	\cite{Sahai1992}	\\
&	&	$	2_3\to1_2	$	&	99.300	&	9	&	OSO	&	37.5	&	1.4	&	\cite{Sahai1992}	\\
&	&	$	2_3\to1_2	$	&	99.300	&	9	&	OSO	&	37.5	&	1.7	&	\cite{Olofsson1998}	\\
&SO$_2$	&	$	3_{1,3}\to 2_{0,2}	$	&	104.029	&	7.7	&	IRAM	&	24	&	0.81	&	\cite{Guilloteau1986}	\\
\hline
\end{tabular}
\tablefoot{* indicates value given is the flux in Jy $\kms$ not the main beam integrated intensity.}
\end{center}
\end{table*}%

\section{Modelling}

\subsection{Modelling procedure}

We perform detailed radiative transfer modelling of the molecular emission lines using an accelerated lambda iteration method code (ALI), which has been previously described and implemented by e.g. \cite{Maercker2008,Schoier2011,Danilovich2014}. ALI is particularly useful in this work as it is able to take into account extensive descriptions of molecular properties --- such as large numbers of energy levels and transitions --- while still fully solving the statistical equilibrium equations and taking temperature and velocity profiles into account.

We assume a smoothly expanding spherical CSE produced by a constant mass-loss rate. The molecules are located in this CSE until they eventually become photodissociated. They are excited by collisions with \h2 molecules and through radiation from the star, the dust, and the cosmic microwave background. ALI input parameters such as the kinetic temperature distribution, dust temperature, and dust optical depth, are taken from CO modelling and, where applicable, are listed in Table \ref{stellarprop}. For R~Dor, R~Cas, IK~Tau, and TX~Cam Maercker et al. (\textit{in prep.}) performed detailed radiative transfer modelling of the CO and \h2O lines and we use their results in our modelling. We based our CO model of W~Hya on the results of \cite{Khouri2014}, but generated a CO model using the same code as in Maercker et al. (\textit{in prep.}) for consistency between the stars.

We calculated the best fit model for each star and molecule using a $\chi^2$ statistic, which we define as
\begin{equation}
\chi^2 = \sum^N_{i=1} \frac{(I_{\mathrm{mod},i} - I_{\mathrm{obs},i})^2}{\sigma_i^2}
\end{equation}
where $I$ is the integrated line intensity, $\sigma$ is the uncertainty in the observations, and $N$ is the number of lines being modelled. We also calculate a reduced $\chi^2$ value such that $\chi^2_\mathrm{red} = \chi^2/(N-p)$ where $p$ is the number of free parameters.

After testing both centrally-peaked and shell-like abundance distributions, we came to the conclusion that the best radial abundance distribution profiles for both SO and \so2 in R~Dor and W~Hya were Gaussian profiles of the form
\begin{equation}\label{abundanceeq}
f = f_p \exp\left( -\left( \frac{r}{R_e} \right)^2\right)
\end{equation}
where $f_p$ is the peak abundance at the inner radius, and $R_e$ is the $e$-folding radius, the radius at which the abundance has dropped by a factor of $1/e$.

In the cases of IK~Tau and R~Cas, we found that a shell model was a better fit to the observed SO lines. As such, we modelled IK~Tau and R~Cas assuming a Gaussian shell for the abundance distribution of the form
\begin{equation}\label{shellabundanceeq}
f = f_p \exp \left( -4 \frac{(r-R_{p})^2}{R_w^2}\right)
\end{equation}
where $f_p$ is the peak abundance, $R_{p}$ is the radial distance of the peak of the distribution from the centre of the star, and $R_w$, is the width of the shell at the $e$-folding radius. Using a shell distribution for both IK~Tau and R~Cas rather than a central Gaussian distribution significantly improved the $\chi^2$ fits of the models. 

Similarly, we can firmly rule out a centrally-peaked model for TX~Cam, as for such a model to fit the archival data we would expect conclusive detections in the HIFI data. As the undetected HIFI lines are of higher energy than the archival detections, a lower abundance in the inner regions of the CSE is expected, than in the outer regions, which points to a shell-like abundance distribution. 

\subsubsection{SO}

For the radiative transfer analysis of SO we include 182 rotational energy levels, denoted $N_J$, up to $N=30$ in the ground and first excited vibrational states. There are 907 radiative transitions. These include pure rotational transitions in the $X^3\Sigma^-$ $v=0$ and $v=1$ states as well as the $v=1\to0$ rovibrational lines. There are 8629 collisional transitions including collisions between all rotational states within a vibrational state, as well as between vibrational states. The rotational energy levels, transition frequencies, and A-values have been adapted directly from the CDMS \citep{Muller2001,Muller2005}. The infrared line list has been computed directly from the rotational levels with the band-head frequency adjusted to give very good agreement with the line positions measured by \cite{Burkholder1987}. 
The vibration-rotation line strengths have been computed in intermediate coupling and have been verified by comparison with the pure rotational line strengths in the CDMS tables. The vibration-rotation transition dipole moment has been taken to be 0.08843 Debye, which yields inverse lifetimes of $A_\mathrm{tot}=3.6$ s$^{-1}$ for the $v=1\to0$ band as computed by \cite{Peterson1990}. 
The collisional rate coefficients for pure rotational transitions were adapted from the He-SO rates computed by \cite{Lique2006} with mass-scaling to H$_2$ as in the smaller data set in the LAMDA database \citep{Schoier2005}. Rates for transitions within $v=1$ were assumed to be identical to those within $v=0$. Crude collision rates for $v=1\to 0$ were scaled in proportion to normalised radiative line strengths for electric-dipole-allowed transitions, with the largest values of the order of \mbox{$1\times 10^{-11}$ cm$^3$ s$^{-1}$}.

\begin{figure*}[t]
\includegraphics[width=\textwidth]{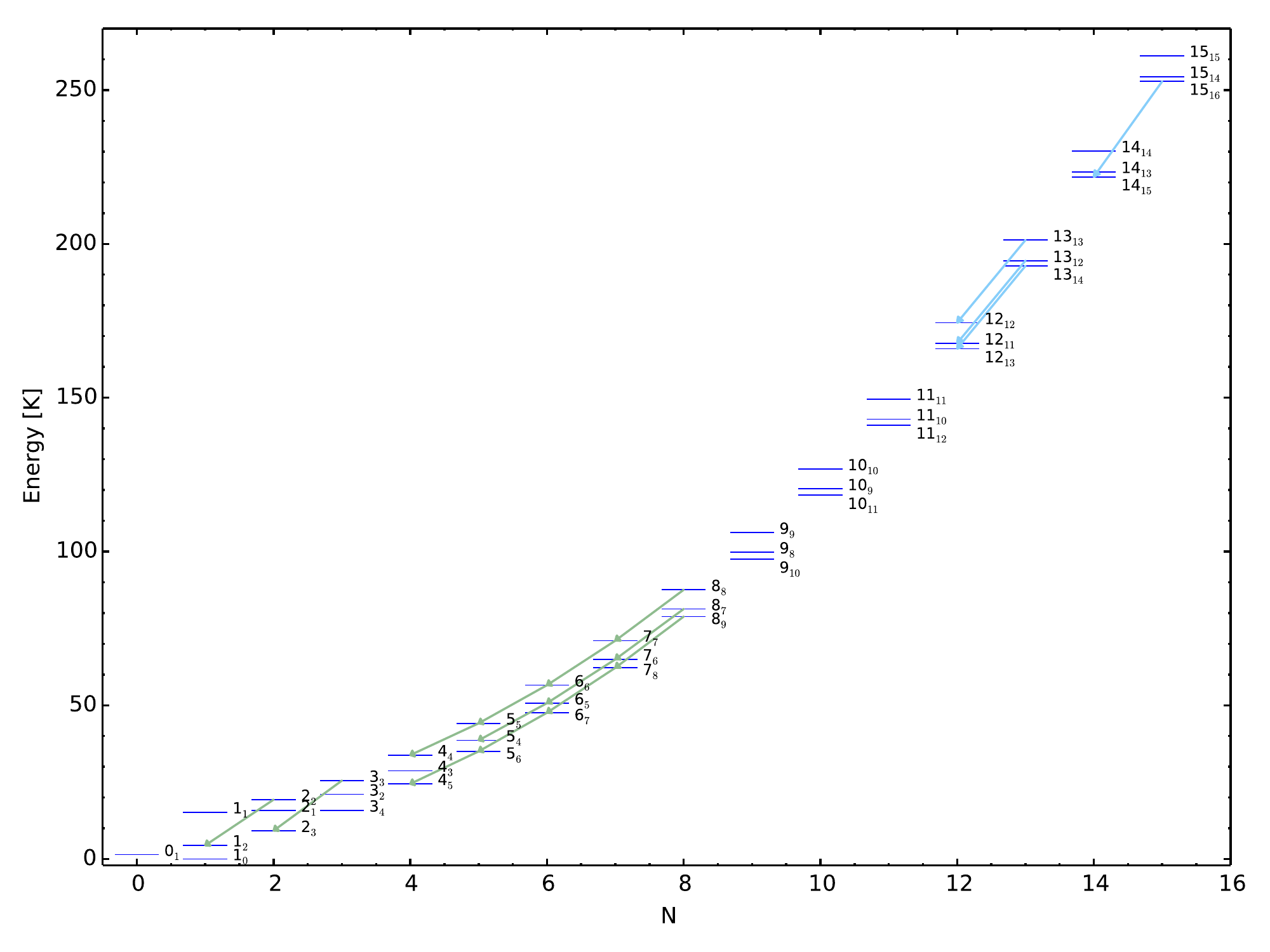}
\caption{SO energy level diagram with levels labelled using the $N_J$ convention. Transitions detected with HIFI and APEX towards R~Dor are indicated in light blue and green, respectively.}
\label{SOeld}
\end{figure*}

In Fig. \ref{SOeld} we include an energy level diagram for SO. Here we have indicated all the transitions of SO detected towards R~Dor with HIFI and APEX. These cover most of the transitions also detected in IK~Tau, R~Cas, W~Hya, and TX~Cam.

For the purposes of modelling the \up{34}SO emission in R~Dor, we used a simpler molecular description than that for \up{32}SO, including the rotational energy levels up to $N=30$, corresponding to those included for \up{32}SO, but only in the ground vibrational state. When adopting the corresponding simpler molecular description for \up{32}SO in the case of R~Dor specifically, we found that the final best fit model only shifted by a few percent between the detailed and simpler descriptions, justifying this approach for \up{34}SO. There was, however, some shift in final model for the other, especially higher mass-loss rate, stars when changing between the detailed and simpler molecular description for SO.

\subsubsection{SO$_2$}\label{so2intro}

Our radiative transfer analysis of \so2 includes 2600 energy levels, denoted $J_{Ka,Kc}$, across the ground vibrational state and the $\nu_1 = 1$ (8.7~$\mic$), $\nu_2 = 1$ (19.3~$\mic$) and $\nu_3 = 1$ (7.3~$\mic$) vibrationally excited states. Levels with energies up to 4830 K and $J = 38$ were included. This gives 15243 radiative transitions, with spectroscopic data taken from the HITRAN database \citep{Rothman2013}, and 15244 collisional transitions. 
The collision rates in the literature for SO$_2$ are inadequate for our purposes. \cite{Green1995} calculated rate coefficients for He-SO$_2$ collisions in the infinite-order sudden approximation for the lowest 50 rotational levels (up to 100 K excitation energy and $J\leq 13$ only). \cite{Cernicharo2011} published rates for H$_2$-SO$_2$ collisions for the lowest 31 rotational levels at low temperatures, 5 to 30 K. The rates for H$_2$ impact were found to be approximately 10 times higher than corresponding rates for He impact. For the much larger number of states in our models, we adopted instead a set of crude collision rates in which the downward rate coefficient is proportional to the radiative line strength and normalised to a total collisional quenching rate of $2.0\times 10^{-10}$ cm$^3$ s$^{-1}$, which is comparable to the highest collision rates found by \cite{Cernicharo2011}. We tested the impact of the chosen collisional transition rates by multiplying the rates, in stages, by up to two orders of magnitude in both directions. We find that such drastic changes had only a very small and barely detectable effect on the resulting models. Hence we conclude that \so2 excitation is radiatively dominated with the choice of collisional transition rates playing only a minor role in the radiative transfer modelling.

\begin{figure*}[t!]
\includegraphics[height=0.955\textwidth, angle=270]{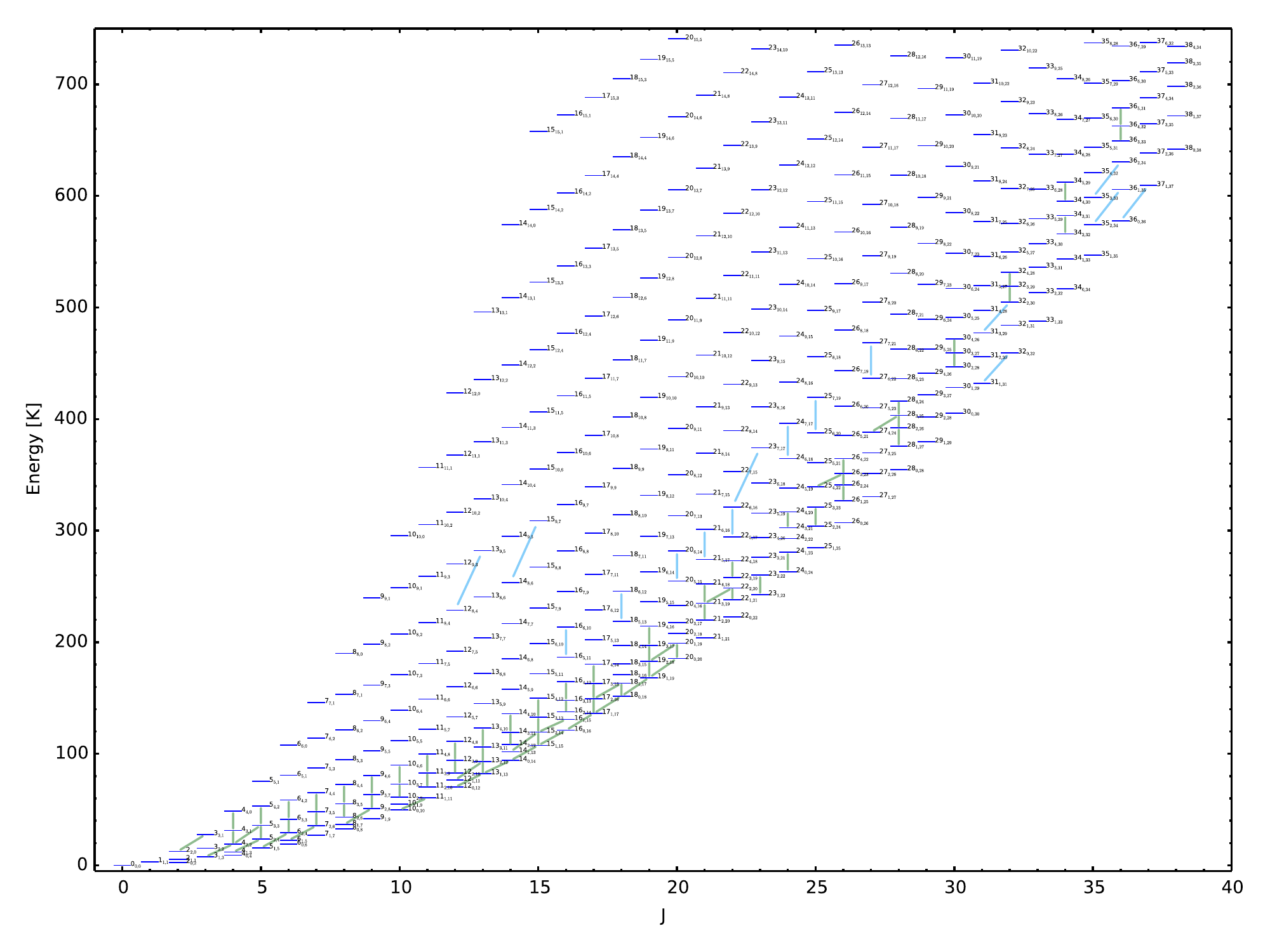}
\caption{\so2 energy level diagram. Levels are labelled $J_{K_a,K_c}$. Transitions detected with HIFI and APEX towards R~Dor are indicated in light blue and green, respectively.}
\label{SO2eld}
\end{figure*}

In Fig. \ref{SO2eld} we include an energy level diagram for \so2. Here we have indicated all transitions of \so2 detected towards R Dor with HIFI and APEX. As can be seen, \so2 has many close energy levels. This leads to a multitude of overlapping transitions, especially in AGB winds with typical expansion velocities of 5 -- 25~$\kms$. The number of levels, transitions and overlaps presents some computational challenges, especially when it comes to fully taking overlapping lines into account or running exhaustive grids. To reduce running time to a manageable interval we restrict the overlaps so that only those within the sampled frequency range, between 200 -- 1200 GHz, are included. This reduced the total number of overlaps by more than an order of magnitude (down to 441 lines participating in overlaps), hence decreasing running time and memory usage. This, however, neglects possible overlaps in pumping lines, which could have a significant effect on some of the lines included in the model. From what tests we were able to run we believe that the overall impact of these omitted lines is relatively minor.

\subsection{R~Dor}

Our modelling is based on the radiative transfer results obtained by Maercker et al. (in prep) for CO in R~Dor. They find a mass-loss rate of $\dot{M} = 1.6\e{-7}~\spy$ and an expansion velocity of $\upsilon_\infty = 5.7~\kms$. They also find an expansion velocity profile following
\begin{equation}
\upsilon(r) = \upsilon_\mathrm{min} + (\upsilon_\infty - \upsilon_\mathrm{min}) \left( 1 - \frac{R_\mathrm{in}}{r}\right)^\beta
\end{equation}
where $\upsilon_\mathrm{min} = 3~\kms$ is taken to be the sound speed at $R_\mathrm{in} = 1.6\e{14}$ cm, the dust condensation radius. $\beta = 1.5$ governs the acceleration of the gas, having the most significant impact in the inner regions, and hence on the excitation of the higher-energy lines. The other relevant stellar properties of R~Dor are listed in Table \ref{stellarprop}.

\begin{table}[tp]
\caption{Stellar properties and input from CO models.}\label{stellarprop}
\begin{center}
\begin{tabular}{lccccc}
\hline\hline
	&	IK~Tau	&	R~Dor	&	 TX~Cam 	&	  W~Hya 	&	 R~Cas 	\\
\hline											
$L_*$ [L$_\odot$] 	&	7700	&	6500	&	8600	&	5400	&	8700	\\
$D$ [pc] 	&	265	&	59	&	380	&	78	&	176	\\
$\upsilon_\mathrm{LSR}$ [$\kms$] 	&	34	&	7	&	11.4	&	40.5	&	25	\\
$T_*$ [K] 	&	2100	&	2400	&	2400	&	2500	&	3000	\\
$R_\mathrm{in}$ [$10^{14}$ cm]	&	2.0	&	1.9	&	2.2	&	2.0	&	2.2	\\
$\tau_{10}$ 	&	1.0	&	0.03	&	0.4	&	0.07	&	0.09	\\
$\dot{M}$ [$10^{-7}\spy$] 	&	 $50$ 	&	 $1.6$ 	&	40	&	 $1$	&	 $8$ 	\\
$\upsilon_\infty$ [$\kms$] 	&	17.5	&	5.7	&	17.5	&	7.5	&	10.5	\\
$\beta$ 	&	1.5	&	1.5	&	2.0	&	5.0	&	2.5	\\
\hline
\end{tabular}
\end{center}
\tablefoot{$\tau_{10}$ is the dust optical depth at $10~\mic$.}
\end{table}%

\subsubsection{SO results}

\begin{figure*}[t]
\center{
\includegraphics[width=0.49\textwidth]{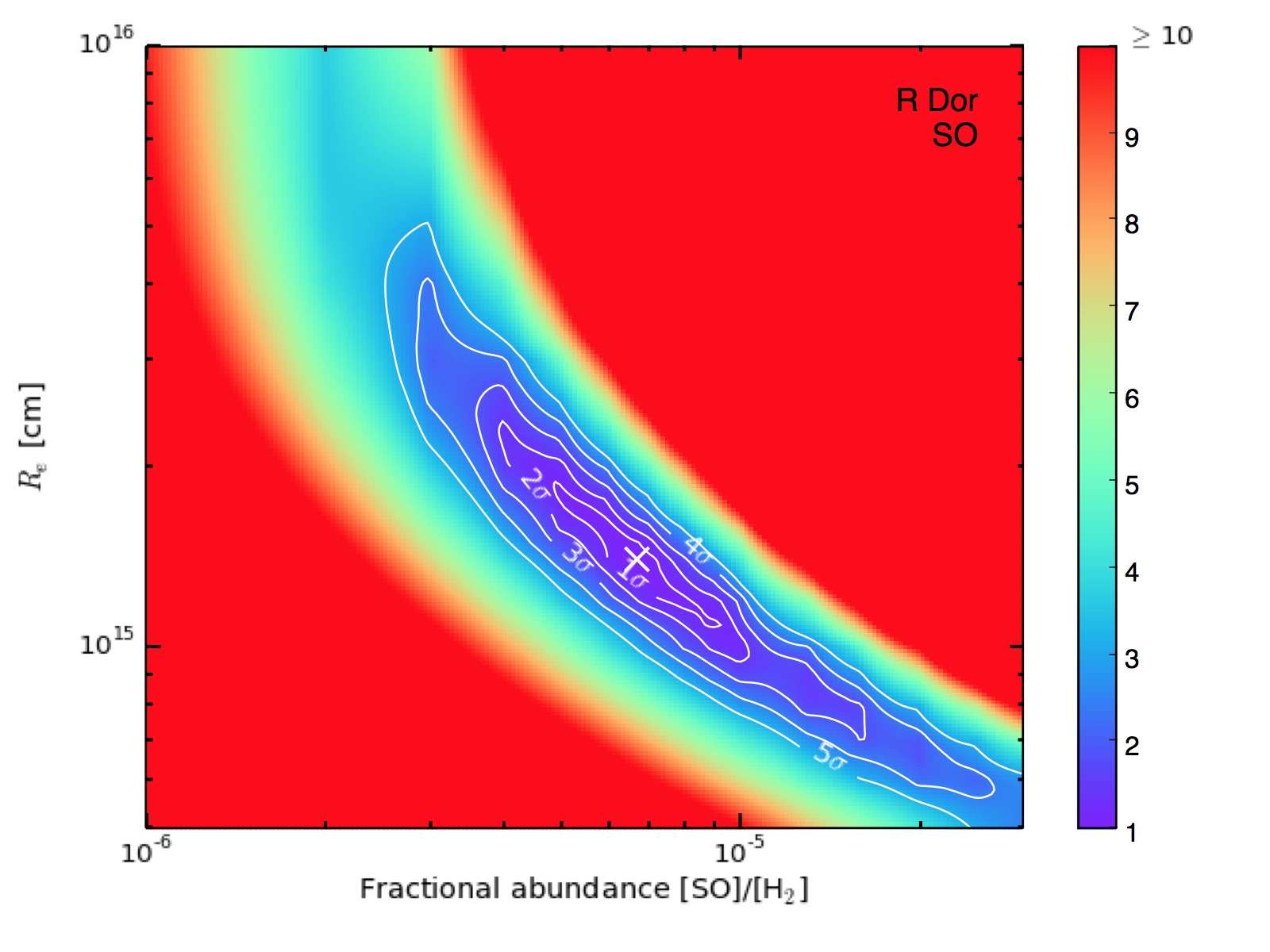}
\includegraphics[width=0.49\textwidth]{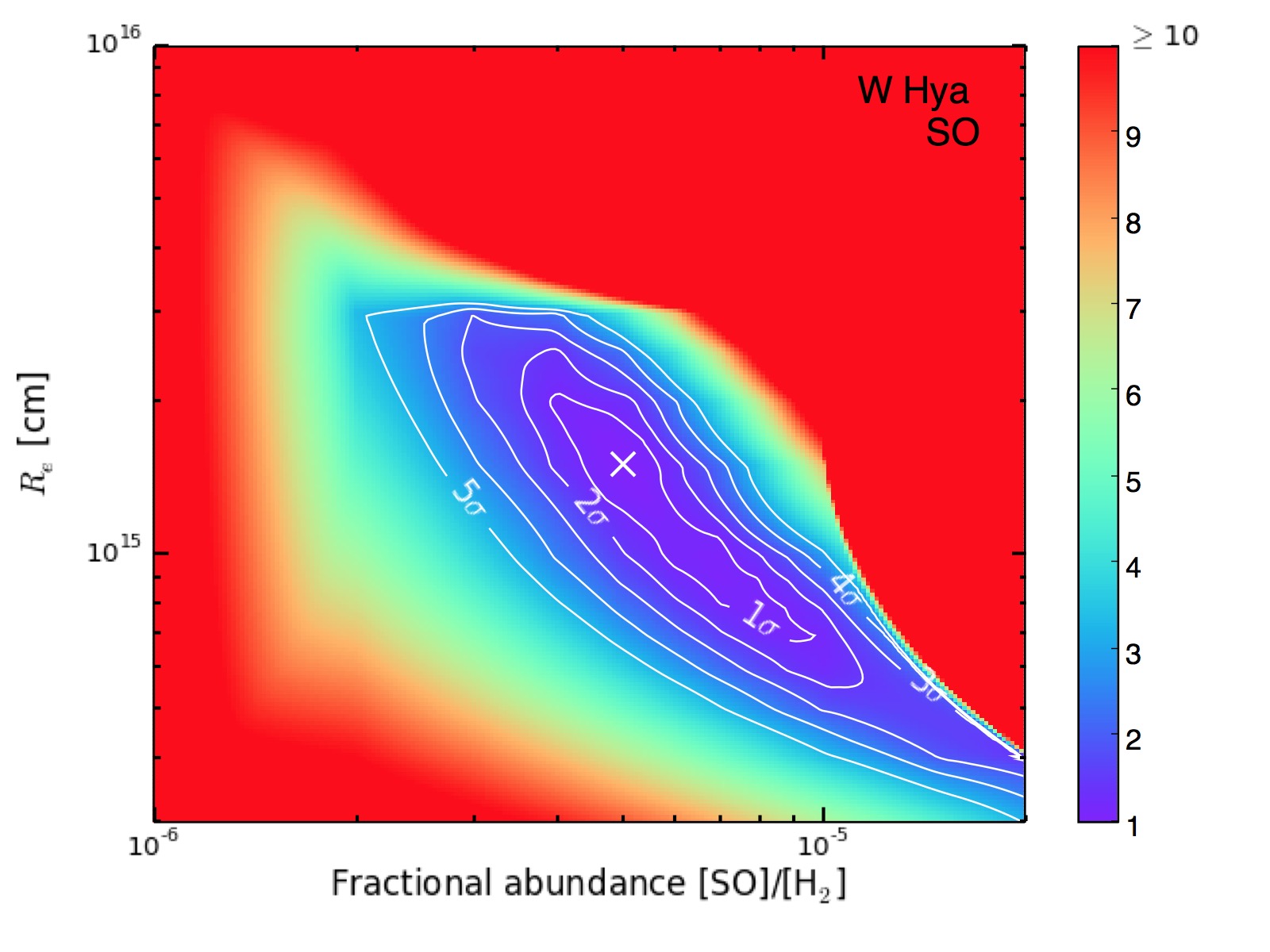}
\includegraphics[width=0.49\textwidth]{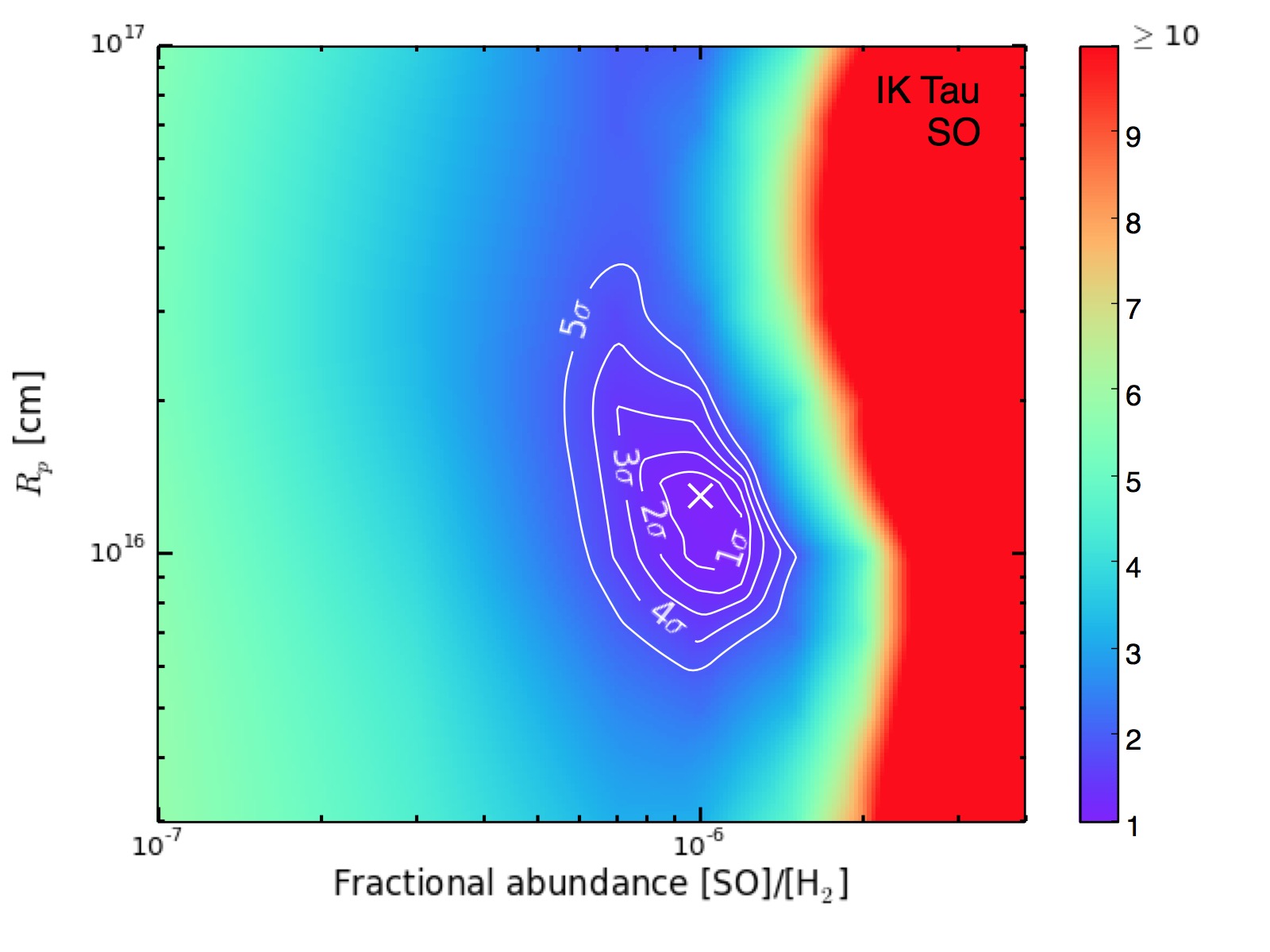}
\includegraphics[width=0.49\textwidth]{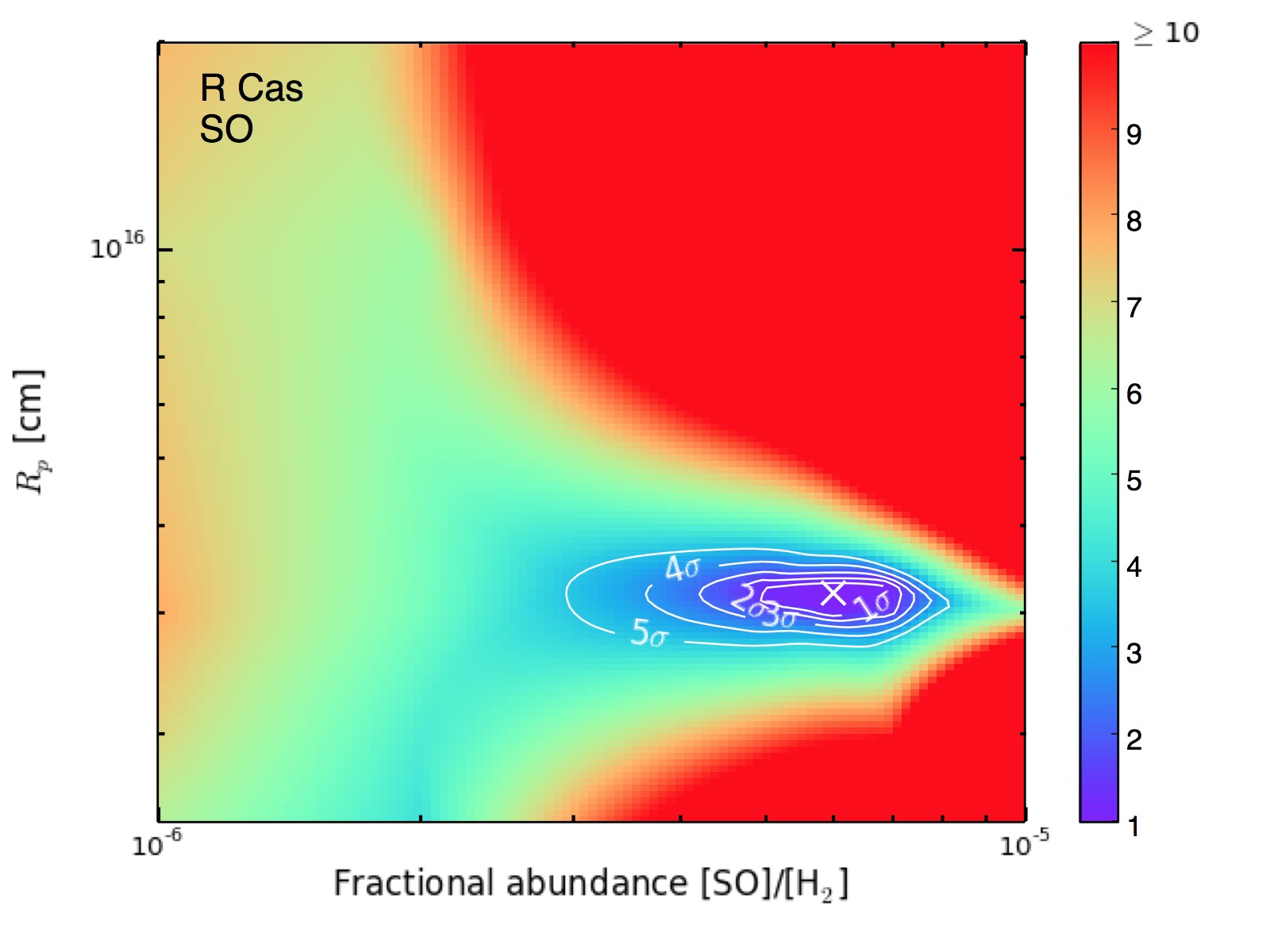}}
\caption{SO $\chi^2$ plots for R~Dor, W~Hya, IK~Tau and R~Cas. The contours show the confidence intervals and the shading represents the $\chi^2_\mathrm{red}$ value for the corresponding model, with the colour-bar indicating multiples of the minimum $\chi^2_\mathrm{red}$ value. The white cross indicates our best-fit model (see Table \ref{resultstab}). For IK~Tau, the slice for which $R_w = 1.8 R_p$ is shown. For R~Cas, the slice for which $R_w = 1.0 R_p$ is shown.}
\label{SOchi2}
\end{figure*}

\begin{figure*}[t!]
\begin{center}
\includegraphics[width=\textwidth]{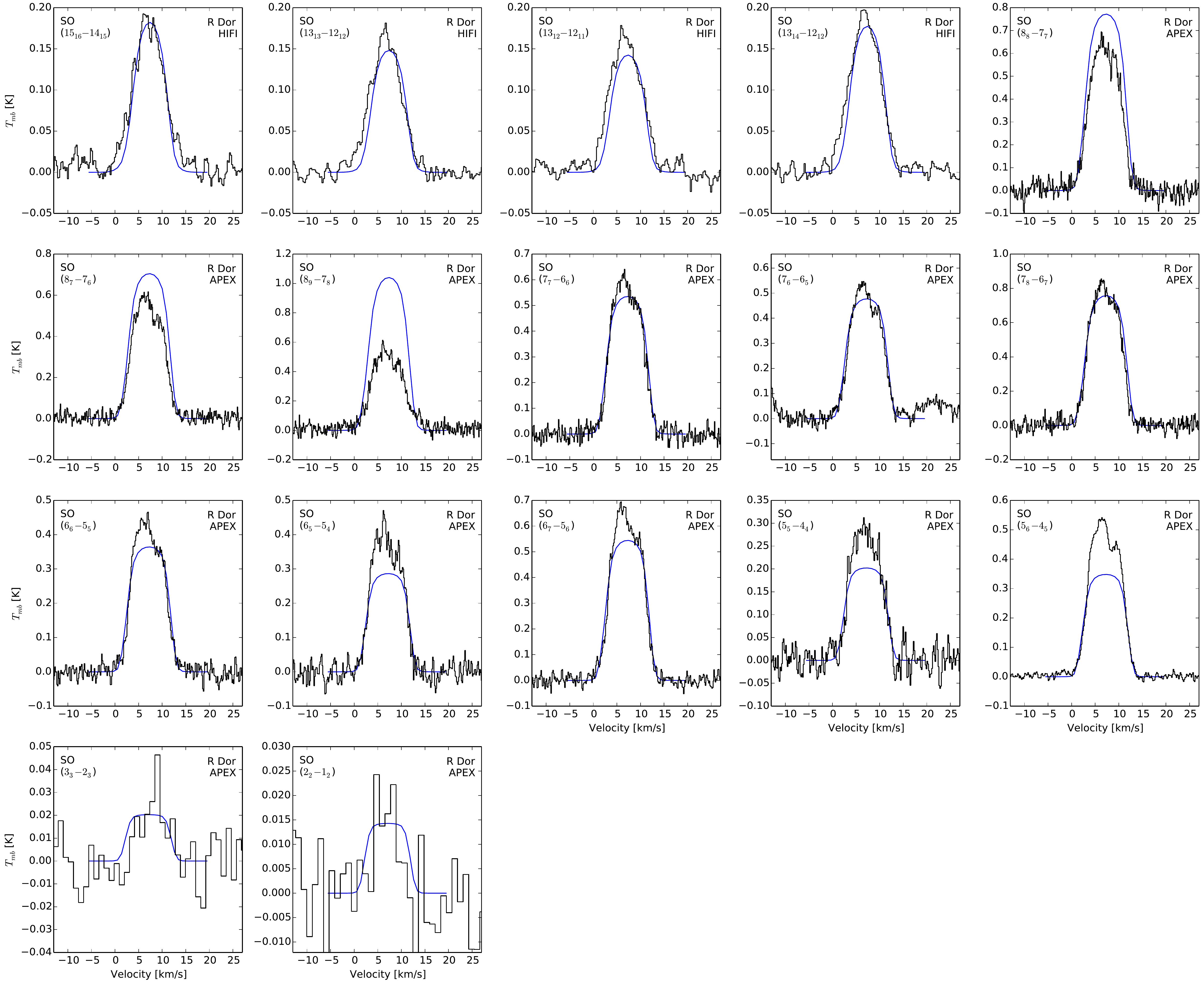}
\caption{SO models (blue lines) and observations (black histograms) for R~Dor.}
\label{rdorsoresults}
\end{center}
\end{figure*}

\begin{figure}[t]
\includegraphics[width=0.24\textwidth]{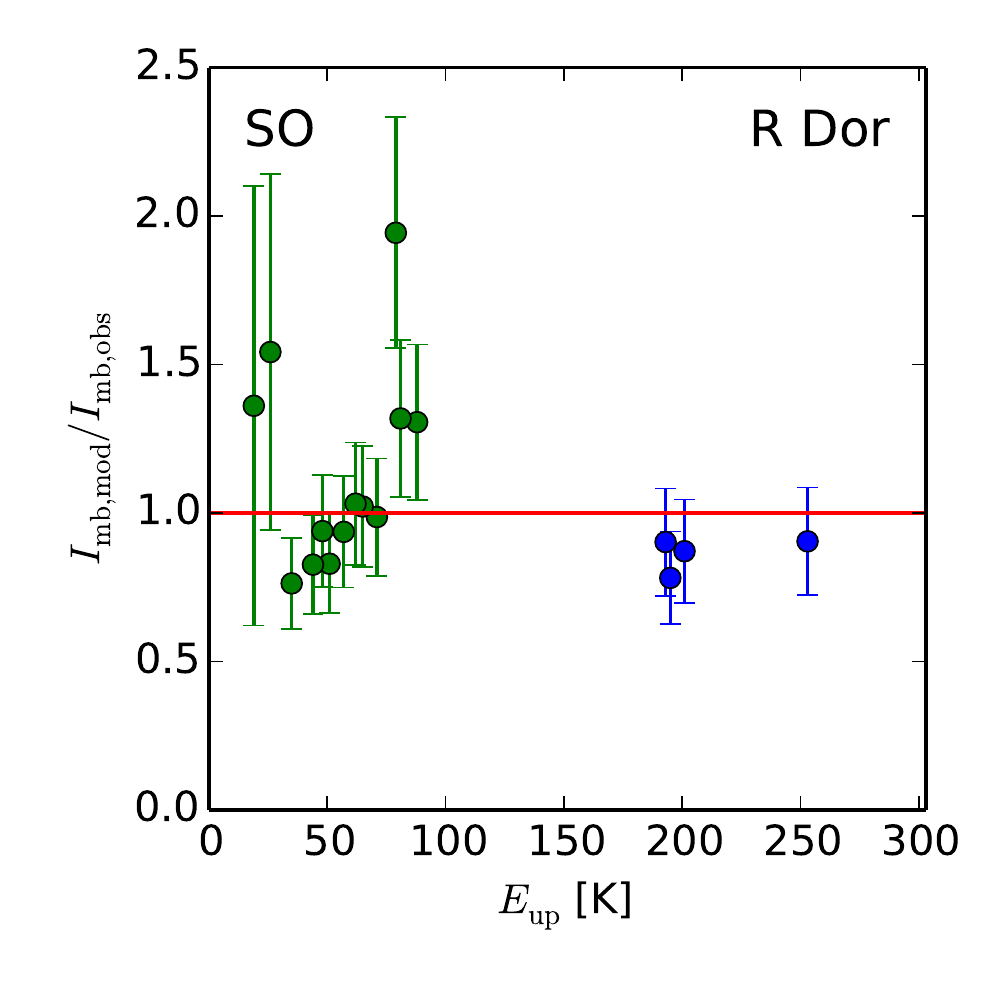}
\includegraphics[width=0.24\textwidth]{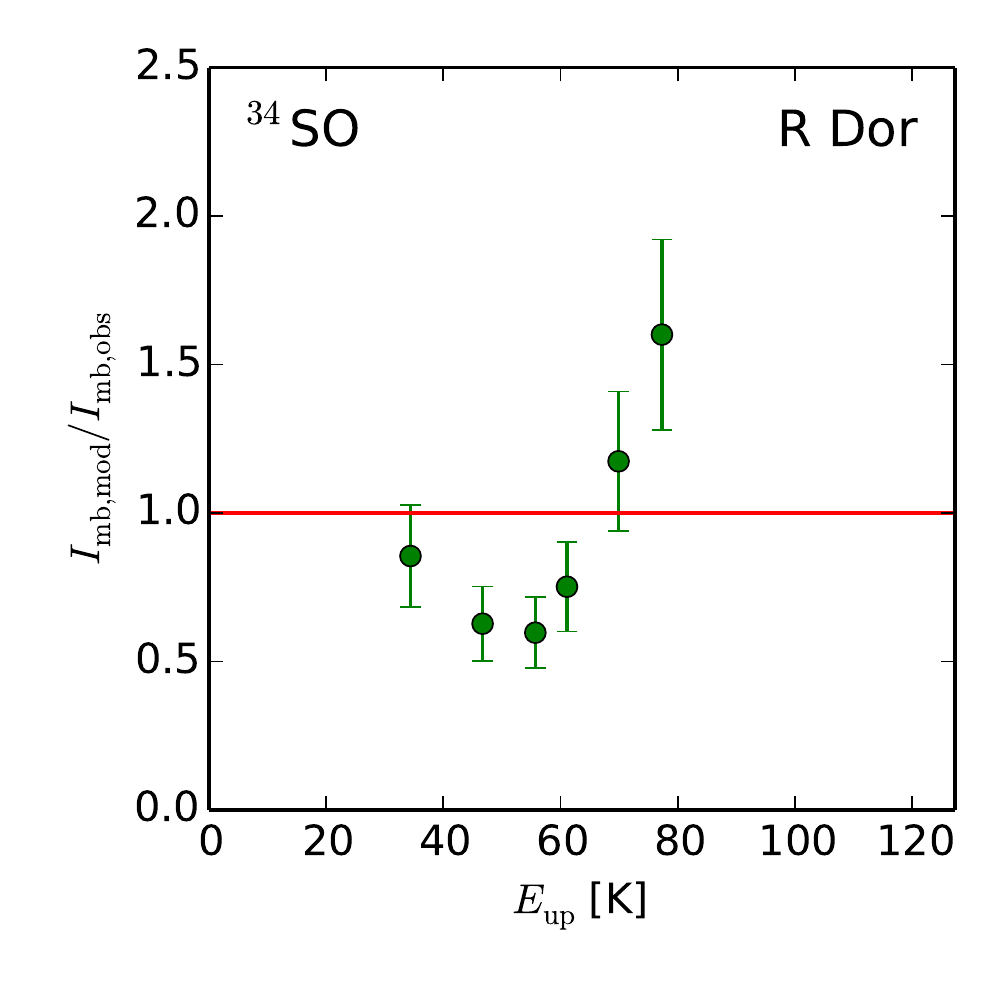}
\includegraphics[width=0.24\textwidth]{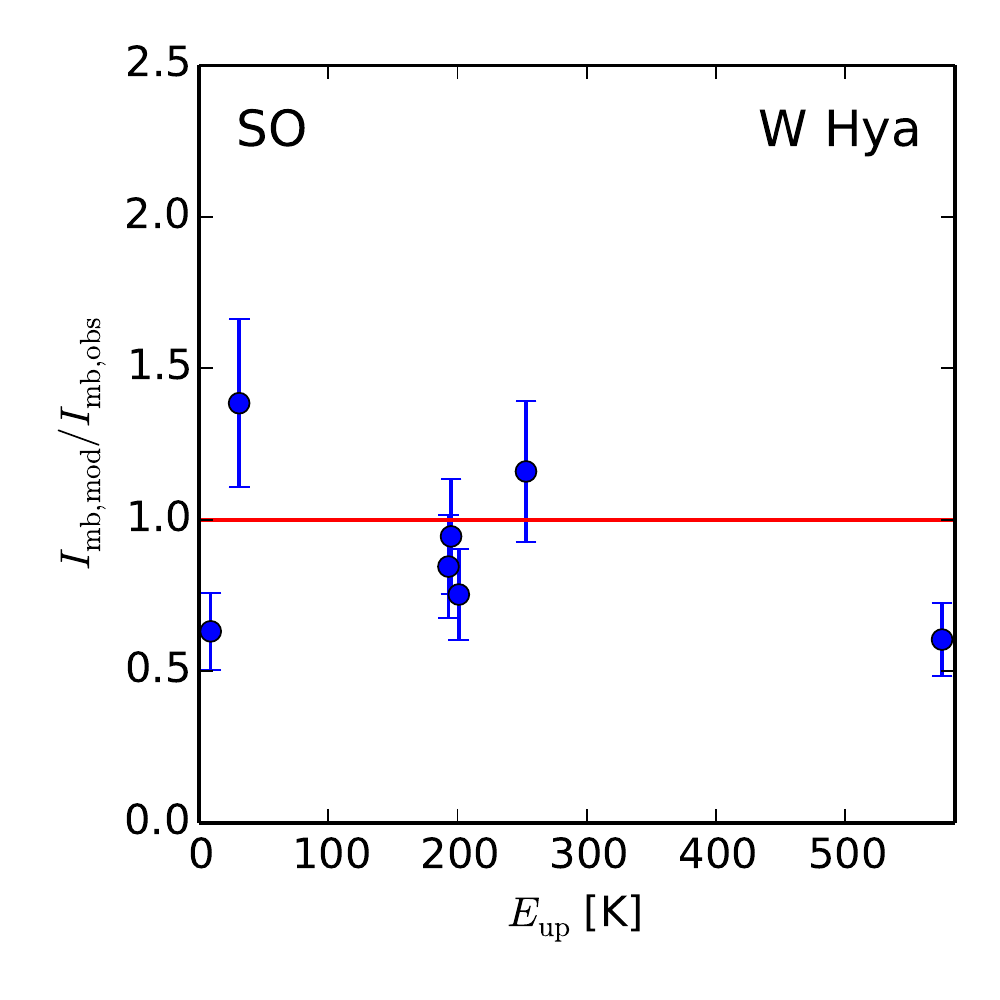}
\includegraphics[width=0.24\textwidth]{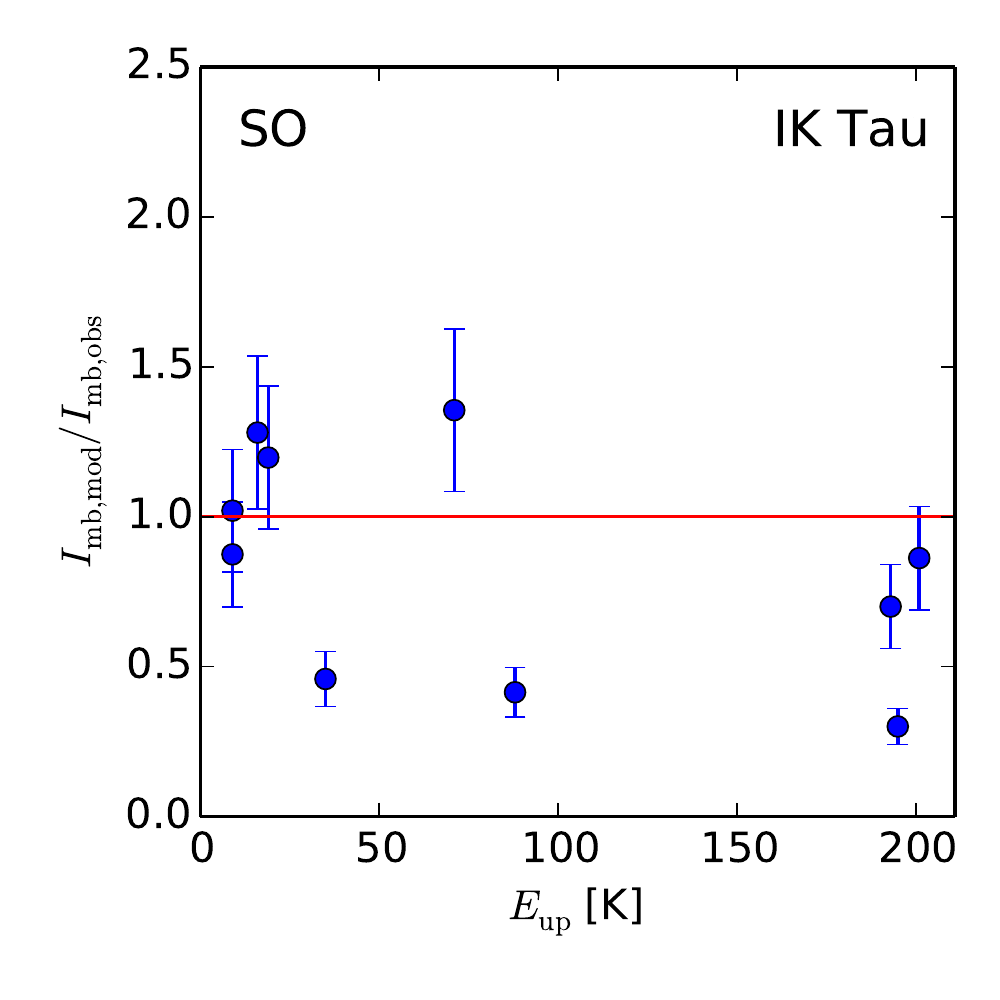}
\includegraphics[width=0.24\textwidth]{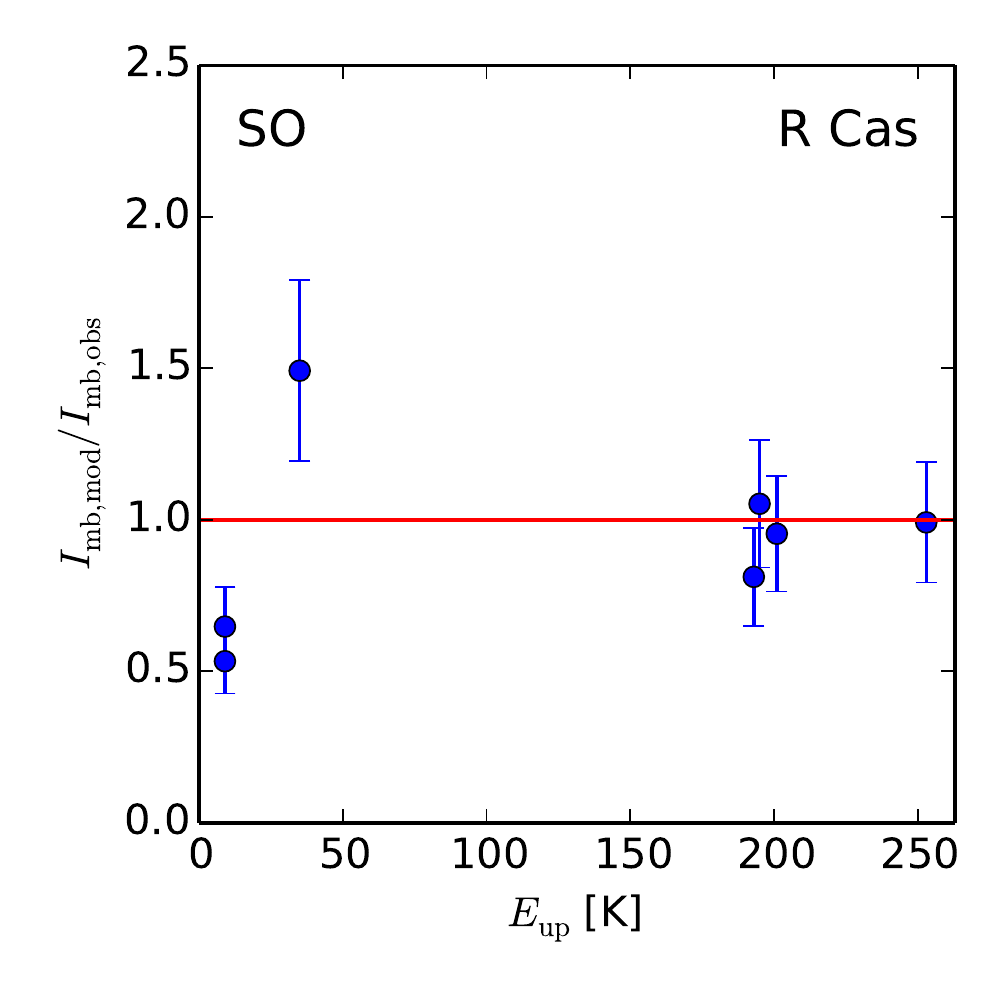}
\includegraphics[width=0.24\textwidth]{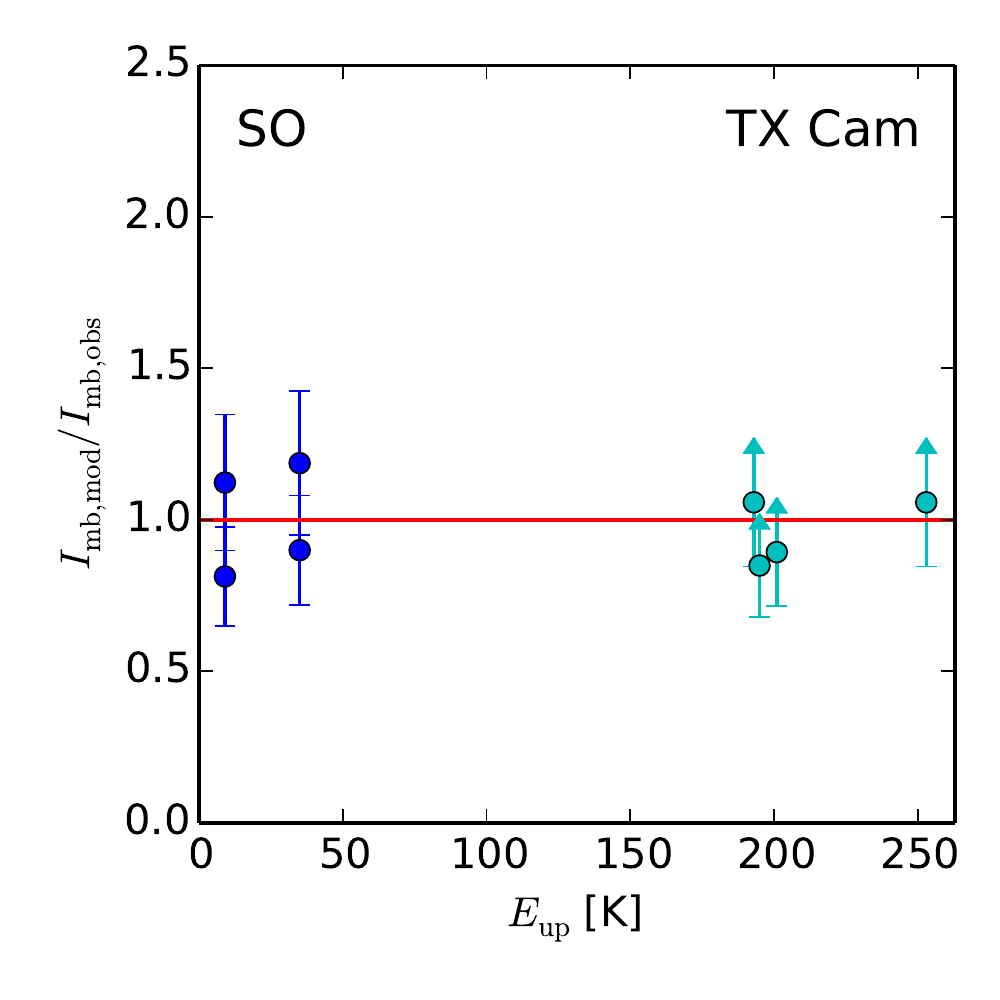}
\caption{SO goodness of fit plots for R~Dor, R~Cas, IK~Tau, and W~Hya. New HIFI lines as well as archival data listed in Table \ref{oldobs} are included. The green points in the R~Dor plots represent the observations from the APEX spectral survey. Undetected HIFI lines are shown as cyan points with arrows, in this case representing lower limits because the vertical axis is the ratio of model integrated intensities to observed integrated intensities or the upper limits thereof.}
\label{SOfits}
\end{figure}

To model the 17 SO lines detected towards R~Dor with APEX and HIFI, we set up a grid sampling different SO abundances and $e$-folding radii. We then ran a finer grid with steps of $0.1\e{-6}$ in abundance and $0.1\e{15}$~cm in $e$-folding radius to find the best possible fit to the observations. The results of our $\chi^2$ analysis can be seen in Fig. \ref{SOchi2}. Our resulting best-fit model, with $\chi^2_\mathrm{red} = 0.90$, has a peak SO abundance relative to \h2 of $(6.7\pm0.9)\e{-6}$ and $e$-folding radius $R_e = (1.4\pm0.2)\e{15}$~cm and is plotted against the observed lines with respect to the LSR velocity in Fig. \ref{rdorsoresults}. A plot illustrating the goodness-of-fit for all the lines is given in Fig. \ref{SOfits}. The abundance profile for SO is plotted in Fig. \ref{abundances} along with \so2 and the CO and \h2O results from Maercker et al. (\textit{in prep.}) for comparison.

One of the detected SO lines, ($8_9\to7_8$), overlaps with \so2($16_{4,12}\to16_{3,13}$) in its wing. We note that this is the only SO line which is significantly over-predicted by the model. Our code is unable to properly take heteromolecular overlaps such as this into account. We suspect that although the \so2 line is much fainter than the SO line (in fact it is difficult to see even in Fig. \ref{rdorso2results} where the \so2 model is overplotted), their interaction likely affects the flux from SO($8_9\to7_8$).

\subsubsection{SO$_2$ results}\label{rdorso2sect}

\begin{figure}[t]
\includegraphics[width=0.5\textwidth]{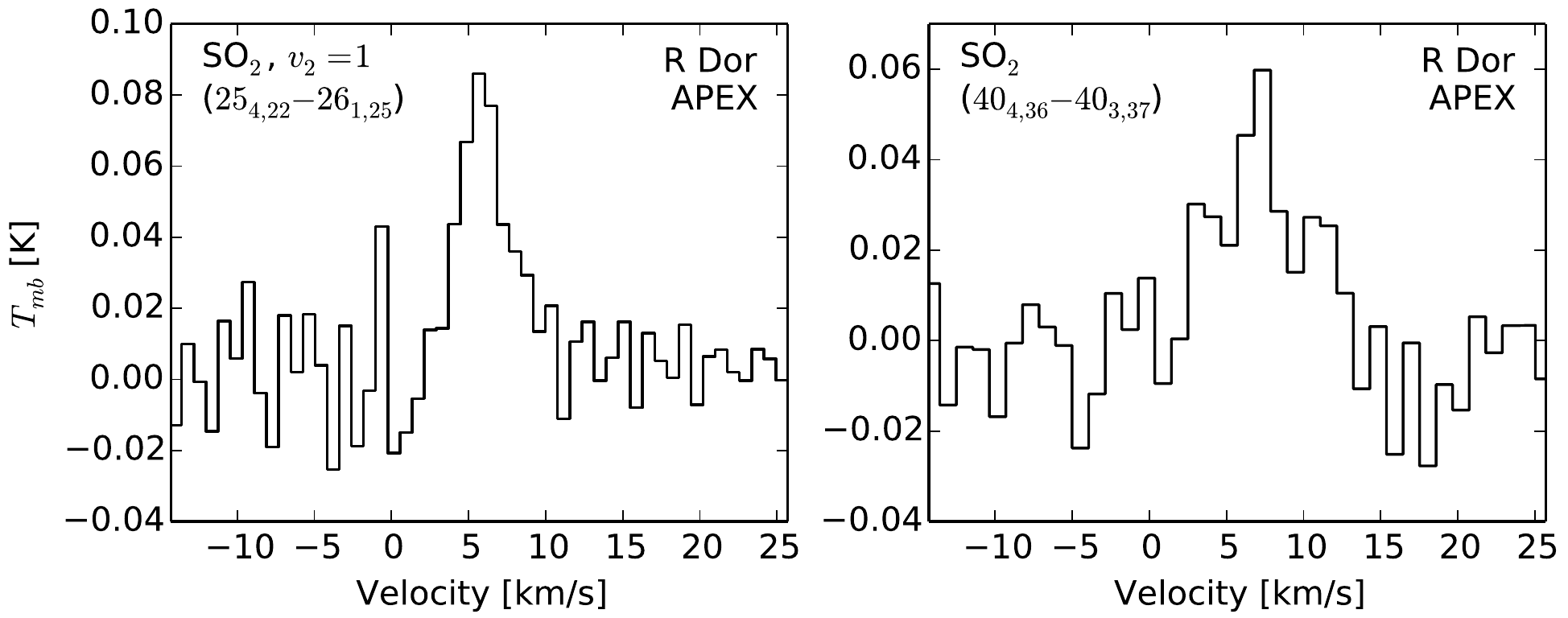}
\caption{\so2 lines excluded from modelling for R~Dor. See text for full explanation.}
\label{rdorleftovers}
\end{figure}

We detected 100 \so2 lines in R~Dor with APEX and HIFI. We exclude the $v_2=1$ ($25_{4,22}\to26_{1,25}$) line at 279.497 GHz from our analysis since it is most likely a maser\footnote{The main evidence for this supposition is that it is in a vibrationally excited state, and that $\Delta K_{a,c}=3$ for this transition. Although this is an allowed transition, it is a very unlikely one under normal circumstances and, if included, our (non-masering) model predicts almost no emission from this transition.}.
We also concluded that it was not computationally viable to model the line with the highest energy level in the ground state, ($40_{4,36}\to40_{3,37}$) at 341.403 GHz, as the number of additional levels and transitions required to fully account for this line represented a significant increase in computation time. (We would have required 3583 levels and 19889 radiative transitions.) The two excluded lines are plotted in Fig. \ref{rdorleftovers}.

\begin{figure}[t]
\center
\includegraphics[width=0.26\textwidth]{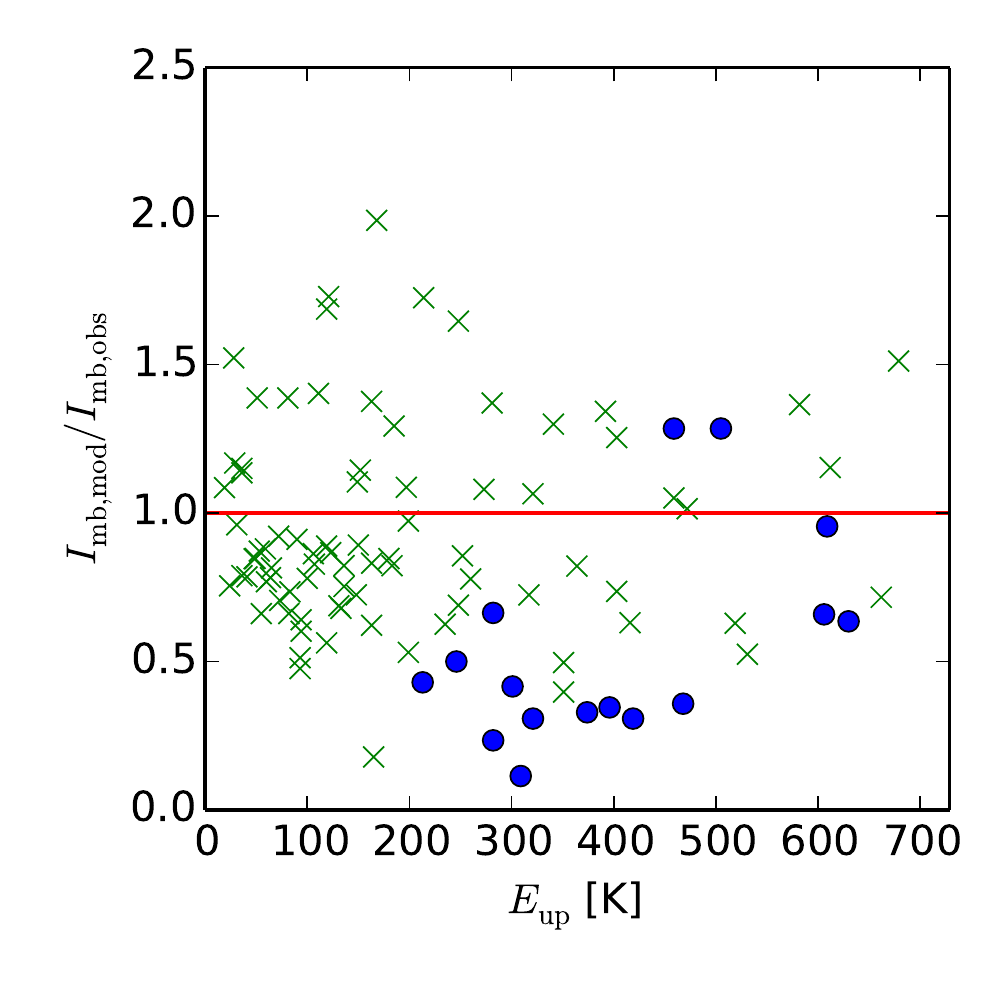}
\includegraphics[width=0.24\textwidth]{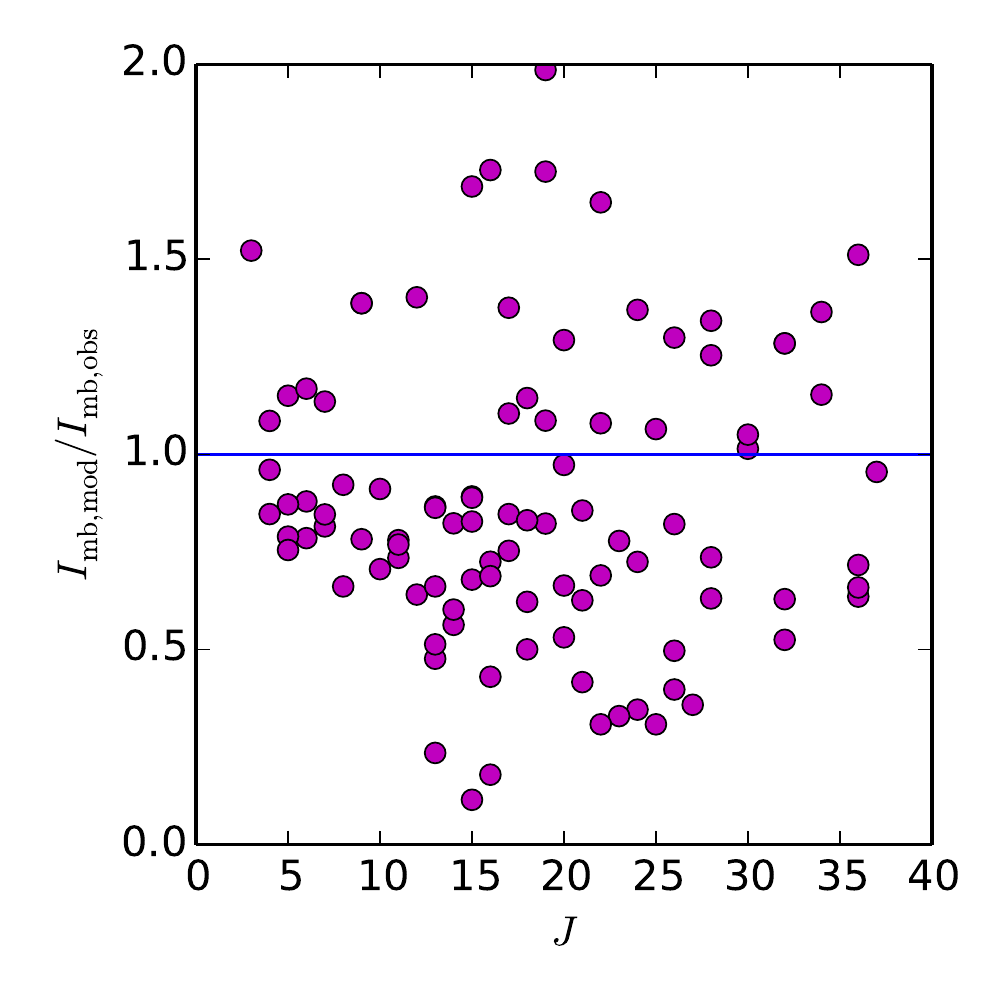}
\includegraphics[width=0.24\textwidth]{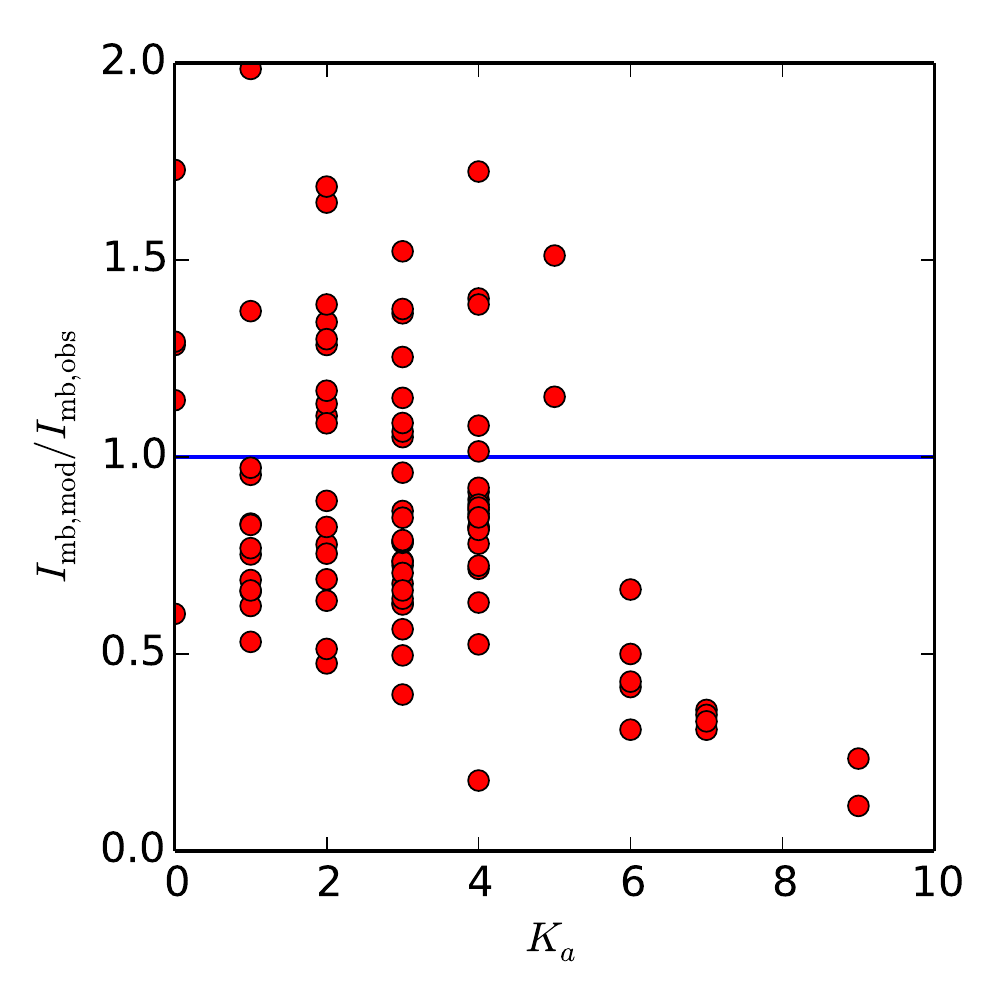}
\caption{\so2 goodness of fit plots for R~Dor. \textit{Top:} Goodness of fit with upper energy level of the transition. HIFI lines are shown as blue points and APEX lines are shown as green crosses. Error bars are excluded to make the plot clearer to read. \textit{Lower left:} Goodness of fit with $J$. \textit{Lower right:} Goodness of fit with $K_a$, a clear downwards trend for $K_a \geq 6$.}
\label{rdorso2fits}
\end{figure}

\begin{figure*}[tp]
\begin{center}
\includegraphics[width=0.49\textwidth]{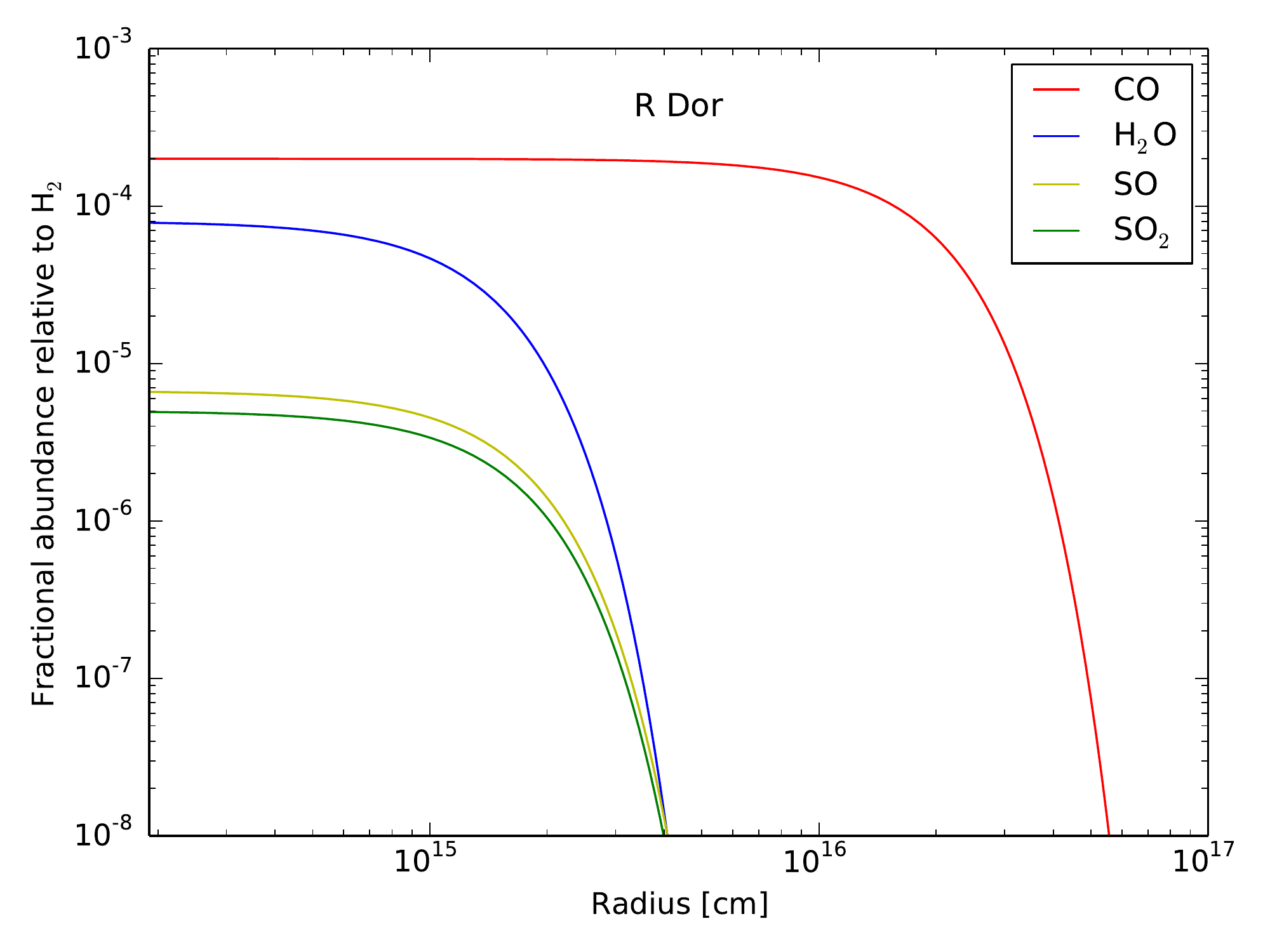}
\includegraphics[width=0.49\textwidth]{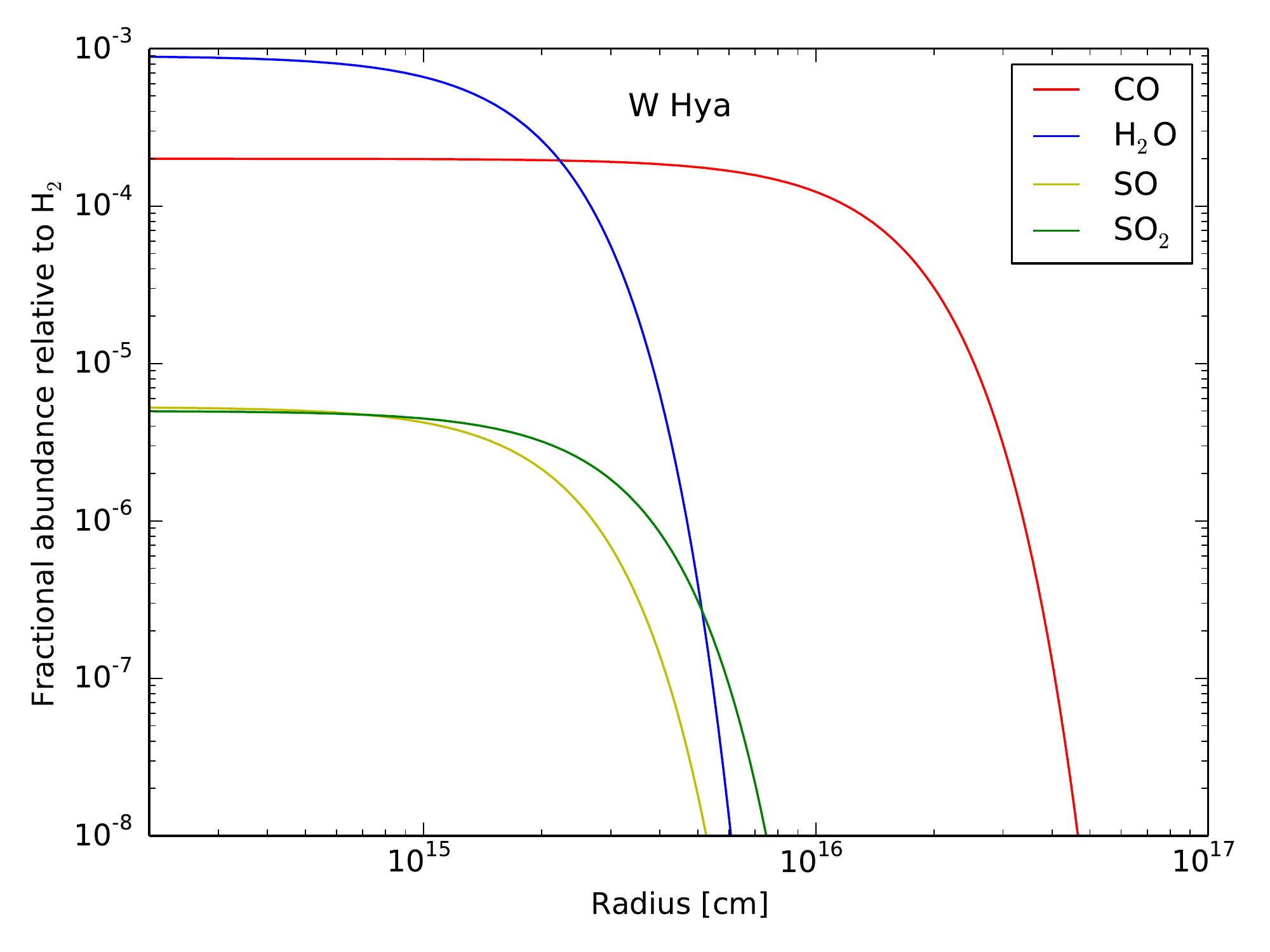}
\includegraphics[width=0.49\textwidth]{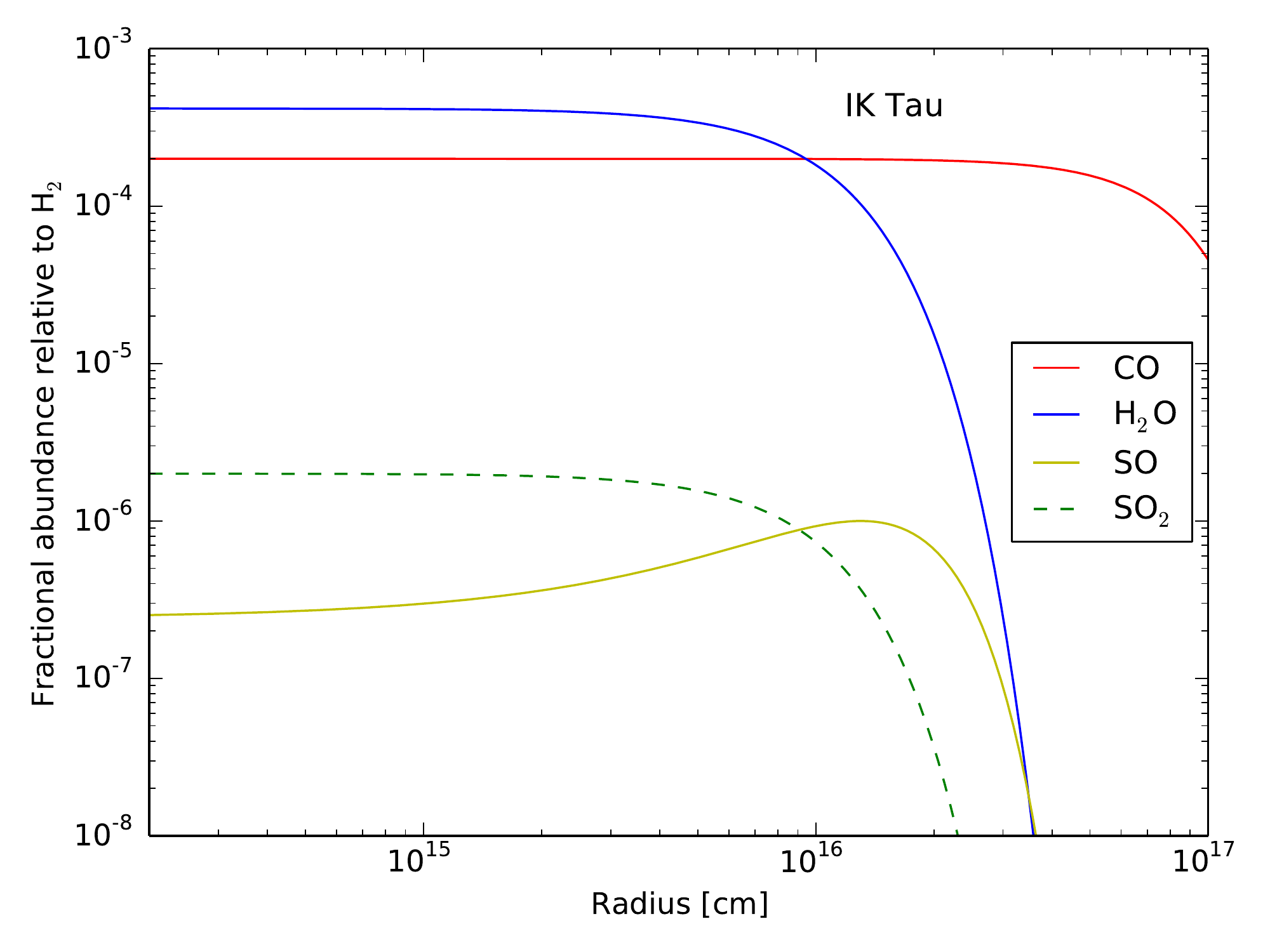}
\includegraphics[width=0.49\textwidth]{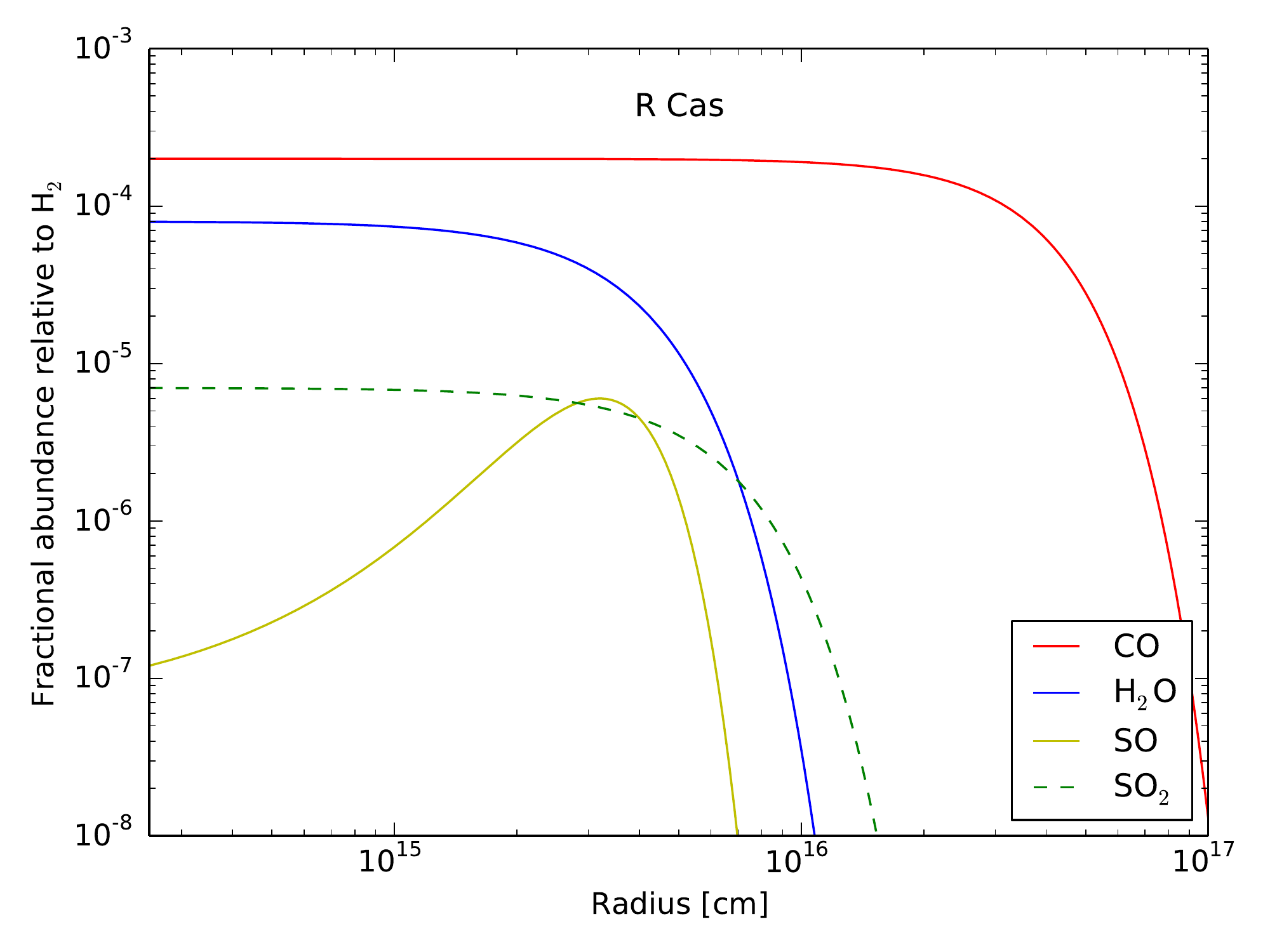}
\caption{Abundance profiles for R~Dor, W~Hya, IK~Tau and R~Cas. The abundances for CO and \h2O are taken from Maercker et al., (\textit{in prep}), except for W~Hya, for which they are taken from \cite{Khouri2014,Khouri2014a}. The dashed line for the \so2 results for IK~Tau and R~Cas indicates that they are tentative.}
\label{abundances}
\end{center}
\end{figure*}

This leaves us with 98 detected \so2 lines with which to constrain our model. Our best fit model has $f_p = 5.0\e{-6}$, $R_e = 1.6\e{15}$ cm and $\chi^2_\mathrm{red} = 3.7$. Due to the significant computational time in running \so2 models, we are unable to provide a comprehensive error analysis as we do for the SO model, hence the lack of formal uncertainties on our results. The model lines are plotted with the observed lines in Fig. \ref{rdorso2results} with goodness of fit shown in Fig. \ref{rdorso2fits}. Overlaps are discussed in detail in Sect. \ref{overlaps}. Fig. \ref{abundances} shows our best-fit abundance profiles for \so2 and SO, along with the results for CO and \h2O from Maercker et al. ({\it in prep.}). 

There is a lot of scatter in the goodness-of-fit plots in Fig. \ref{rdorso2fits}. There is no trend in goodness-of-fit with upper energy level or $J$, but the observed lines that are most strongly under-predicted by the model are those lines for which the upper energy level has quantum number $K_a \geq 6$ (see lower right plot in Fig. \ref{rdorso2fits}). This corresponds to the lines further away from the ``backbone" of $K_a = 0,1$ energy levels in the energy level diagram in Fig. \ref{SO2eld}. We suspect this could be partially due to our exclusion of overlaps for lines outside of the observed frequency range (see Sect. \ref{so2intro}). When testing models with and without overlaps enabled, we note that some lines that do not participate in overlaps can still be strongly affected by the inclusion (or not) of overlaps in our model. For example \so2($27_{7,21}\to 27_{6,22}$) at 657.885 GHz was one such line, with the model predicting weaker emission by a factor of a few when overlaps were omitted. Unfortunately, due to computational limitations, it is not feasible to properly include overlaps in a full radiative transfer analysis, as discussed in Sect. \ref{so2intro}. It should also be noted that the R~Dor data were taken over a long observational campaign (see Sect. \ref{obs}), so any variability in \so2 line brightnesses with pulsation period may contribute to the scatter.

\subsubsection{Overlapping lines}\label{overlaps}

\begin{table*}[t]
\caption{Overlapping lines in R~Dor}\label{olaptbl}
\begin{center}
\begin{tabular}{lllll}
\hline\hline
Primary line	&	Frequency	&	Secondary line	&	Frequency	&	Notes	\\
\hline
\so2 $32_{0,32}\to 31_{1,31}$	&	571.553	&	\so2 $32_{2,30}\to 31_{3,29}$	&	571.532	&	Two distinct peaks	\\
\so2 $13_{1,13}\to12_{0,12}$	&	251.200	&	\so2 $8_{3,5}\to8_{2,6}$	&	251.211	&	Two distinct peaks	\\
SO $8_9\to7_8$	&	346.528	&	\so2 $16_{4,12}\to16_{3,13}$	&	346.524	&	SO line strongly dominates, \so2 line in SO wing	\\
\so2 $24_{7,17}\to26_{6,18}$	&	659.898	&	\so2 $40_{1,39}\to40_{0,40}$	&	659.886	&	Primary line dominates, secondary appears in wing*	\\
\so2 $6_{3,3}\to6_{2,4}$	&	254.281	&	\so2 $24_{2,22}\to24_{1,23}$	&	254.283	&	Unresolved overlap of two lines of similar strength	\\
\so2 $32_{3,29}\to 32_{2,30}$	&	300.273	&	\so2 $24_{8,16}\to25_{7,19}$ ($\nu_2=1$)	&	300.280	&	Secondary line not detected\\
\so2 $15_{4,12}\to15_{3,13}$	&	357.241	&	\so2 $37_{4,34}\to 38_{1,37}$ ($\nu_2=1$)	&	357.230	&	Secondary line not detected 	\\
\so2 $12_{3,9}\to12_{2,10}$	&	237.069	&	\so2 $26_{3,23}\to25_{4,22}$ ($\nu_2=1$)	&	237.062	&	Secondary line not detected \\
\so2 $7_{3,5}\to7_{2,6}$	&	257.100	&	\so2 $8_{3,5}\to8_{2,6}$ ($\nu_2=1$)	&	257.099	&	Lines coincide very closely; not distinguishable	\\
\so2 $13_{2,12}\to12_{1,11}$	&	345.339	&	H\up{13}CN $4\to3$	&	345.340	&	Lines not distinguishable in profile	\\
\hline
\end{tabular}
\end{center}
\tablefoot{* The \so2 ($40_{1,39}\to40_{0,40}$) line is not included in our final model. See discussion in Sect. \ref{overlaps}}.
\end{table*}%

Table \ref{olaptbl} contains an inventory of known line overlaps for the presented lines. In our radiative transfer modelling, we are able to take into account overlaps which occur between two lines of the same molecule --- i.e. two \so2 lines. (For computational purposes we only include \so2 overlaps in the range 200 GHz -- 1.2 THz. Note, however, that all possible homomolecular overlaps are taken into account for SO in all modelled stars.) However, if there is a line overlap between two lines generated by different molecules, we are unable to properly treat this, as our code only allows for the modelling of one molecular species at a time. In R~Dor we observe two such heteromolecular overlaps. The first between the SO($8_9\to7_8$) and \so2($16_{4,12}\to 16_{3,13}$) lines, where the much weaker \so2 line appears in the wing of the bright SO line, and the second between \so2($13_{2,12}\to12_{1,11}$) and H\up{13}CN($4\to3$), where the two lines coincide very closely so as to be indistinguishable. Based on our model, we expect approximately half the flux to be due to the H\up{13}CN($4\to3$) transition, which would agree with the H\up{13}CN($3\to2$) line also covered by the APEX survey. However, without modelling H\up{13}CN, it is not possible to fully gauge the impact of this overlap on our model. 

The remaining line overlaps for lines modelled in this paper are homomolecular. 

Three of the line pairs that are treated as overlapping in the code consist of a bright primary line in the vibrational ground state and a very weak secondary line in the $\nu_2=1$ vibrationally excited state. As can be seen in Fig. \ref{rdorso2results}, these secondary lines are not detectable above the noise in our observations, but are taken into account in our modelling.

The \so2 ($24_{7,17}\to26_{6,18}$) line at 659.898 GHz overlaps with the \so2 ($40_{1,39}\to40_{0,40}$) at 659.886 GHz and we would expect the latter to have an effect on the former. However, the \so2 ($40_{1,39}\to40_{0,40}$) line falls outside of the range of energy levels we included in our model. As noted in Sect. \ref{so2intro}, it was not feasible to include a larger number of higher energy levels, hence this particular overlap is not taken into account in our modelling.


\subsubsection{Isotopologue results}\label{isoresults}

\begin{figure}[tp]
\begin{center}
\includegraphics[width=0.5\textwidth]{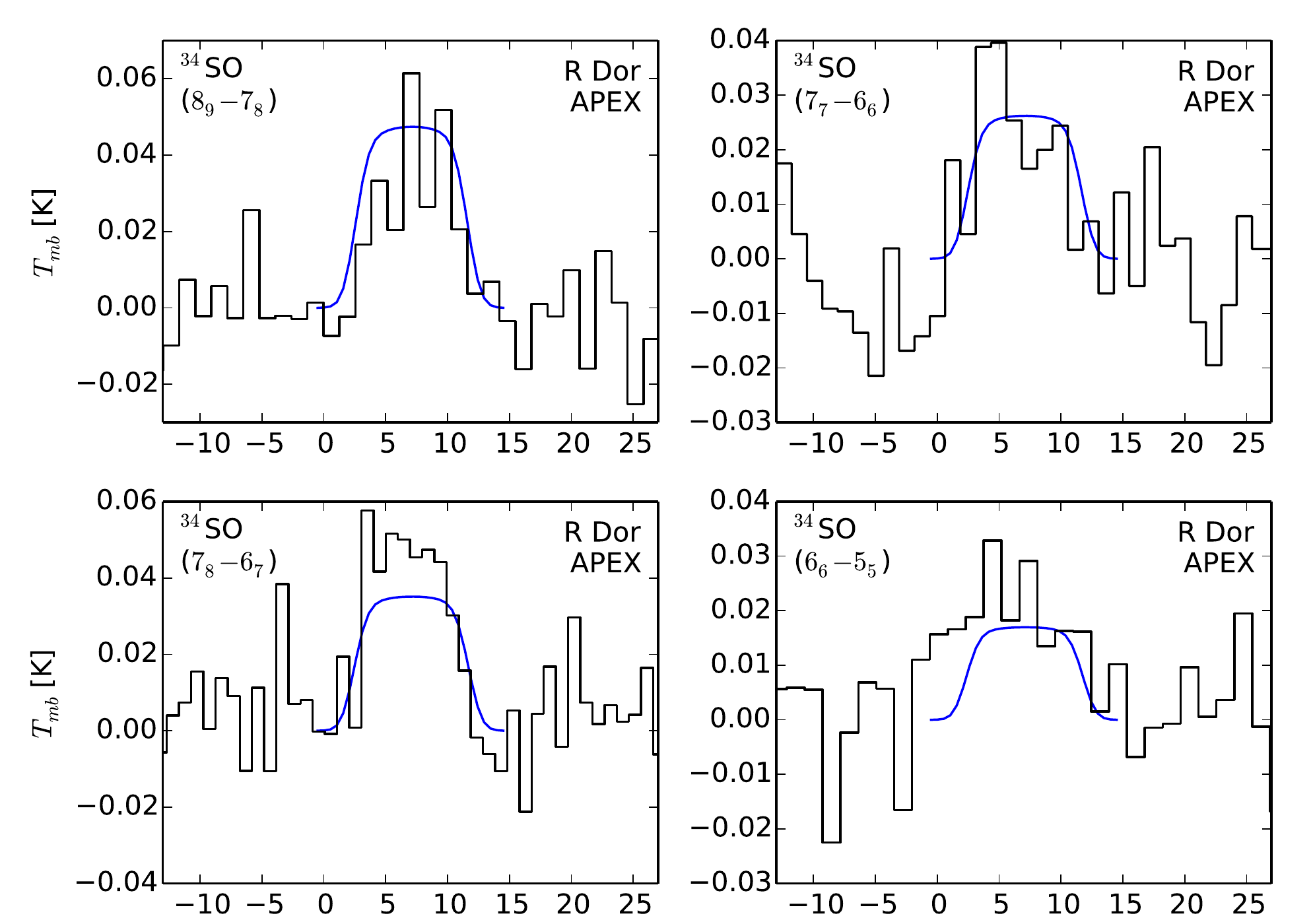}
\includegraphics[width=0.5\textwidth]{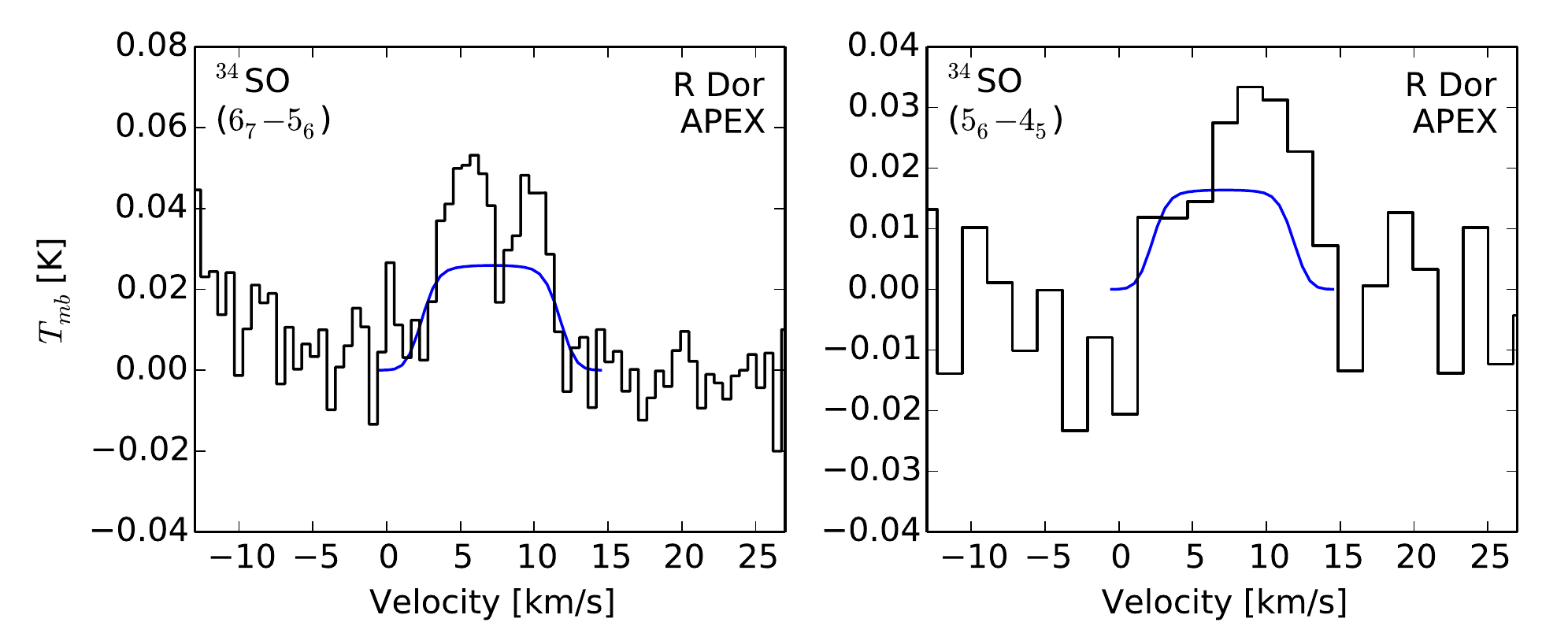}
\caption{\up{34}SO model (blue lines) and observations (black histograms) for R~Dor.}
\label{34so}
\end{center}
\end{figure}

Based on the analysis of 6 \up{34}SO lines, and assuming the same $e$-folding radius as found for \up{32}SO, we find a \up{34}SO abundance of $(3.1\pm0.8)\e{-7}$ in a best fit model that has $\chi^2_\mathrm{red} = 1.4$. This gives a \up{32}SO/\up{34}SO ratio of $21.6\pm8.5$. The best fit model is shown in Fig. \ref{34so}. The goodness of fit plot showing the ratio between the model and observed integrated intensities is shown in Fig. \ref{SOfits}.

Modelling \up{34}\so2 in the same detailed manner as we have modelled \up{32}\so2 is impractical given the computational time required, the complexity of the molecular data file, and the low number of detected lines. However, all of the \up{32}\so2 lines we modelled are optically thin, so we can approximate the \up{32}\so2/\up{34}\so2 ratio by comparing the intensity ratios of two lines of the same transition. The best \up{34}\so2 transition for this purpose is $20_{0,20}\to19_{1,19}$. Comparing the integrated intensities for this transition, we find a \up{32}\so2/\up{34}\so2 ratio of $21.6\pm12.1$, in good agreement with the result from \up{34}SO modelling.

The solar system value of \up{32}S/\up{34}S is 22.5 \citep{Cameron1973} and \cite{Kahane1988} found a value of 20.2 for the carbon star CW Leo using SiS isotopologues, both in agreement with our results.

While a detailed model of \up{34}\so2 would be extremely time consuming, a detailed model of SO\up{18}O would not be computationally feasible. Due to the asymmetry of the two oxygen atoms, SO\up{18}O has approximately double the number of energy levels and transitions as \so2, when looking at the same energy range, meaning that an SO\up{18}O molecular data file would have to be approximately twice the size of our already very large \so2 file to probe a similar range of energies. 
The more complex energy level structure also means it is not possible to directly compare lines between SO\up{18}O and \so2, even when the transitions have the same quantum numbers. For \up{34}\so2 and SO\up{18}O we present the (tentative) detections in Fig. \ref{leftoverisotopologues}.

\subsection{Other M stars}\label{othersoresults}

We model SO and \so2 line emission for the remaining stars using HIFI observations, as listed in Table \ref{hifiobs}, and archival observations with different ground-based instruments, as listed in Table \ref{oldobs}. Several of these older observations probe energy levels significantly lower than the HIFI observations, allowing us to better constrain the size of the emitting molecular envelope. This is particularly important for IK~Tau, where only the three $N = 13\to 12$ SO lines were detected with HIFI, as these are emitted from a similar region of the CSE.

The stellar parameters used in our SO and \so2 models, taken from CO model results, are listed in Table \ref{stellarprop}.

\subsubsection{W~Hya}

\begin{figure}[t]
\includegraphics[width=0.5\textwidth]{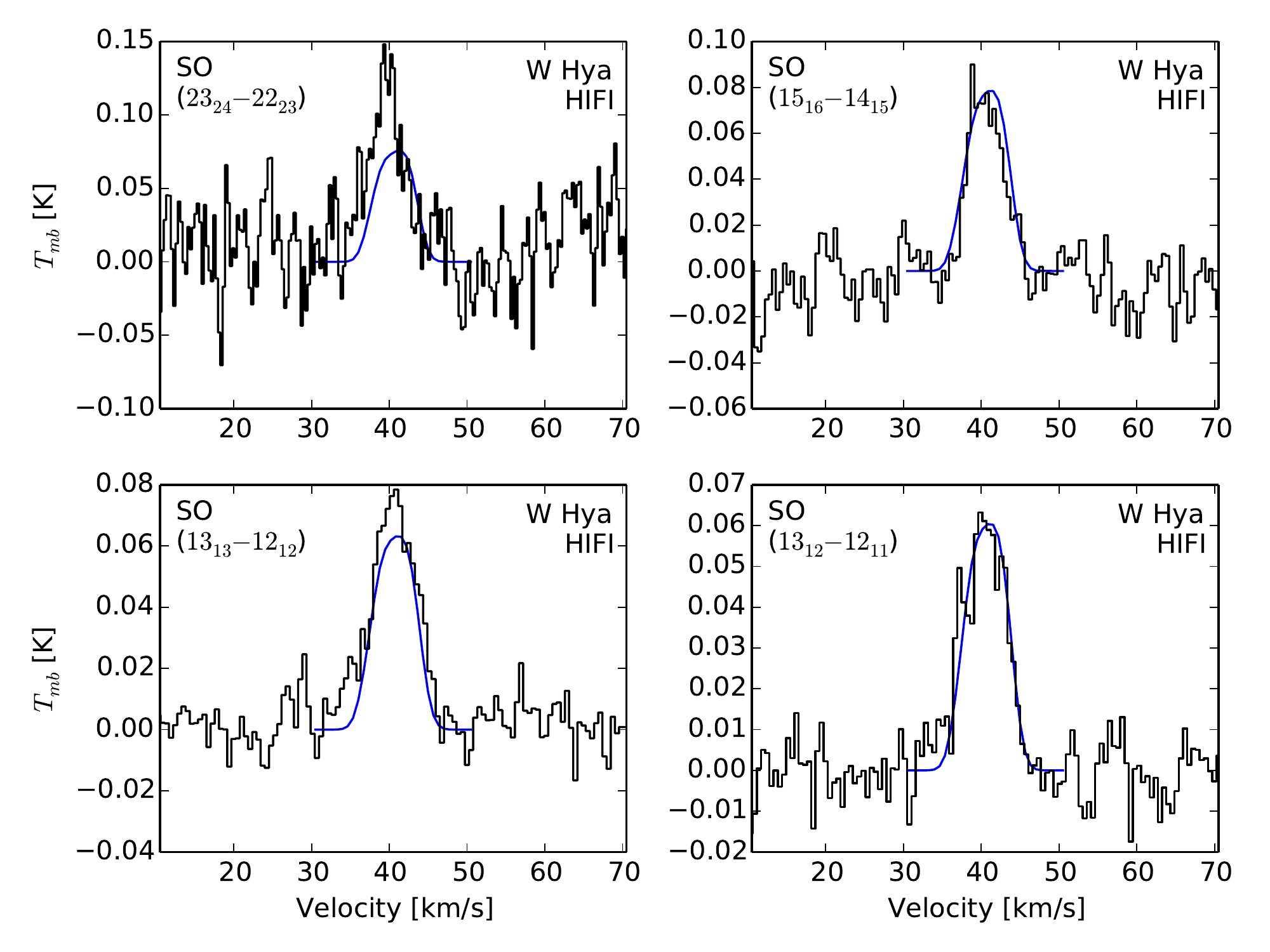}
\includegraphics[width=0.5\textwidth]{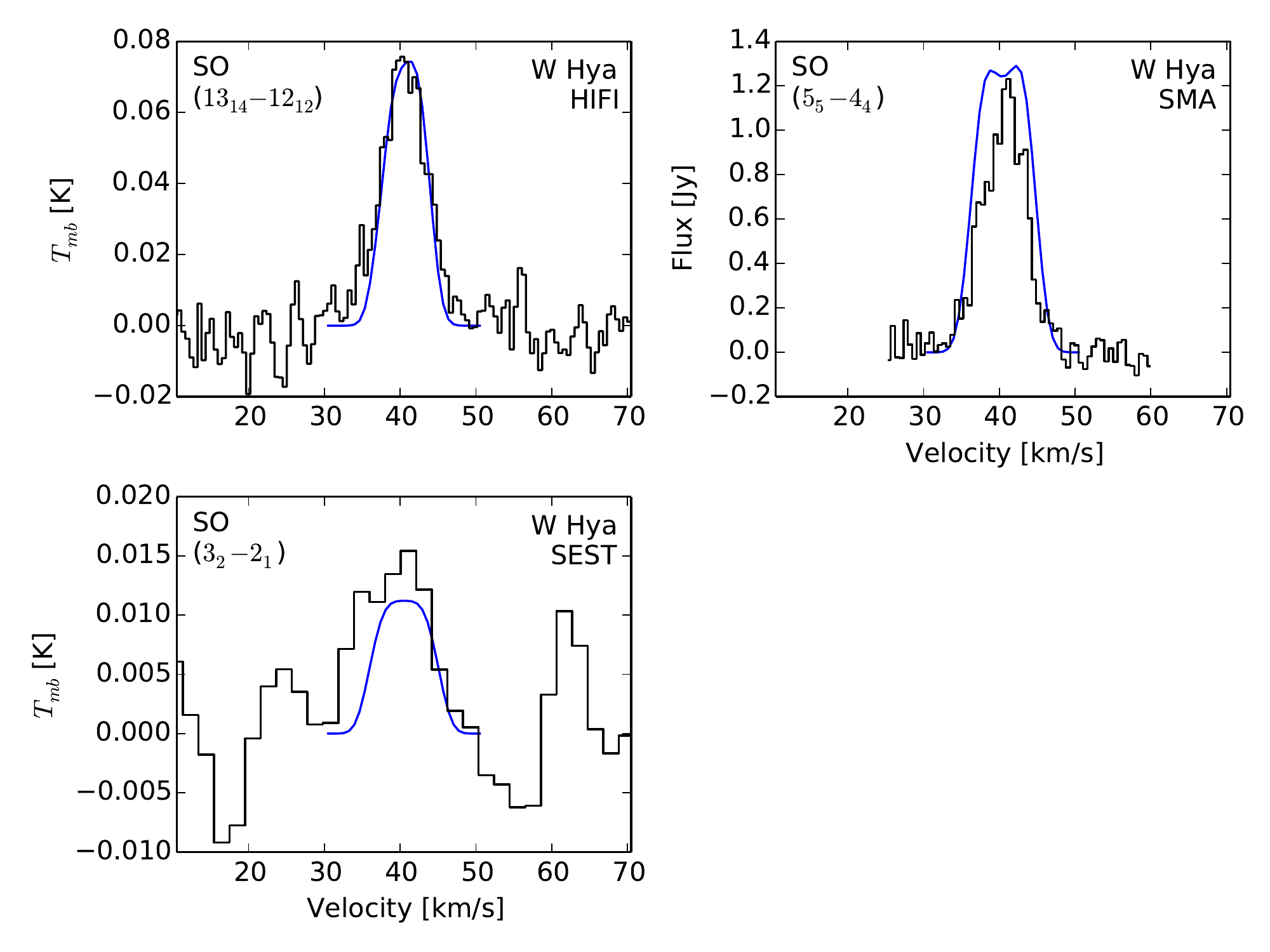}
\caption{Models (blue lines) and observations (black histograms) for SO towards W~Hya.}
\label{whyaSOlines}
\end{figure}

In the case of W~Hya we find an SO model that fits the data well using the Gaussian abundance distribution given in Eq. \ref{abundanceeq}. We found $f_p = (5.0\pm1.0)\e{-6}$ and $R_e = (1.5\pm0.5)\e{15}$ cm, with $\chi^2_\mathrm{red} = 2.57$. This result is qualitatively similar to that of R~Dor. As with R~Dor, this suggests that SO in the CSE of W~Hya is formed close to the star and is not found in a shell around the star as might be expected if it were a photodissociation product of another molecule such as \h2S. The HIFI observations and model line plots for SO are shown in Figure \ref{whyaSOlines}. The corresponding $\chi^2$ plot is shown in Fig. \ref{SOchi2}.

\begin{figure*}[t]
\includegraphics[width=\textwidth]{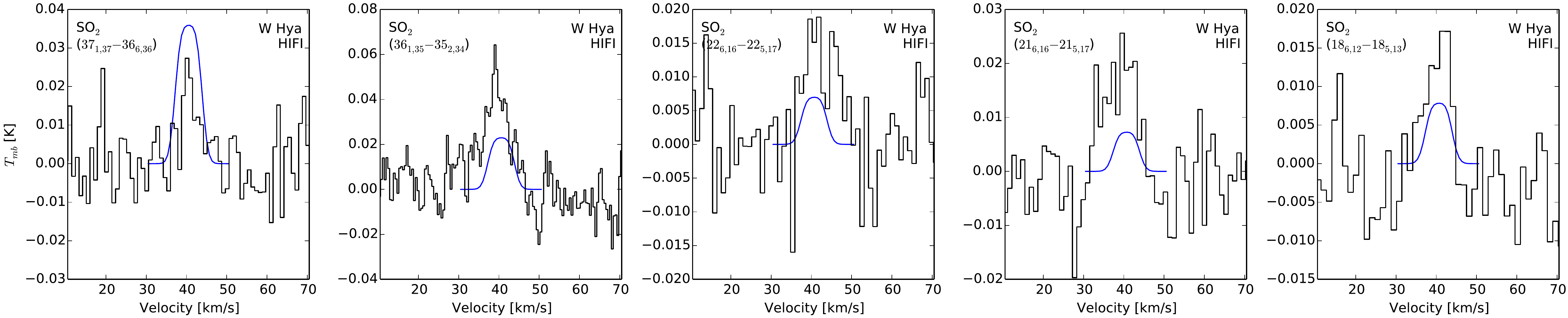}
\caption{Models (blue lines) and observations (black histograms) for \so2 towards W~Hya.}
\label{whyaSO2lines}
\end{figure*}

The HIFI observations and model line plots for \so2 towards W~Hya are shown in Figure \ref{whyaSO2lines}. The main difficulty we had in fitting an \so2 model was finding a model which fit the two highest-energy lines. As can be seen in Table \ref{hifiobs}, the \so2($37_{1,37} \to 36_{0,36}$) and \so2($36_{1,35} \to 35_{2,34}$) lines are only $\sim3$~K apart in upper energy level. Also we note that the lower-energy line is almost a factor of 3 brighter than the higher-energy line. Our model invariably predicts a smaller difference in intensity with the higher-energy line being the brighter. The same is true for R~Dor, however, in R~Dor the detected lines reflect this (although the model fit is not perfect). This phenomenon is probably due in part to the noise in our observations but could also reflect a problem with our molecular description of \so2. In this case, the most likely cause is the cut-off in included energy levels at $J=38$. The variation in these lines cannot be due to variations in brightness due to stellar pulsations as both lines were observed simultaneously (and, indeed, all the \so2 lines in W~Hya were observed within two days). In any case, the apparently outlying line of ($37_{1,37} \to 36_{0,36}$) strongly contributes to the poorly fitting model we find for \so2 in W~Hya. We are able to find a better fit by excluding this line, but do not have a strong basis for doing so, hence we leave it in.

Our best fit model for \so2 has $f_p = 5.0\e{-6}$, based on a small grid with steps of $0.5\e{-6}$, and $R_e = 3.0\e{15}$~cm, based on a small grid with steps of $0.5\e{15}$~cm. This model has $\chi^2_\mathrm{red}=5.7$. We also test an \so2 model using the parameters we found for SO. That model is not a significantly worse fit with almost the same $\chi^2_\mathrm{red}$.

The abundance distributions for SO and \so2, along with the CO and \h2O abundance distributions from \cite{Khouri2014,Khouri2014a} for comparison, are shown in Fig. \ref{abundances}.




\subsubsection{IK~Tau}\label{iktauresults}

\begin{figure}[t]
\includegraphics[width=0.5\textwidth]{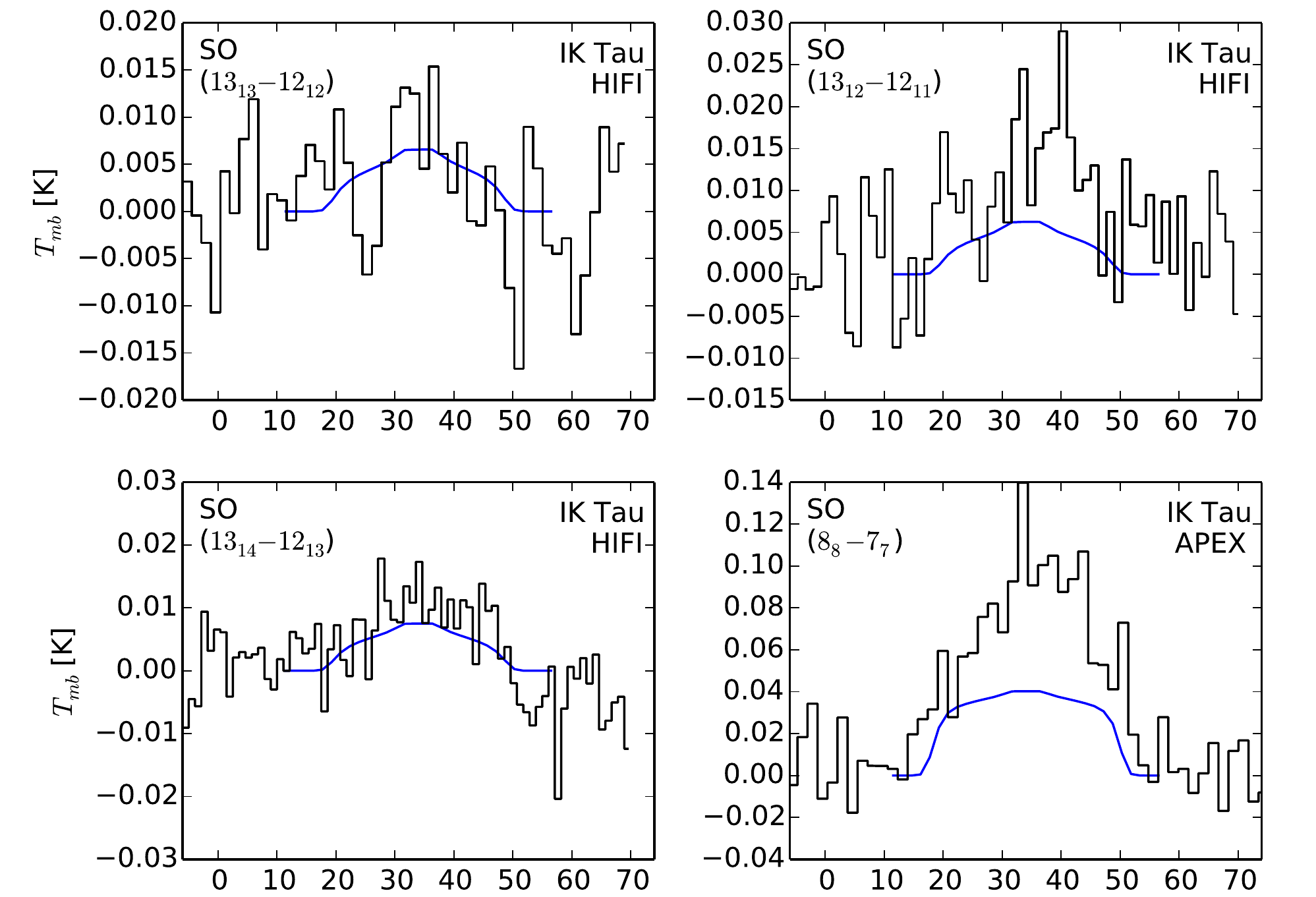}
\includegraphics[width=0.5\textwidth]{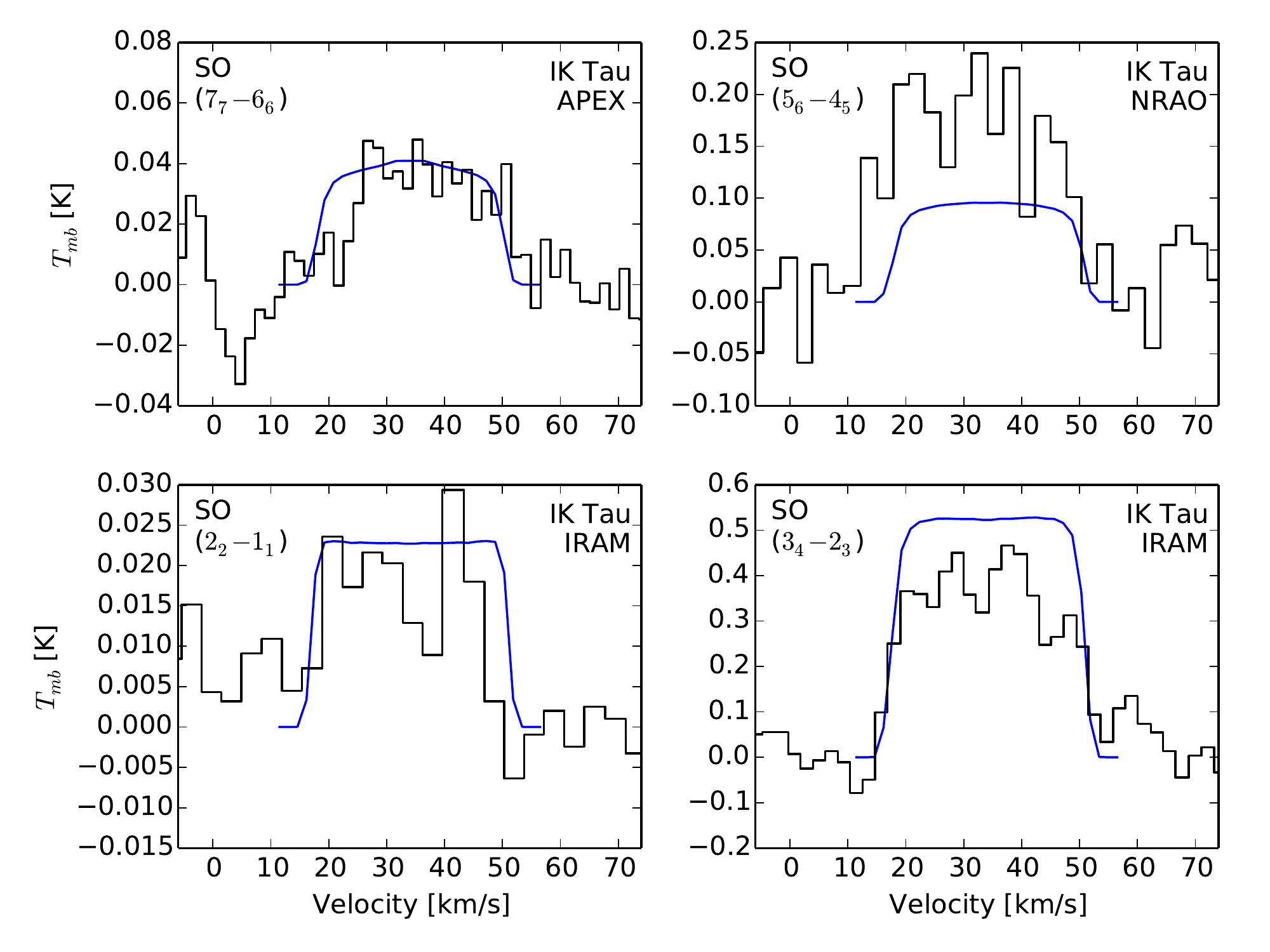}
\center{\includegraphics[width=0.25\textwidth]{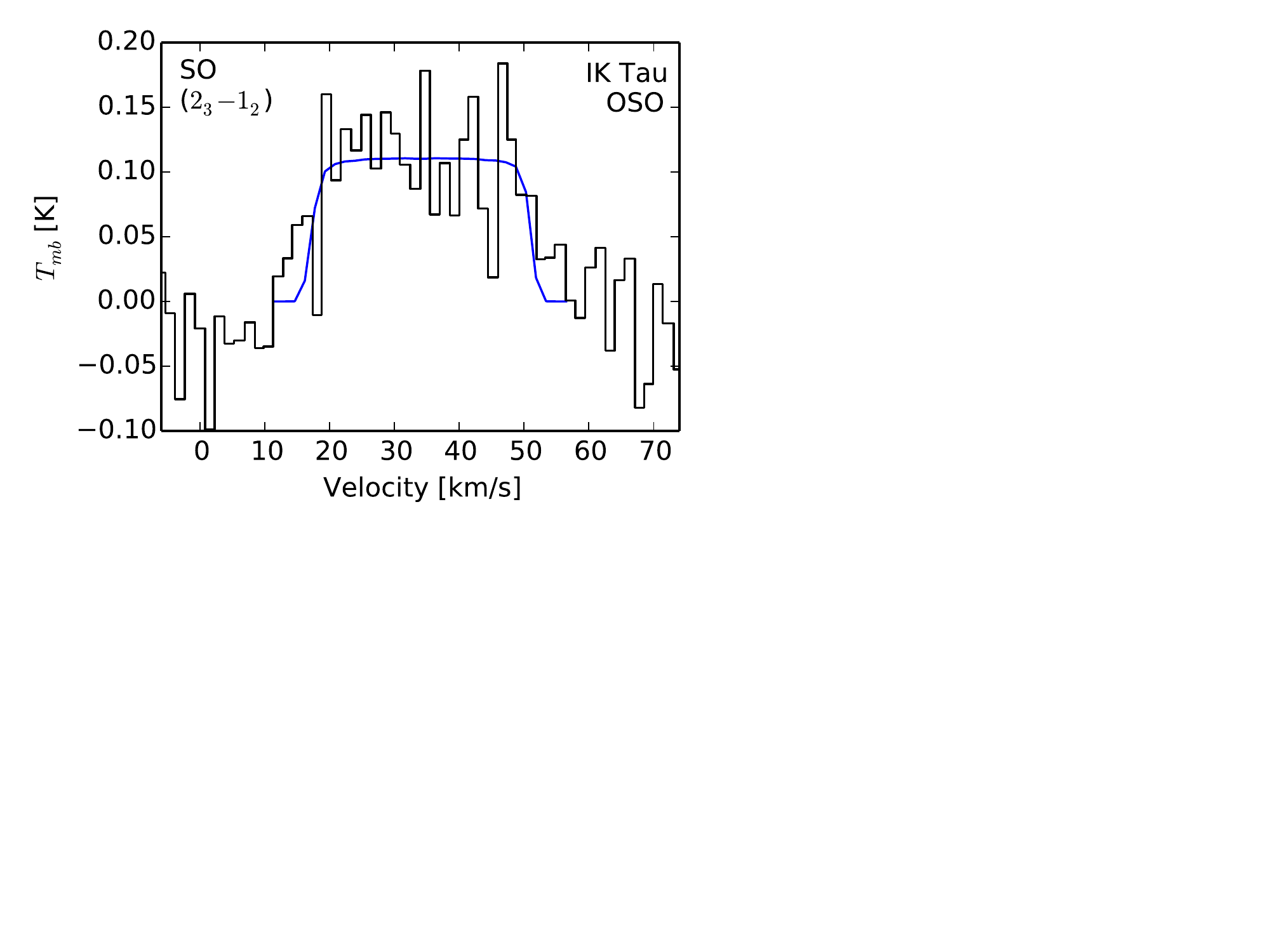}}
\caption{SO models (blue lines) and observations (black histograms) for IK~Tau.}
\label{iktauSOlines}
\end{figure}

When we try to fit the SO IK~Tau observations with a centrally peaked Gaussian distribution, we cannot constrain the $e$-folding radius with the available data. The $\chi^2$ analyses of centrally-peaked Gaussian models point towards very large $e$-folding radii, significantly larger (by more than half an order of magnitude) than the half-abundance radius Maercker et al. (in prep.) found for the corresponding CO envelope. Since it is highly unlikely that the SO envelope is more extensive than that of CO, we conclude that a centrally-peaked Gaussian distribution is unlikely for SO in IK~Tau. Instead, we run a three-parameter grid across $f_p$, $R_p$, and $R_w$ (see Eq. \ref{shellabundanceeq}) to find the best model. We find $f_p = (1.0\pm0.2)\e{-6}$, $R_p = (1.3\pm 0.2)\e{16}$~cm, and $R_w = 1.8R_p$ (which we gridded in steps of $0.2R_p$), with $\chi^2_\mathrm{red} = 4.67$ and the resultant lines are shown in Fig. \ref{iktauSOlines}.

The $\chi^2$ plot for SO in IK~Tau is shown in Fig. \ref{SOchi2}. IK~Tau has a significantly larger $\chi^2$ value for the best fit model (compared with R~Dor and W~Hya) because of some noisy observations. This is also seen in the goodness of fit plot in Fig. \ref{SOfits}. In comparison, R~Dor and W~Hya have brighter and more uniform line observations, making it easier to find a good model fit.

\citet{Decin2010} perform a radiative transfer analysis of IK~Tau in a way that is similar to our method. They find an SO abundance distribution that is similar to our shell-like distribution, but with an increased abundance in the inner region. They find an abundance at $200R_*$ (which corresponds to about $3\e{15}$~cm) of $\sim2\e{-7}$ using two lines to fit the model. This did not change significantly in the follow up in \cite{Decin2010a} which included one of the HIFI lines as well. Our model results give a corresponding abundance about a factor of 2 higher at the same radius but using a different shape for the abundance distribution. We also use 10 lines with a broader range of energy levels to constrain the model. 

In the case of \so2 in IK~Tau we are unable to include overlaps as we do for R~Dor and W~Hya due to the larger expansion velocity of the circumstellar gas around IK~Tau. The larger expansion velocity means there are a larger number of overlaps (since the lines are about three times wider than for R~Dor) which quickly become computationally infeasible to fully account for.

\begin{figure}[t!]
\begin{center}
\includegraphics[width=0.5\textwidth]{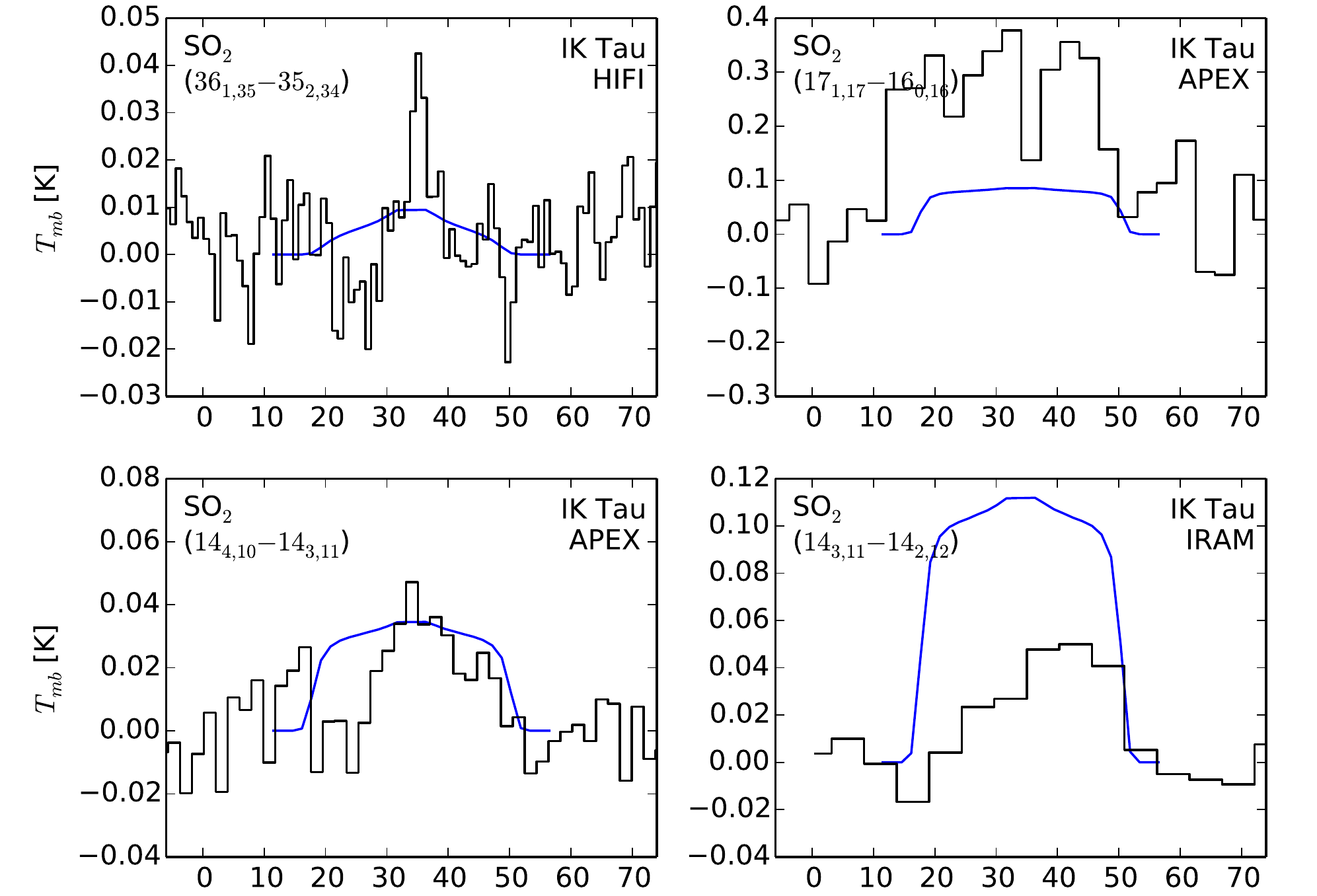}
\includegraphics[width=0.5\textwidth]{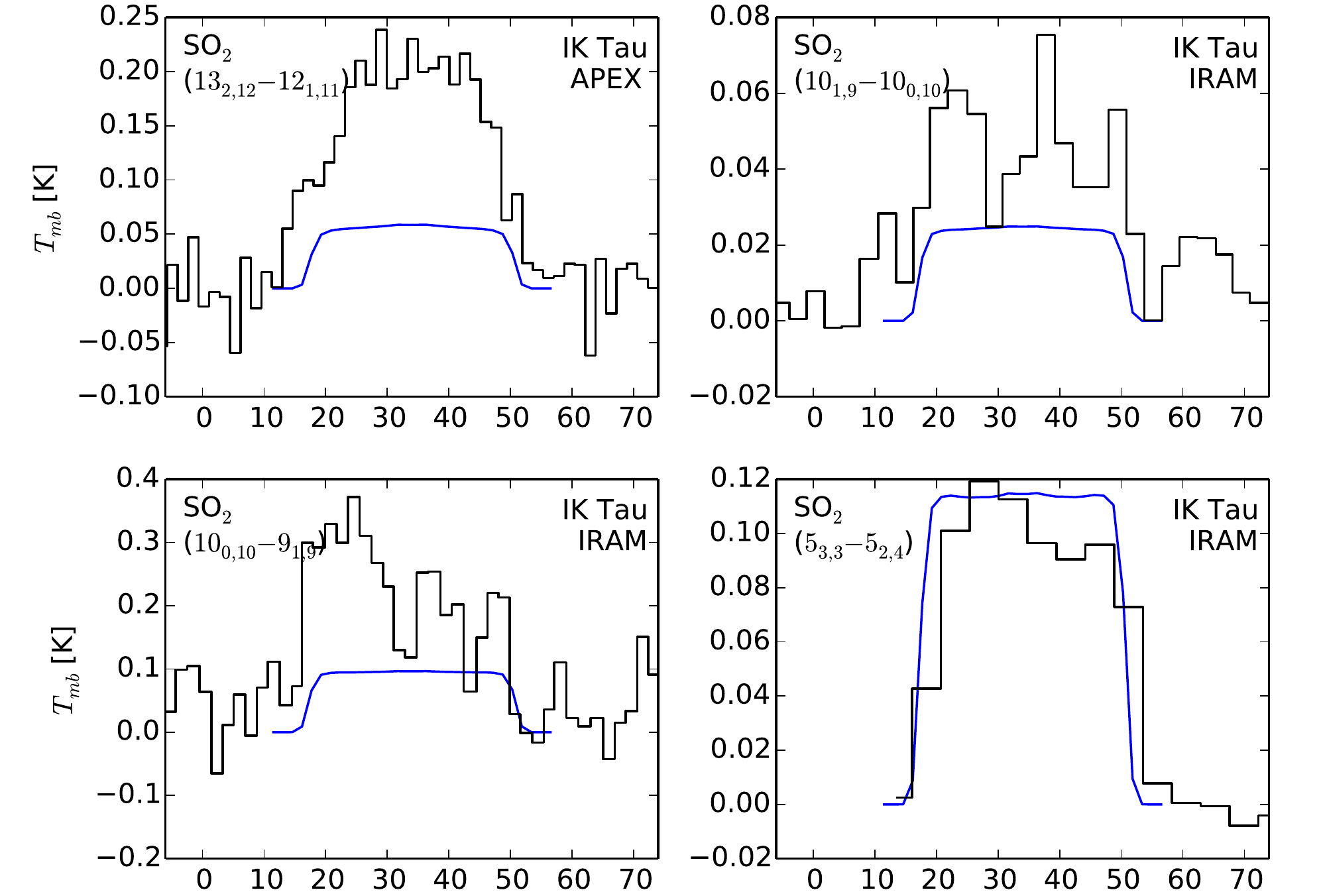}
\includegraphics[width=0.5\textwidth]{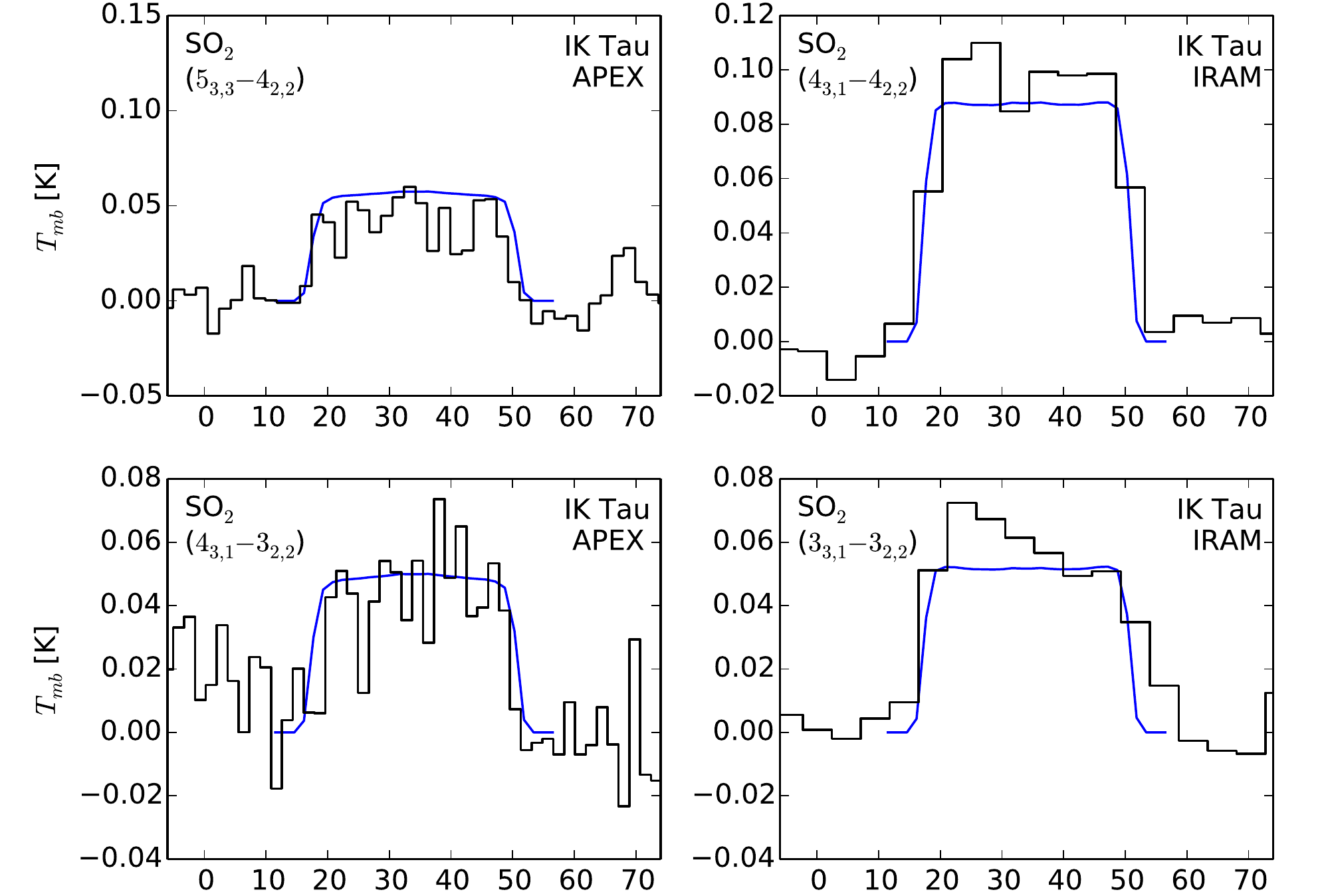}
\includegraphics[width=0.5\textwidth]{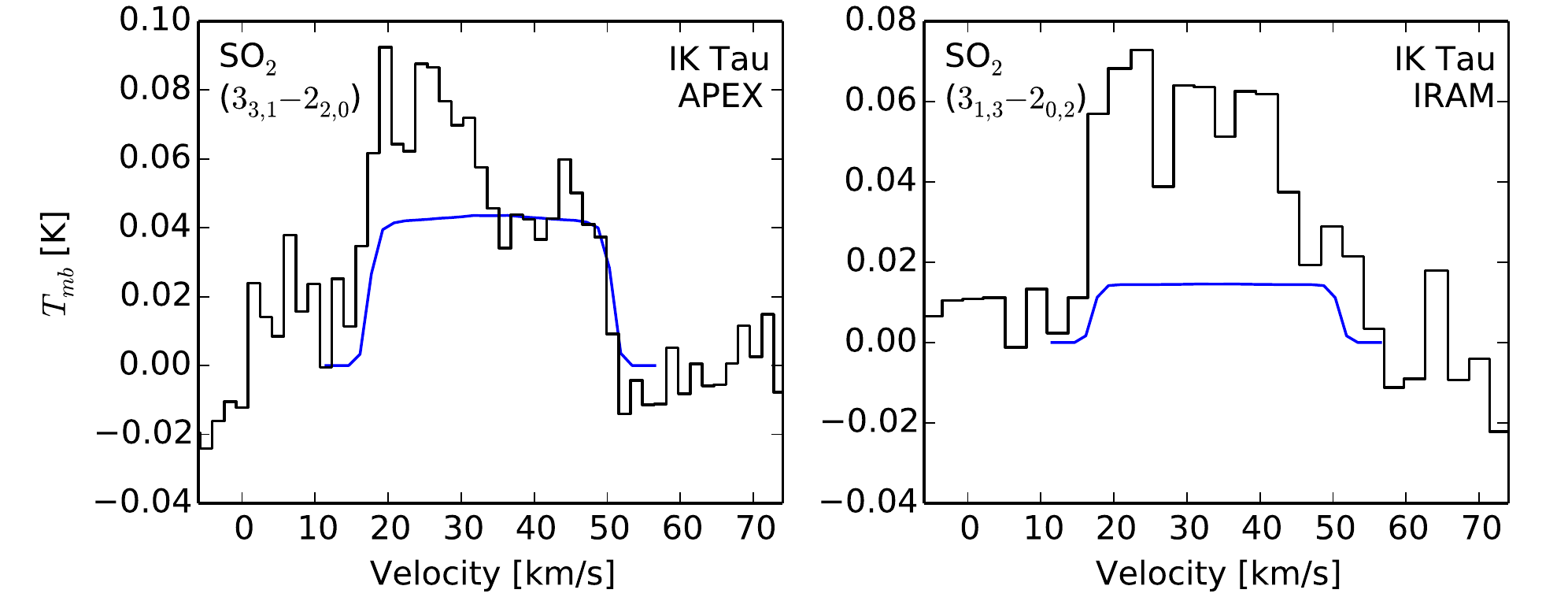}
\caption{\so2 model (blue line) and observations (black histograms) of IK~Tau. For details on the archival observations, see Table \ref{oldobs}.}
\label{iktauso2lines}
\end{center}
\end{figure}

We could not find a consistent model for IK~Tau that matched all the available observed \so2 lines. In particular, there was a very large scatter in goodness-of-fit for the lines with upper energy levels of 136~K or less (which is all of the lines other than the one HIFI observation). There was no way to simultaneously fit all these observed lines well. A centrally-peaked Gaussian model matches the data reasonably well --- particularly the HIFI line, which according to the best shell model should have been a non-detection --- and much better than the shell model. A Gaussian model with $e$-folding radius located at the peak of the SO distribution is a better fit than a model with the SO distribution parameters, but we find that decreasing the $e$-folding radius to $R_e=1\e{16}$~cm gives a better fit again. We cannot constrain the $e$-folding radius better than by a factor of $\sim2$, however, because of the large scatter in the lower-energy lines. The model we present in this paper, plotted in Fig. \ref{iktauso2lines}, has a peak \so2 abundance $f_p = 2\e{-6}$, and $R_e=1\e{16}$~cm. This model has $\chi^2_\mathrm{red} = 18.4$, the high value reflecting the poor overall fit.
The large scatter in the IK~Tau \so2 lines could be due to variability in line brightness with pulsation period. The data we used were observed at different times corresponding to different phases of pulsation. For example, the brightest lines, $(17_{1,17} \to 16_{0,16})$ and $(13_{2,12} \to 12_{1,11})$, were observed less than two weeks apart close to maximum brightness in 2006. The most under-predicted line, $(14_{3,11} \to 14_{2,12})$, was observed four months later when the star was approaching minimum brightness. On the other hand, the most well-fit lines --- those with $J=5,4,3$ as can be seen in Fig. \ref{iktauso2lines} --- were variously taken close to minimum and maximum brightness, so perhaps it is the higher $J$ lines which are most strongly affected. {Future monitoring of these lines observationally would allow us to confirm whether the effect on the higher-$J$ lines is really due to variability over a pulsation period.}

The abundance distributions for SO and \so2 in IK~Tau, along with the CO and \h2O abundance distributions from Maercker et al. (\textit{in prep.}) for comparison, are shown in Fig. \ref{abundances}. In general, we do not consider our \so2 results for IK~Tau conclusive. A more rigorous model which is properly able to take overlaps into consideration and which perhaps includes more lines in the intermediate to high energy range (with upper energy level $>136$~K) is recommended.


\cite{Decin2010} have similar issues modelling the \so2 in IK~Tau, especially with the ($17_{1,17}\to16_{0,16}$) and ($13_{2,12}\to12_{1,11}$) lines which we also strongly under-predict, as can be seen in Fig. \ref{iktauso2lines}. When they exclude these two lines, \cite{Decin2010} find a high inner abundance of \so2, in general agreement with our results. The poor fit of our model could be a result of unusual structure in the CSE of IK~Tau or could be a result of not being able to properly consider overlaps in the \so2 model. The higher wind velocity 
would also lead to more overlapping lines overall --- including in regions we have not observed --- which could have an effect on the overall energy distribution between all molecular energy levels. 



\citet{Kim2010} use a combination of Monte-Carlo radiative transfer modelling, to find CSE properties, and LTE formulations, to determine SO and \so2 abundance for IK~Tau. For SO they find fractional abundances in the range 3 to $8\e{-7}$ and for \so2 their fractional abundances were in the range $4\e{-6}$ to $1\e{-5}$.
Their results are not accompanied by clear abundance distributions, making them difficult to compare with our results. Nevertheless, their SO result is very close to our peak abundance for SO, while their \so2 result is much higher than we found.

\subsubsection{R~Cas}

\begin{figure}[t]
\includegraphics[width=0.5\textwidth]{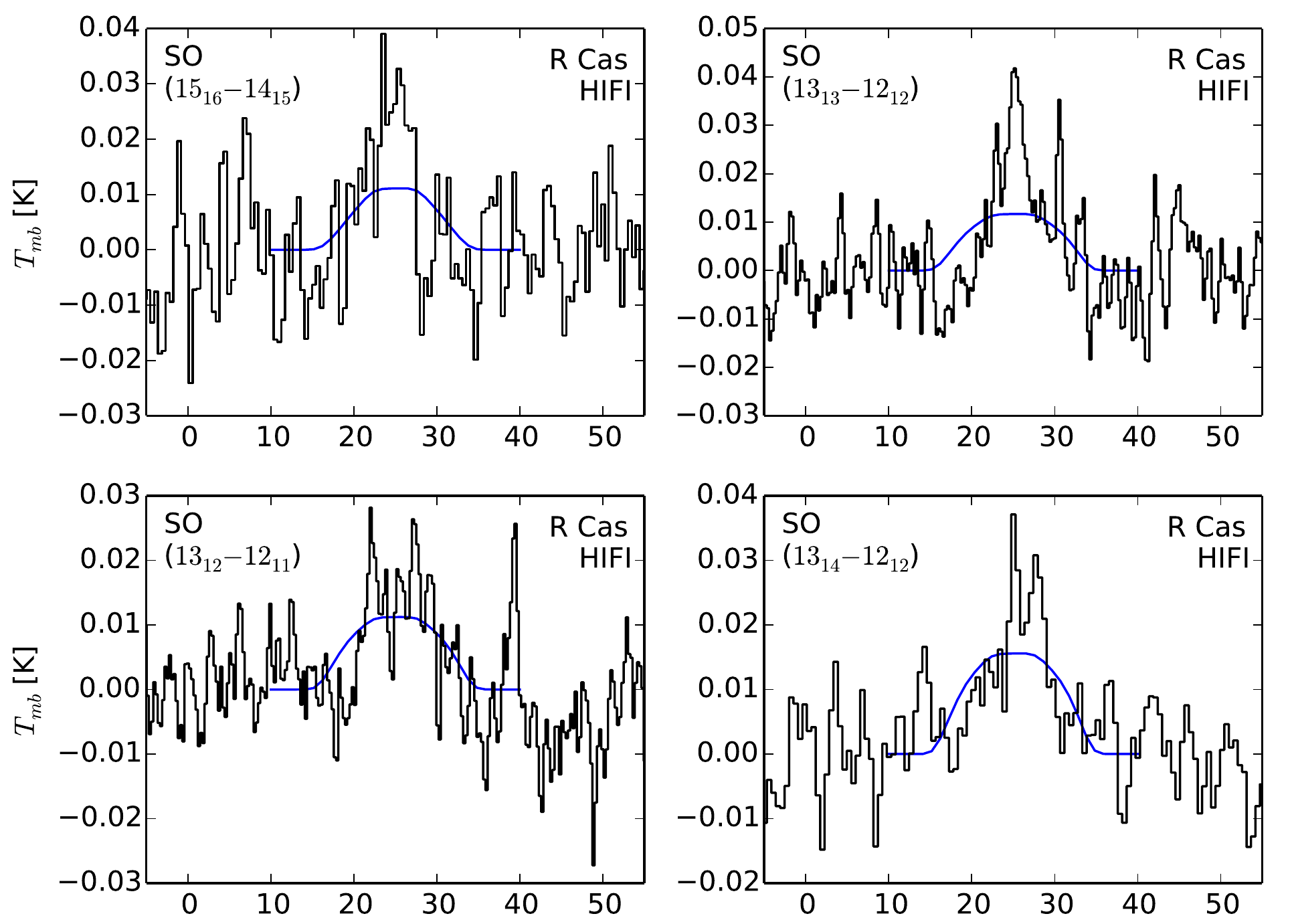}
\includegraphics[width=0.5\textwidth]{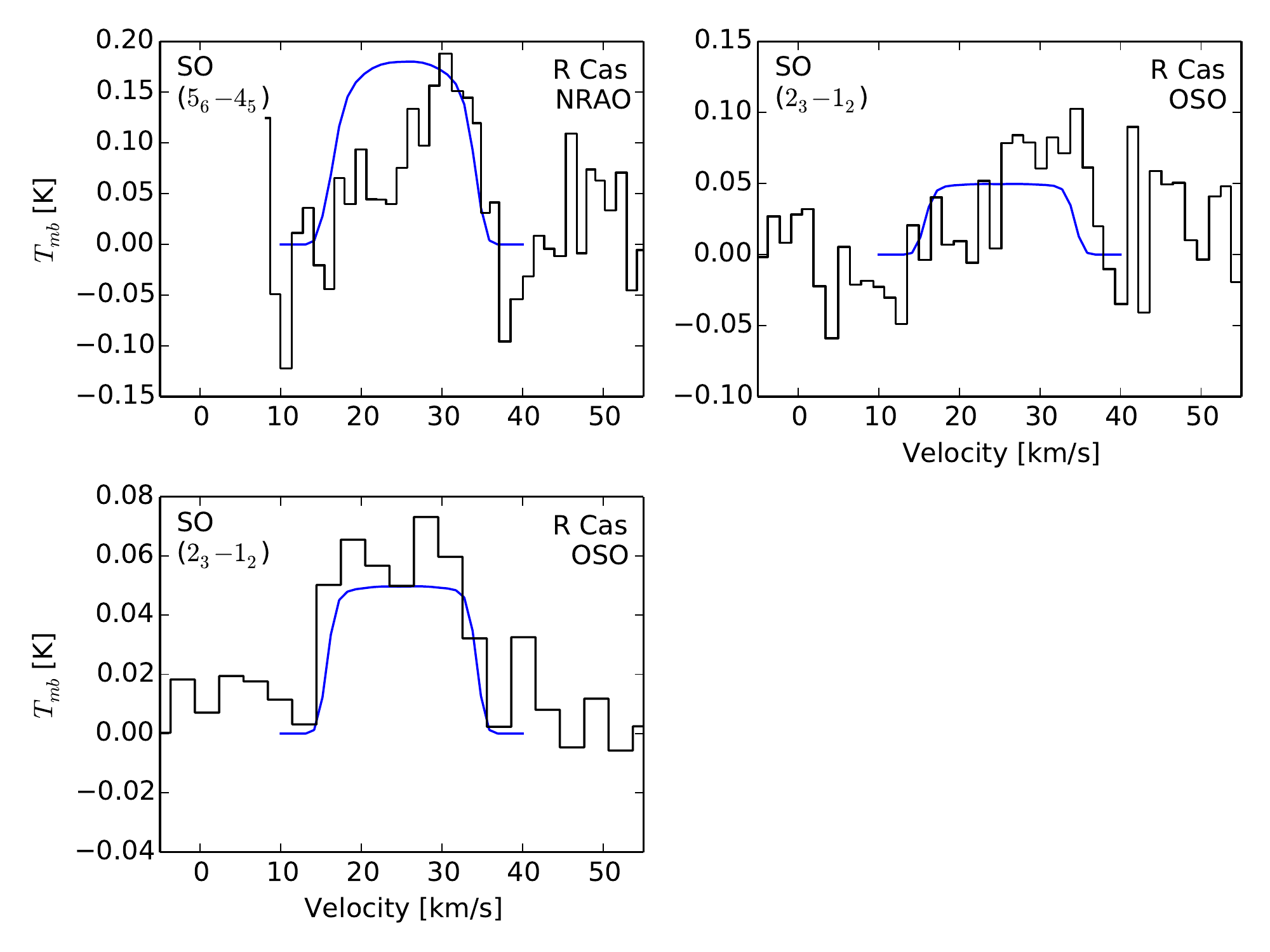}
\caption{SO models (blue lines) and observations (black histograms) for R~Cas.}
\label{rcasSOlines}
\end{figure}

As with IK~Tau, we find that a model with a centrally peaked Gaussian distribution of SO does not match the observed data. We again run a three-parameter grid to find the best shell-model fit to the data and find $f_p = (6.0\pm1.2)\e{-6}$, $R_p = (3.2\pm0.3)\e{15}$~cm, and $R_w = 1.0R_p$~cm (gridded in steps of $0.2R_p$), with $\chi^2_\mathrm{red} = 3.12$. The resultant model lines are shown in Fig. \ref{rcasSOlines} with the observations. The $\chi^2$ plot for SO in R~Cas is shown in Fig. \ref{SOchi2} and the goodness of fit plot is included in Fig. \ref{SOfits}.

\begin{figure}[t]
\begin{center}
\includegraphics[width=0.26\textwidth]{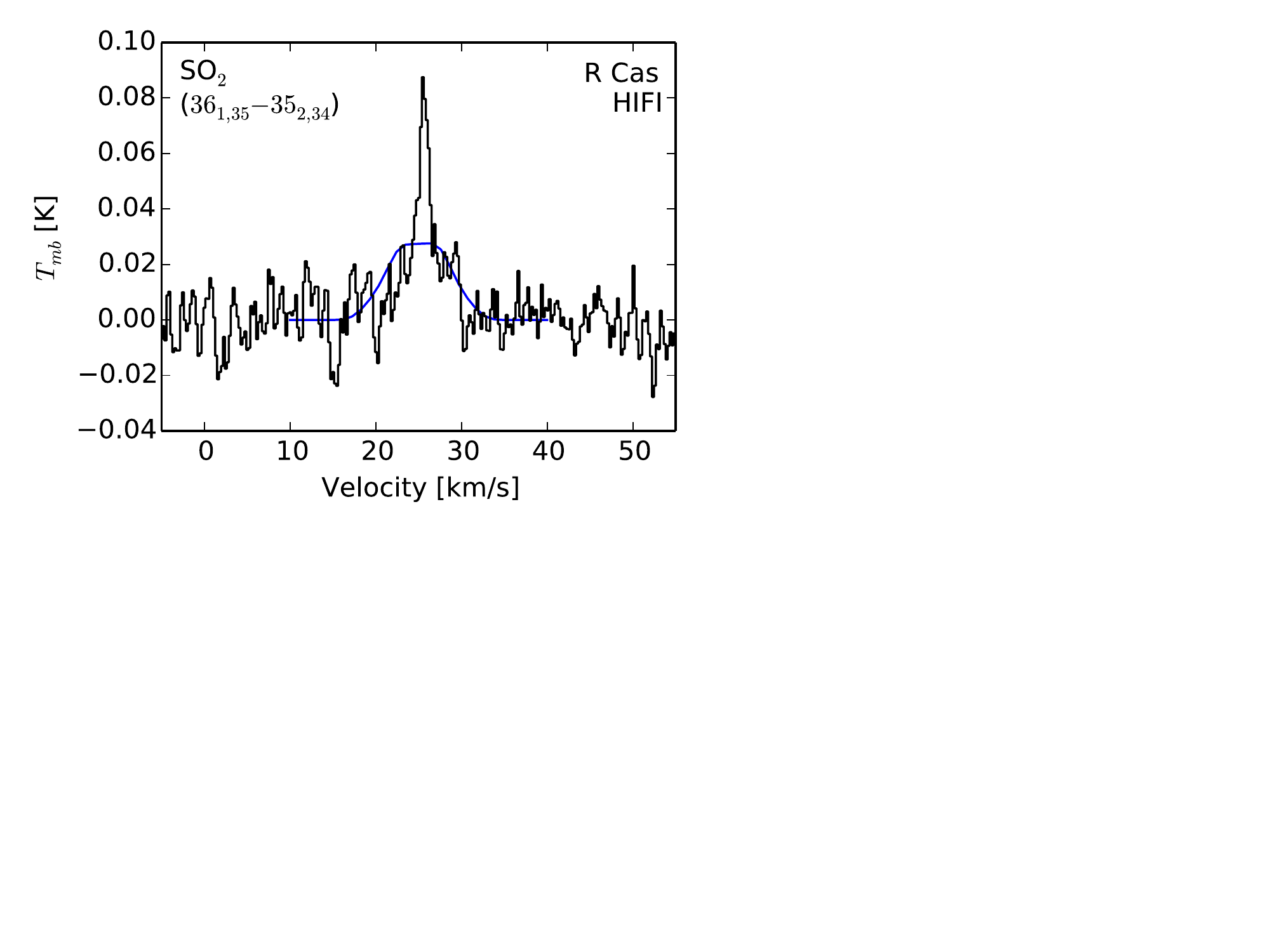}
\end{center}
\caption{\so2 model (blue line) and observation (black histogram) for R~Cas.} 
\label{rcasSO2}
\end{figure}

As there are only 2 \so2 lines observed towards R~Cas, we are only able to find an approximate model for \so2. As with IK~Tau, a shell-like model based on the R~Cas SO results does not fit the \so2 observations. Our best model has $f_p = 7\e{-6}$ (best within steps of $1\e{-6}$) and $R_e = 6\e{15}$~cm (best within steps of $1\e{15}$~cm). In Fig. \ref{rcasSO2} we plot the HIFI detection with our model.
An abundance plot for SO, \so2, CO and \h2O towards R~Cas is shown in Fig. \ref{abundances}.

We note that the HIFI detection in Fig. \ref{rcasSO2} has a central narrow peak, much narrower than the gas expansion velocity, This skews the overall integrated line intensity somewhat. Interestingly, IK~Tau has a similar narrow peak in the same transition line (see Fig. \ref{iktauso2lines}), also at approximately the stellar velocity. R~Dor does not have such a peak and W~Hya may have one which is significantly less bright with respect to the rest of the emission line. The cause of this feature is unclear.

\subsubsection{TX~Cam}

\begin{figure}[t]
\includegraphics[width=0.5\textwidth]{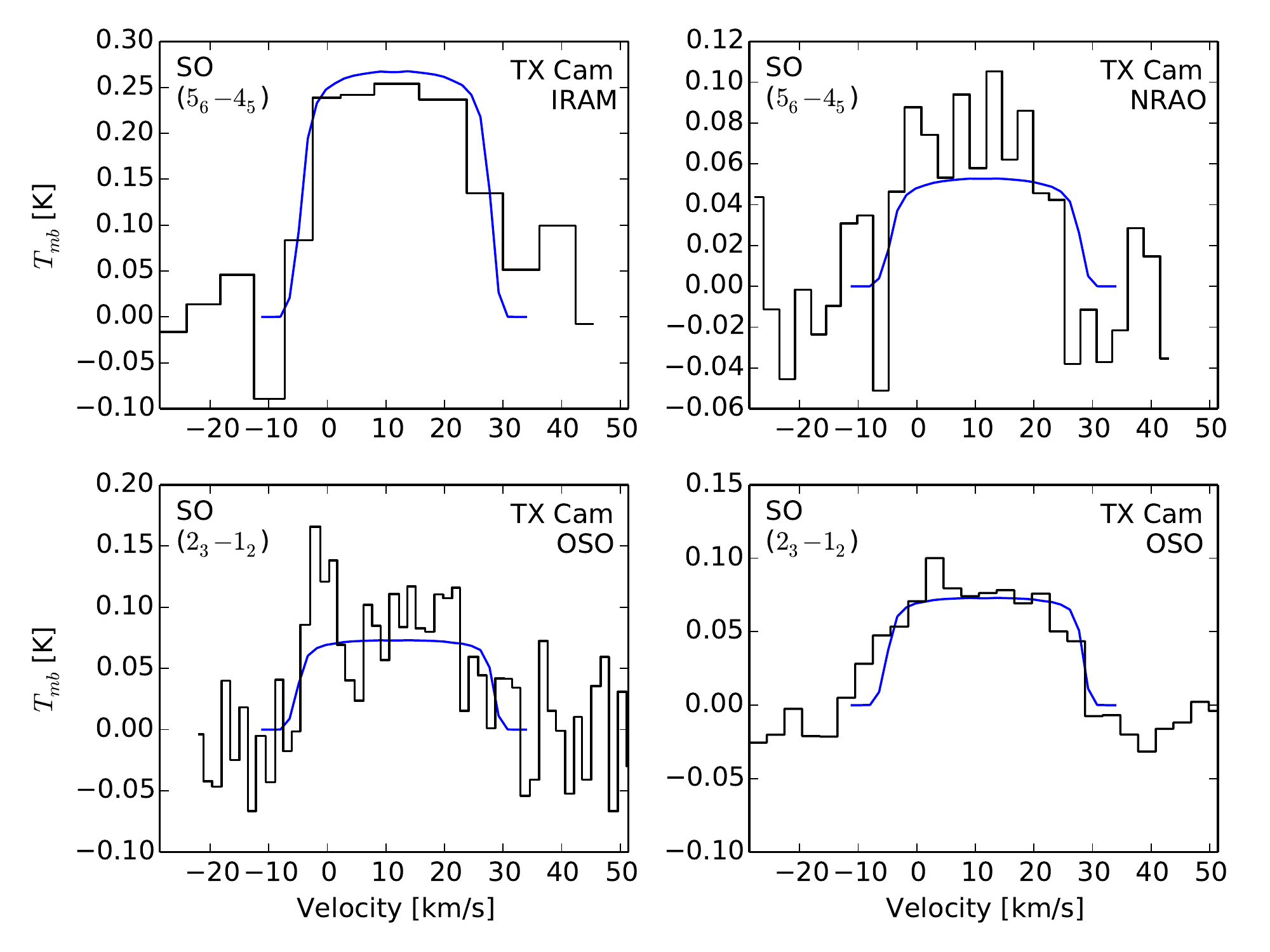}
\caption{SO models (blue lines) and observations (black histograms) for TX~Cam. We only plot the archival SO detections, not the non-detections from HIFI.}
\label{txcamSOlines}
\end{figure}

In the case of SO towards TX~Cam, we do not have sufficient constraints to run a full grid and perform a $\chi^2$ analysis as we did for the other stars. Instead we aim to fit the archival lines and find the largest $R_w$ allowed by the HIFI non-detections. The best model with these assumptions has $f_p = 1.7\e{-6}$, $R_p = 1.4\e{16}$~cm and $R_w = 1.6R_p$. We plot the detected lines with our model in Fig. \ref{txcamSOlines}. The goodness of fit, represented by the ratio between model integrated intensities and observed integrated intensities, is plotted in Fig. \ref{SOfits}. 
We stress that the dearth of observational results leaves our model poorly constrained and this is just one possible model that fits the available data. We can, however, rule out a centrally peaked model, as in that case we would expect the HIFI lines to have been detected, given the constraints on the fit from the archival lines.

There were no \so2 lines detected towards TX~Cam.

\begin{table*}[tp]
\caption{SO and \so2 model results}\label{resultstab}
\begin{center}
\begin{tabular}{lccccc}
\hline\hline
	&	IK~Tau	&	R~Dor	&	 TX~Cam	&	  W~Hya 	&	R~Cas 	\\
\hline											
$f_{p, \mathrm{SO}}$ $\e{-6}$ 	&	 $1.0\pm0.2$ 	&	$6.7\pm0.9$ 	&	 1.7*	&	 $5.0\pm1.0$ 	&	 $6.0\pm1.2$ 	\\
$R_{e,\mathrm{SO}}$ [$\e{15}$ cm] 	&	 - 	&	 $1.4\pm0.2$ 	&	 -	&	 $1.5\pm0.5$	&	 - 	\\
$R_{p,\mathrm{SO}}$ [$\e{15}$ cm] 	&	 $13\pm2$ 	&	 - 	&	 14*	&	- 	&	 $3.2\pm0.3$	\\
$R_{w,\mathrm{SO}}$ [$\times R_{p,\mathrm{SO}}$] 	&	1.8	&	 - 	&	 1.6*	&	-	&	1.0	\\
$\chi^2_\mathrm{red}$ (SO)	&	4.7	&	0.9	&	 -	&	2.6	&	3.1	\\
$N_\mathrm{SO}$	&	10	&	17	&	 4 (4)	&	7	&	7	\\
\hline											
$f_{p,\mathrm{SO}_2}$ $\e{-6}$ 	&	0.86	&	5.0	&	-	&	5.0	&	7	\\
$R_{e,\mathrm{SO}_2}$ [$\e{15}$cm]	&	10	&	1.6	&	-	&	3.0	&	6	\\
$\chi^2_\mathrm{red}$ (\so2)	&	14.3	&	3.7	&	-	&	5.7	&	-	\\
$N_{\mathrm{SO}_2}$ 	&	14	&	98	&	0	&	5	&	2	\\ 
\hline
\end{tabular}
\end{center}
\tablefoot{$N$ is the number of lines used to constrain our models. The uncertainties are for the 90\% confidence level. The number in brackets for $N$ is the number of upper limits used in addition to the detected lines. Values marked with a * indicate an upper-limit model.}
\end{table*}%

\section{Discussion}

\subsection{SO distribution}\label{sodistdisc}

\begin{figure}[t]
\includegraphics[width=0.5\textwidth]{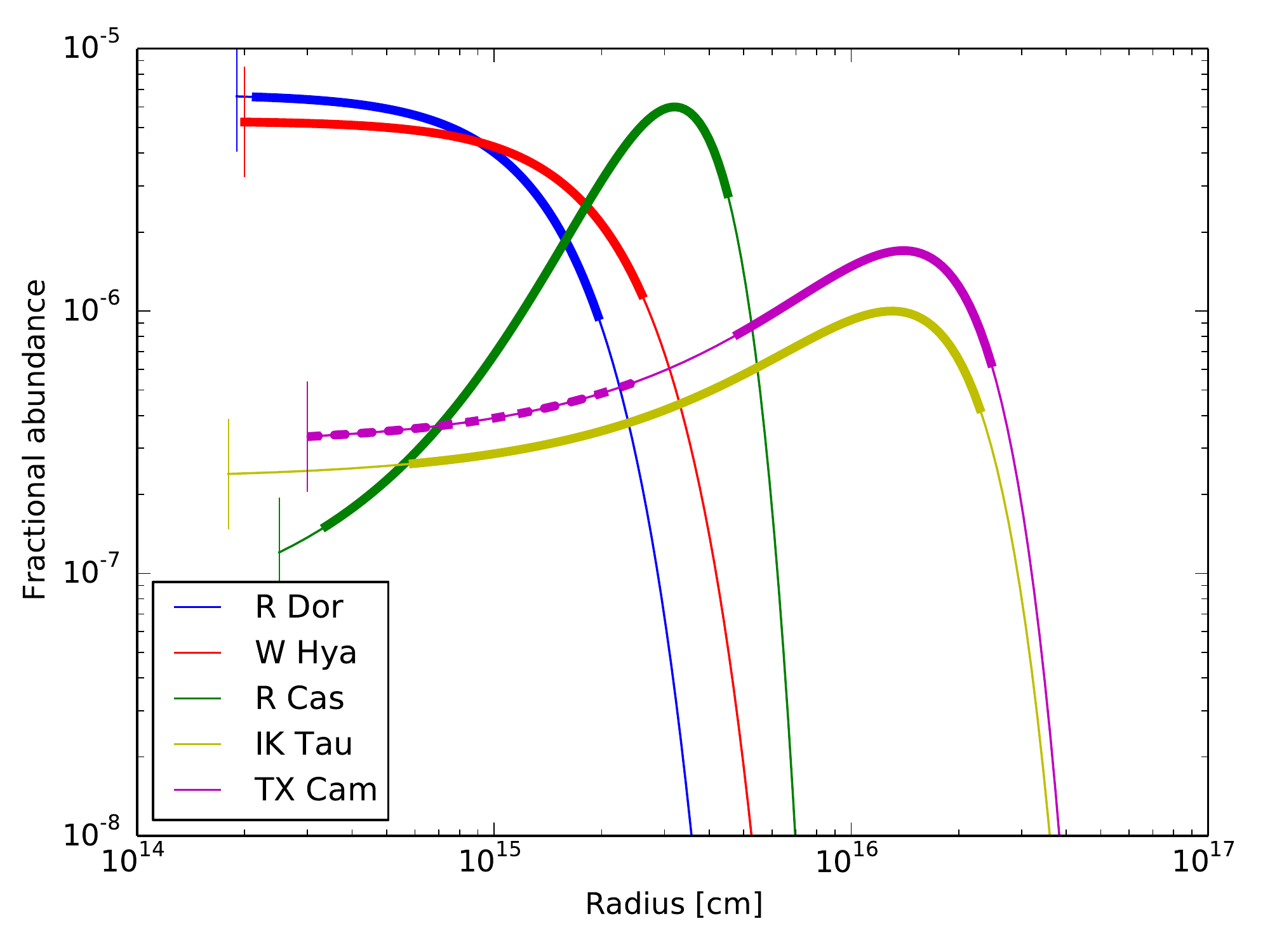}
\caption{SO abundance distributions for all stars modelled. The vertical lines represent the dust condensation radii, where our models stop. The thicker sections of the curves represent the area probed by our observations and for TX Cam the thick dashed line is the area probed by the upper limits imposed by the HIFI non-detections.}
\label{allSOabundances}
\end{figure}

Our results for circumstellar SO are summarised in Table~\ref{resultstab}. In Fig. \ref{allSOabundances} we plot the circumstellar SO abundance profiles of the stars we modelled. We also show the radial range probed by the available observational data for each line with a thicker line. These ranges were found by considering the brightness distributions for each emission line and the radii at which these fall to half of their maximum values. 
It is interesting to note that for R~Cas, IK~Tau, and TX~Cam, the three stars with shell-like SO distributions, the location of the peak is found progressively further out with increasing mass-loss rate. The two low mass-loss rate stars, however, both seem to have centrally peaked SO distributions, which could be interpreted as shells with peaks close to the star, especially if the peaks are near or within our inner radii. Looking at the three stars with shell-like SO distributions, there also seems to be a trend of decreasing SO abundance with increasing mass-loss rate (or with the radius of peak SO abundance).

\begin{figure}[t]
\includegraphics[width=0.5\textwidth]{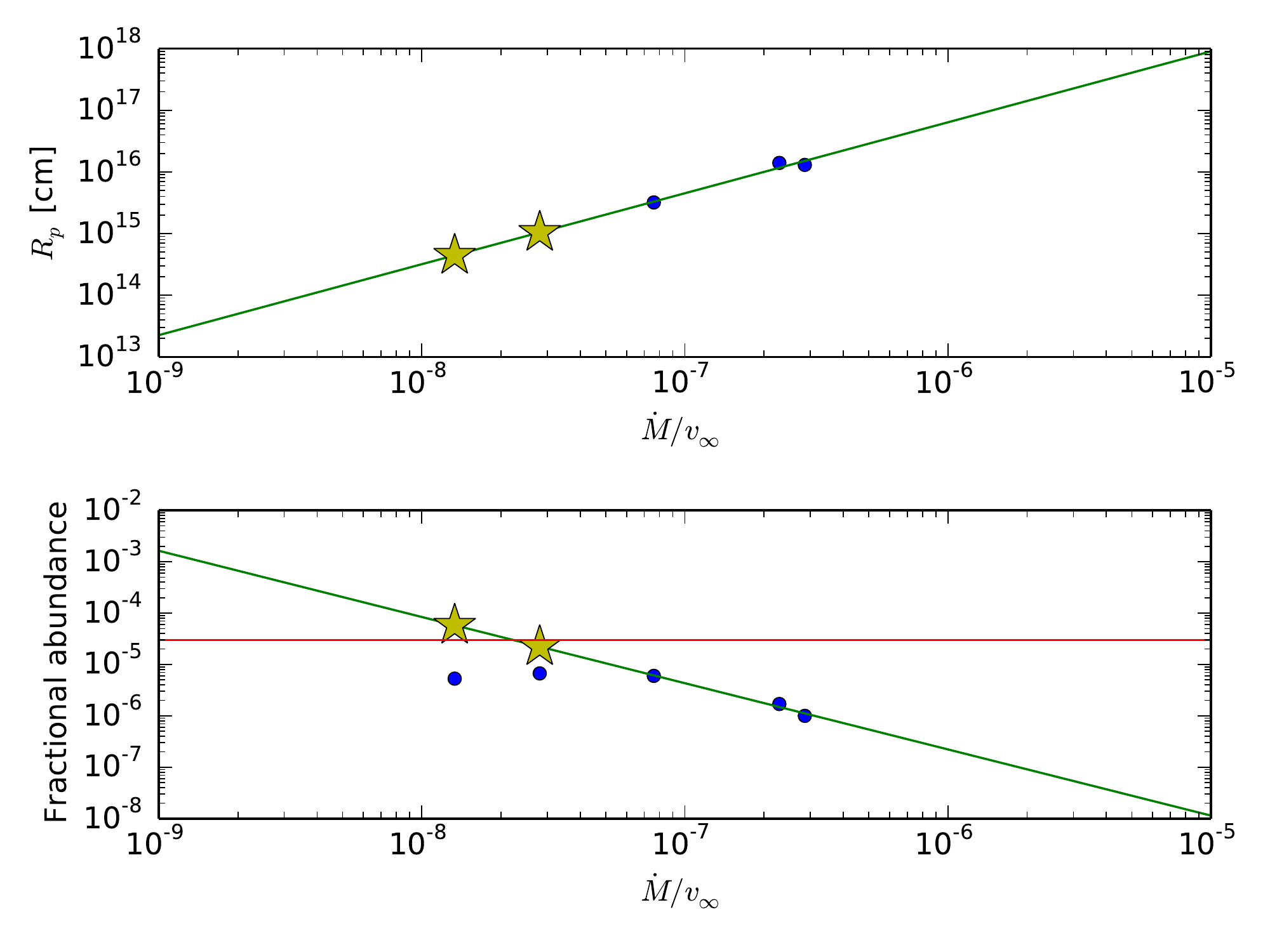}
\caption{Trends in SO peak radius (\textit{top}) and fractional abundance (\textit{bottom}) against the circumstellar density measure, $\dot{M}/\upsilon_\infty$. The green lines are the trends fitted to the three higher mass-loss rate stars (R Cas, TX Cam and IK Tau), while the yellow stars are the predicted locations of the lower mass-loss rate stars (R Dor and W Hya) based on the trend. In the fractional abundance plot, the red line shows the hard limit for SO abundance based on solar S abundance and the blue points in line with the yellow stars represent the real abundance values for R Dor and W Hya.}
\label{SOtrends}
\end{figure}

In Fig. \ref{SOtrends} we plot the peak positions against the wind density, ${\dot{M}}/{\upsilon_\infty}$, and fit a power law to the three higher mass-loss rate stars (R~Cas, TX~Cam and IK~Tau). The results for these stars are well fit by a power law
\begin{equation}
R_p \propto \left( \frac{\dot{M}}{\upsilon_\infty} \right)^{\alpha_R}
\end{equation}
with $\alpha_R=1.15\pm0.24$. We extend the power law to predict the peak positions for R~Dor and W~Hya, were they also to fit this trend. This predicts the peak in SO abundance for R~Dor to lie at $1.0\e{15}$~cm, which is close to the $R_e$ we found, and for W~Hya the predicted SO peak lies at $4.4\e{14}$~cm, about three times smaller than the $e$-folding radius. Running models with shell-like distributions at the predicted peaks, we found they could not provide as good a fit for either R~Dor or W~Hya as the star-centred Gaussian models. In general, the best shell models had at least twice the $\chi^2$ values of the best star-centred Gaussian models. Given the region probed by our observations as shown in Fig. \ref{allSOabundances}, it is not surprising that changing the inner abundance would have an effect on the model fit.

We performed a similar fit for the peak abundance values against density for the three highest mass-loss rate stars. The results for these stars  are well fit by a power law
\begin{equation}
f_p \propto \left( \frac{\dot{M}}{\upsilon_\infty} \right)^{\alpha_f}
\end{equation}
with $\alpha_f=-1.29\pm0.17$, Fig.~\ref{SOtrends}. Doing a similar extrapolation to predict the peak abundance values for R~Dor and W~Hya based on the power law, we find fractional abundance predictions of $2.2\e{-5}$ for R~Dor and $5.8\e{-5}$ for W~Hya. Both of these are higher than the values we find from our modelling and in the case of W~Hya this represents more sulphur than should be available, i.e. it exceeds the solar and ISM abundances (see below). Because the SO abundance cannot increase with decreased mass-loss rate indefinitely, there must be a maximum SO abundance set by the abundance of sulphur. 



\begin{figure}[t]
\includegraphics[width=0.5\textwidth]{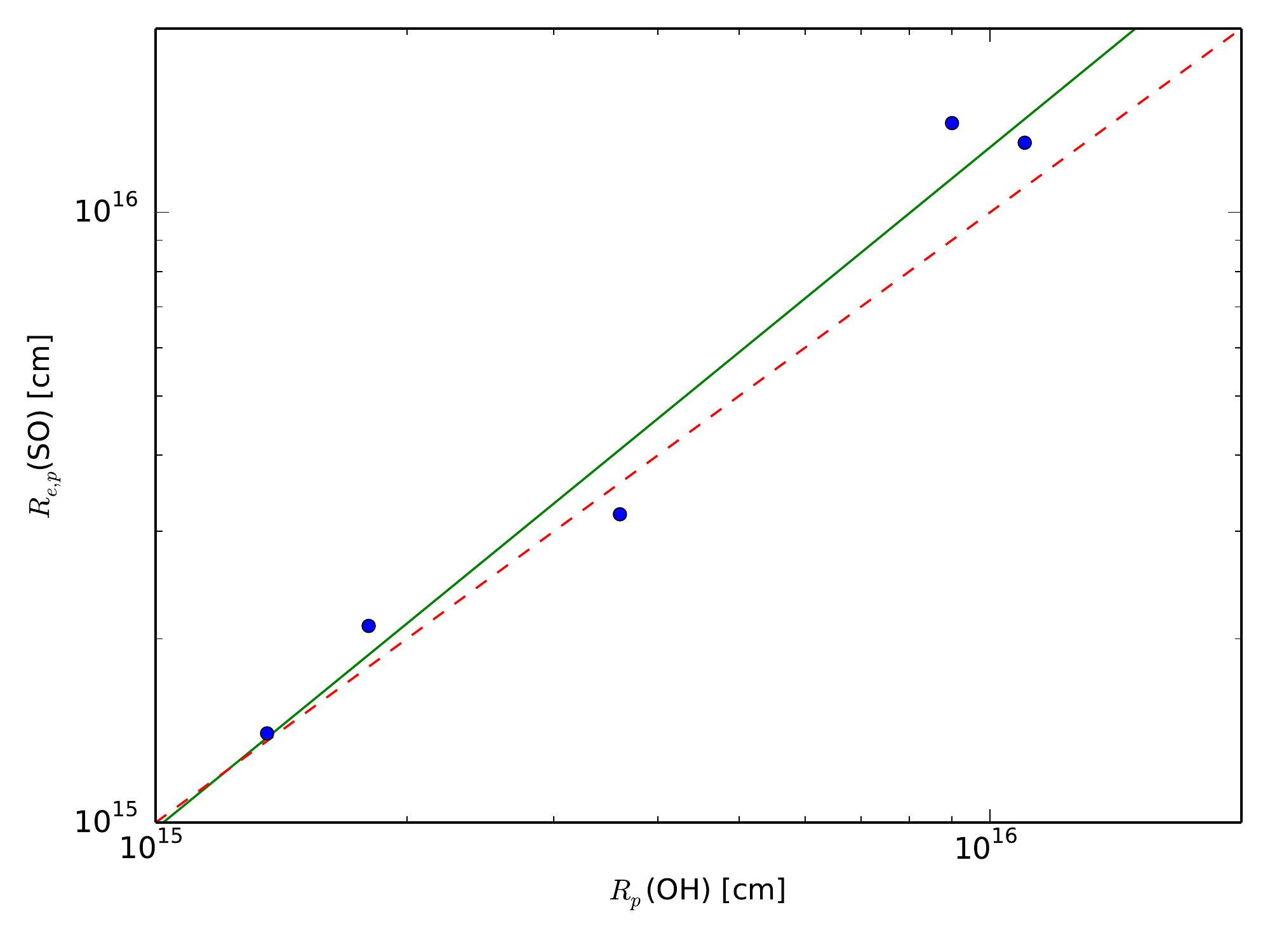}
\caption{The peak abundance radii of SO (for R~Cas, TX~Cam, and IK~Tau) and the $e$-folding radii of SO (for R~Dor and W~Hya), plotted against the $e$-folding radii of \h2O found by Maercker et al (\textit{in prep.}) and \citet[][for W~Hya]{Khouri2014a}. The solid green line is the best fit to the data and the dashed red line traces a 1:1 relationship.}
\label{SOvsH2O}
\end{figure}

The shell-like distributions of SO for the three stars with the highest mass-loss rates means that circumstellar chemistry, most likely related to photodissociation, must play an important role, in these cases. It is therefore interesting to compare with results on photodissociation for other species. H$_2$O is particularly interesting here, since, as will be discussed below, SO (and also SO$_2$) may owe its origin to the presence of circumstellar OH, which in turn is a photodissociation product of H$_2$O. \cite{Netzer1987} predict a peak OH radius that scales with both mass-loss rate and expansion velocity. Their formulation has been used by \cite{Maercker2008,Maercker2009,Schoier2011,Danilovich2014} and others to define the $e$-folding radius of \h2O, since OH is a photodissociation product of \h2O and peaks in abundance where \h2O drops off. In Fig. \ref{SOvsH2O} we plot our SO peak abundance radii or $e$-folding radii (as relevant) against the \h2O $e$-folding radii of the same stars found by Maercker et al. (\textit{in prep.}) and \citet[][for W~Hya]{Khouri2014a}. We find a strong correlation, which is close to being 1:1. The chemistry of SO will be discussed in Sect.~\ref{s:chem}.

\subsection{SO$_2$ distribution}


Our results for circumstellar SO$_2$ are summarised in Table~\ref{resultstab}. For R~Dor and W~Hya we find circumstellar \so2 distributions of similar size and abundance to the circumstellar SO distributions. Our results for the higher mass-loss rate stars, however, are less clear. For both IK~Tau and R~Cas, the best models are centrally-peaked Gaussian distributions of \so2, rather than the shell-like models we found for SO. For IK~Tau, we were unable to constrain the $e$-folding radius better than by a factor of 2 and for R~Cas we only had two observations, but in both cases shell models similar to the corresponding SO models are ruled out. 
Our \so2 results for IK~Tau and R~Cas suggest that \so2 is formed in the inner regions irrespective of the mass-loss rate.
It appears that \so2 is formed more favourably than SO in the inner regions. There does not appear to be a strong correlation between $e$-folding radius and mass-loss rates or \h2O $e$-folding radii, as we found for SO, but this might become clearer if we add more \so2 observations to our models, especially for R~Cas and W~Hya. We emphasise that our SO$_2$ results for the higher mass-loss rate stars, R~Cas and IK~Tau, are particularly uncertain.




\subsection{Comparisons with chemical models}
\label{s:chem}

There exist two studies of the abundances of SO and SO$_2$ in the extended atmospheres of AGB stars \citep{Cherchneff2006,Gobrecht2016}. In both cases the effects of shock-induced chemistry, due to pulsational motion, are included. \citet{Cherchneff2006} has chosen TX~Cam as the representative star, and the modelling covers the region from 1 to 5 stellar radii, $R_*$, which means that the outer reach of their model is approximately an order of magnitude smaller than our inner radii. For their model star with C/O = 0.75, they find an SO abundance at $5R_*$ of $3\e{-7}$, and an SO$_2$ abundance several orders of magnitude lower than this. \citet{Gobrecht2016} use IK~Tau as the example, and extend the calculations to 10 stellar radii (the outer radius of their model then approximately meets the inner radius used by us). Their model focuses on the shock chemistry in this region, much of which varies with the pulsation phase. Their abundances of SO and \so2 are much lower than we observe, at $\sim10^{-8}$ and $\sim2\e{-9}$, respectively. This means that the predicted SO and \so2 abundances close to the star, at least for these higher mass-loss rate stars, are substantially lower than we derive for our sample stars.

The shell-like SO distributions for the higher mass-loss rate stars suggest a circumstellar origin. \citet{Willacy1997} describe circumstellar chemical models of four M-type AGB stars, including R~Dor, TX~Cam, and IK~Tau. Their models differ from ours in terms of CSE parameters, for example taking the inner radius to be $2\e{15}$~cm, about an order of magnitude larger than our inner radii. They assume that all the sulphur is carried by H$_2$S that is eventually photodissociated. SO is subsequently formed through the following reactions
\begin{eqnarray}
\mathrm{S} + \mathrm{OH} \to \mathrm{SO} + \mathrm{H}\label{soform}\\
\mathrm{SH} + \mathrm{O} \to \mathrm{SO} + \mathrm{H}
\end{eqnarray}
which are favoured depending on the availability of OH and SH, respectively, and with Eq. \ref{soform} dominating at the lower gas temperatures in the CSE. Following this, SO can be destroyed through
\begin{equation}
\mathrm{SO} + \mathrm{OH} \longleftrightarrow \mathrm{SO}_2 + \mathrm{H}\label{so2form}
\end{equation}
and hence form SO$_2$. Unfortunately, they only visualise their model results for TX~Cam, the least well-constrained star of our sample. Nevertheless, the location of the peak of the SO distribution that we find for TX~Cam agrees quite well with their predicted peak location. Our peak abundance is about 50\% higher than theirs, but this must be considered to be within the errors. Their \so2 distribution for TX Cam peaks at roughly the same radius as SO, but with a peak abundance about an order of magnitude lower than that of SO (as we do not have any \so2 detections for TX~Cam, we cannot compare this directly). 
The molecular column densities they list for R~Dor, IK~Tau, and TX~Cam are, in general, a few orders of magnitude lower than those predicted by our models.

In conclusion, our SO results for the outer CSE are reasonably consistent with the results of \citet{Willacy1997} for the higher mass-loss rate objects, although they do not predict a peak abundance that decreases with mass-loss rate. Thus, an origin through OH is likely, a result that is further strengthened by the correlation between SO and H$_2$O sizes that we found. However, we note that neither \citet{Cherchneff2006} nor \citet{Gobrecht2016} predict high abundances of H$_2$S in the upper atmosphere. For the lower mass-loss rate objects, where the SO abundance is high close to the star, the models of \citet{Cherchneff2006} and \citet{Gobrecht2016} fail by more than two orders of magnitude to reproduce our estimated abundances. They even fail to reproduce the inner SO abundances for the higher mass-loss rate stars.

In the case of SO$_2$ we find no evidence for a photo-induced circumstellar origin along the \citet{Willacy1997} model for any of our objects. Once again, we caution that the results for R~Cas and IK~Tau are uncertain. The SO$_2$ abundances that we estimate for R~Dor and W~Hya are an order of magnitude higher than those predicted by \citet{Cherchneff2006} and \citet{Gobrecht2016}.

\subsection{Sulphur chemistry: can we account for all the sulphur?}\label{sulphuraccount}


AGB stars and their progenitors do not produce S via nucleosynthesis. As such, the quantity of sulphur available to form molecules in the CSE of an AGB star is fixed and not dependent on the stage of evolution or mass of the star in question. \cite{Rudolph2006} find an S/H ratio in the ISM of $\sim10^{-5}$ in the solar neighbourhood and \cite{Lodders2003} indicate a solar S/H abundance of $1.5\e{-5}$. All stars in our sample are at distances $<\,400$~pc and hence can be assumed to trace a similar S/H abundance. In this work we refer to the fractional molecular abundance with respect to \h2. Hence, assuming all hydrogen is in the form of \h2 in the CSE, we will take the S/\h2 ratio to be $\sim 2$ to $3\e{-5}$, which represents the maximum total amount of sulphur that should be found in an AGB star.

For R~Dor and W~Hya we find combined SO and \so2 abundances of $\sim 1.2\e{-5}$ and $\sim 1.0\e{-5}$, respectively. Hence, in these cases most of the sulphur is locked up in SO and \so2 within the inner regions of the CSE and within the errors. This result is consistent with the non-detections (or low-level emission) for CS and SiS in the APEX spectral scan of R~Dor, and no reported detections of these species towards W~Hya. In the case of R~Cas, the combined SO and  \so2 abundances in the mid-CSE is $\sim 1.4\e{-5}$, suggesting that these two species carry all the sulphur, but here the uncertainty on the \so2 abundance is substantial. For the high mass-loss rate object IK Tau, the combined SO and \so2 abundance is well below that of sulphur.

In general, higher mass-loss rate stars also show definitive detections of other S-bearing molecules. For example, SiS was detected in several carbon and M-type stars by \citet{Schoier2007} and \cite{Danilovich2015a}. \citet{Schoier2007} reported circumstellar SiS abundances of $4\e{-7}$, $4\e{-7}$, and $1\e{-7}$ for R~Cas, TX~Cam, and IK~Tau, respectively, suggesting that SiS is less abundant than SO and \so2 by up to an order of magnitude, at least in the outer CSE. It should be noted that \citet{Schoier2007} assume a Gaussian distribution of SiS centred on the star, but find their model fit greatly improved when they include a high-abundance inner component, which could represent the SiS reservoir before depletion through dust condensation. \citet{Decin2010}, who model IK~Tau in detail, also find evidence of depletion of SiS.

Another S-bearing molecule, CS, has mainly been detected in carbon stars rather than M-type stars. CS has been detected and modelled in IK~Tau by \citet{Kim2010} and \citet{Decin2010}, detected in TX~Cam and IK~Tau by \citet[][with non-detections in R~Cas and W~Hya]{Bujarrabal1994} and \citet[][with a non-detection in R~Cas]{Lindqvist1988}. Derived CS abundances for M-type stars have generally been low, in the range $\sim10^{-8}$ to $\sim5\e{-8}$ \citep[see][for examples]{Bujarrabal1994,Decin2010}.

\h2S is considered as a parent species of sulphur in the chemical modelling of \citet{Willacy1997}. However, \h2S has not been widely detected in AGB stars other than in OH/IR stars. For example, in the HIFISTARS project \h2S was only detected in AFGL~5379 \citep{Justtanont2012}, despite being in the observed range for all stars except TX~Cam, while \cite{Justtanont2015} detected \h2S in all OH/IR stars observed with SPIRE and some observed with PACS. In a study of 25 stars, \cite{Ukita1983} detected \h2S only in OH231.8+4.2 aka the Rotten Egg Nebula. \cite{Omont1993} detected \h2S in several high mass-loss rate stars, including several OH/IR stars. Of the stars we modelled, \cite{Ukita1983} did not detect \h2S in W~Hya, R~Cas, and TX~Cam, but it was detected in IK~Tau by \cite{Omont1993} and De Beck et al. (\textit{in prep.}).  This suggests that \h2S may require high densities to form, or may be able to survive longer in the CSEs of high mass-loss rate stars, or that the excitation conditions are such that the emission is only bright enough in the very high mass-loss rate stars to be detectable. 
We note that \cite{Gobrecht2016} predict a fairly rapid decline of \h2S inside of the dust condensation radius (which is where our models start) even for IK~Tau, which is a relatively high mass-loss rate object.

To check what the lack of detections predicts in terms of \h2S abundances, we run radiative transfer models for R~Dor and IK~Tau to find upper limits for the \h2S abundances based on the non-detections in HIFI, the non-detection in APEX for R~Dor, and the SMA detection in IK~Tau by De Beck et al. (\textit{in prep.}). We used the ortho-\h2S molecular data file available on LAMDA\footnote{The Leiden Atomic and Molecular Database, found at \texttt{http://home.strw.leidenuniv.nl/$\sim$moldata/}} \citep{Schoier2005} which includes the lowest 45 rotational energy levels, 139 radiative transitions with frequencies taken from JPL\footnote{\tt http://spec.jpl.nasa.gov/} and 990 collisional transitions taken from \cite{Dubernet2009} for temperatures from 5--1500~K. 
For IK~Tau, 
using the detection from De Beck et al. (\textit{in prep.}) and the HIFI upper limit to also constrain the envelope size, we find a small envelope with $R_e\simeq4\e{14}$~cm and $f_p \simeq 4\e{-6}$. This is consistent with a rapid destruction of \h2S.
For R~Dor, using both non-detections and assuming the $R_e$ we find for SO, we find an upper limit on the abundance of $f_p \lesssim 2.5\e{-7}$. If we instead use the \h2S envelope size found for IK~Tau, the abundance upper limit increases slightly to $f_p \lesssim 6\e{-7}$. In any case, these results limit the possibility that \h2S is a significant S-carrier in the inner CSE, certainly for the low mass-loss rate objects.


None of the molecules discussed thus far have been found in sufficient quantities towards higher mass-loss rate AGB stars to account for the full amount of expected sulphur. It is possible that the remaining sulphur is locked up in dust or left as atomic S or locked up in molecules that are difficult to detect for various reasons, such as the spectral region they are most likely to emit in, as is the case with HS. Both \cite{Cherchneff2006} and  \cite{Willacy1997} predict a rapid decline of HS with radius as it is consumed by various chemical processes (although we note that the two studies make predictions for different regions around the star). The only detection of HS in the literature is through ro-vibrational lines identified by \cite{Yamamura2000} 
towards R~And (an S-type AGB star). 
They estimate a molecular abundance of \mbox{HS/H $\sim 1\e{-7}$}, which is well below the sulphur limit. 
There have been no other detections of circumstellar HS, although it has been detected in the ISM \citep[see e.g.][]{Neufeld2015}.


To fully study the issue of sulphur in the CSEs of AGB stars of different mass-loss rates, a more thorough investigation including more molecular species --- such as SiS, CS, and \h2S in addition to SO and \so2 --- across a larger sample of stars is needed.

\section{Conclusions}

We present new APEX observations of a very large number of SO and \so2 lines towards the low mass-loss rate M-type AGB star R~Dor. Combining these data with higher-frequency observations from \textit{Herschel}/HIFI, we compute comprehensive radiative transfer models to determine the molecular abundances and distributions of the two molecules. For R~Dor we find a Gaussian abundance distribution centred on the star, with a peak SO fractional abundance of $(6.7\pm0.9)\e{-6}$ and $e$-folding radius of $(1.4\pm0.2)\e{15}$~cm, and an \so2 fractional abundance of $5.0\e{-6}$ and $e$-folding radius of $1.6\e{15}$~cm. Our \up{34}SO model assumes the same $e$-folding radius as for \up{32}SO and we find an abundance of $(3.1\pm0.8)\e{-7}$. This gives an \up{32}SO/\up{34}SO ratio of $21.6\pm8.5$, which is in agreement with previous results from other nearby stars.

We also model SO in four other M-type AGB stars that were observed as part of HIFISTARS: IK~Tau, TX~Cam, W~Hya, and R~Cas. For TX~Cam for we are only able to provide an upper limit model since there are no SO lines detected with HIFI. Of these four stars only W~Hya has a similar SO distribution to R~Dor. The other three stars, all of which have higher mass-loss rates, are best fit with shell-like abundance distributions. We find that the radial position of the peak of the distributions increases with mass-loss rate, while the peak abundances decrease. The location of the peaks of the SO distributions correlates with the photodissociation of \h2O into OH (itself partly dependent on mass-loss rate), suggesting that the production of SO depends on the availability of OH to participate in the formation process.

We are only able to model \so2 in an additional three stars, IK~Tau, W~Hya, and R~Cas, owing to the dearth of detections towards TX~Cam. For W~Hya we find an \so2 distribution similar to SO in abundance and envelope size. We have some difficulty fitting an \so2 model to observations for IK~Tau and ultimately find an uncertain model which differs in shape from the SO distribution. For R~Cas the \so2 model is also very uncertain because there are only two detected lines.

Overall, the circumstellar SO and \so2 abundances are much higher than predicted by chemical models of the extended stellar atmosphere. These two species may also account for all the available sulphur in the lower mass-loss rate stars. The S-bearing parent molecule appears not to be \h2S. The \so2 models for the higher mass-loss rate stars are less conclusive, but suggest an origin close to the star for this species. This is not consistent with present chemical models. The combined circumstellar SO and \so2 abundances are significantly lower than that of sulphur for these higher mass-loss rate objects.

To better constrain the behaviour of sulphur we need more observations of SO and \so2, as well as other S-bearing species. Observations of a larger sample of stars will also allow us to confirm the trends we see in the SO abundance distributions.

\begin{acknowledgements}
TD and KJ acknowledge funding from the Swedish National Space Board. HO acknowledges financial support from the Swedish Research Council.

This publication is based on data acquired with the Atacama Pathfinder Experiment (APEX). APEX is a collaboration between the Max-Planck-Institut f\"ur Radioastronomie, the European Southern Observatory, and the Onsala Space Observatory.

HIFI has been designed and built by a consortium of institutes and university departments from across Europe, Canada and the United States under the leadership of SRON Netherlands Institute for Space Research, Groningen, The Netherlands and with major contributions from Germany, France and the US. Consortium members are: Canada: CSA, U.Waterloo; France: CESR, LAB, LERMA, IRAM; Germany: KOSMA, MPIfR, MPS; Ireland, NUI Maynooth; Italy: ASI, IFSI-INAF, Osservatorio Astrofisico di Arcetri-INAF; Netherlands: SRON, TUD; Poland: CAMK, CBK; Spain: Observatorio Astron\'omico Nacional (IGN), Centro de Astrobiolog\'ia (CSIC-INTA). Sweden: Chalmers University of Technology - MC2, RSS \& GARD; Onsala Space Observatory; Swedish National Space Board, Stockholm University - Stockholm Observatory; Switzerland: ETH Zurich, FHNW; USA: Caltech, JPL, NHSC.

\end{acknowledgements}

%

\bibliographystyle{../aa}
\bibliography{Sulphur}

\appendix

\section{R~Dor plots}

Our best fit model lines for \so2 in R~Dor are plotted along with the corresponding observations in Fig. \ref{rdorso2results}. For more details see Sect. \ref{rdorso2sect}.

The tentative detections of the isotopologues \up{34}\so2 and SO\up{18}O from the APEX survey towards R~Dor are plotted in Fig. \ref{leftoverisotopologues}.

\begin{figure*}[t]
\begin{center}
\includegraphics[width=\textwidth]{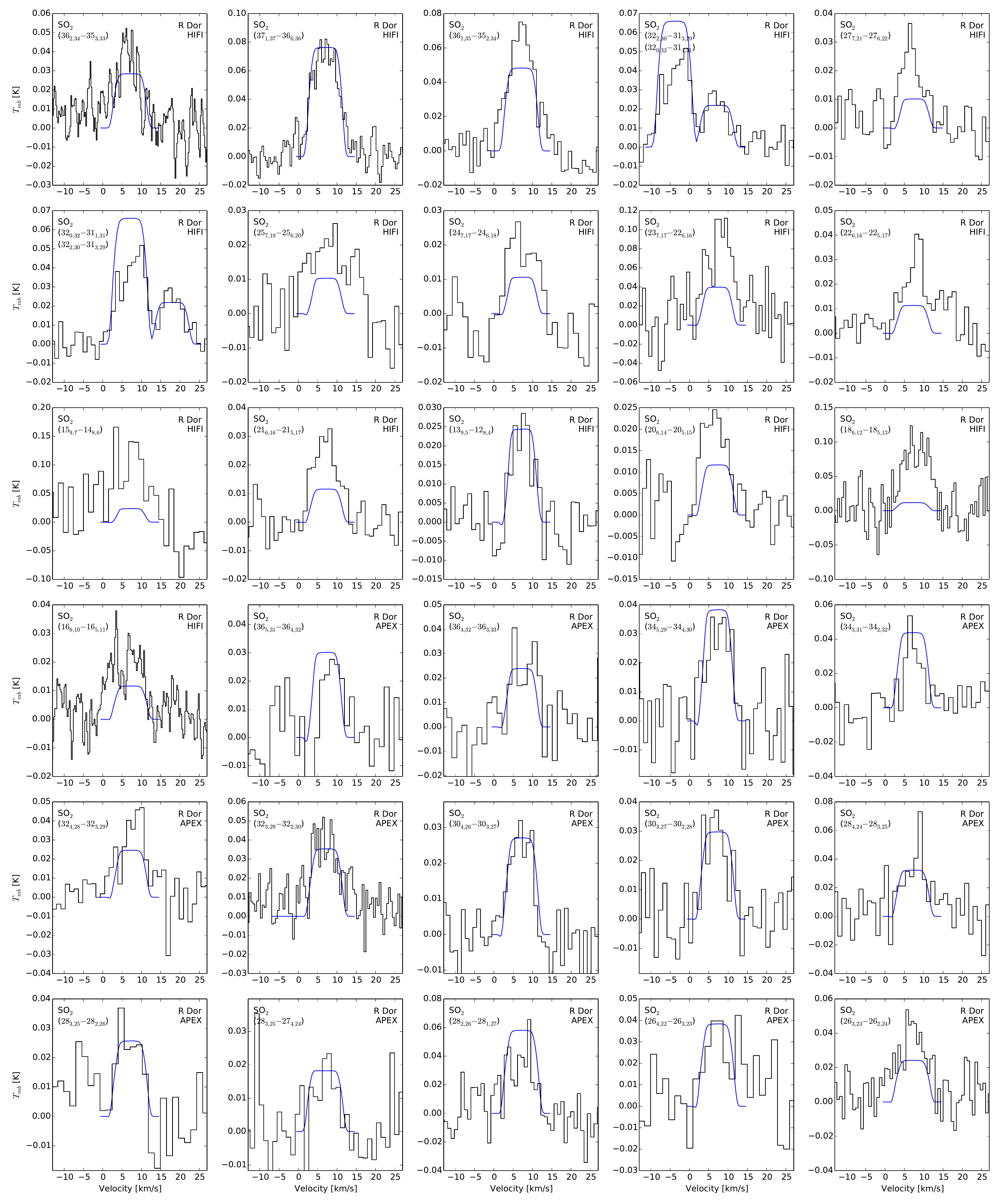}
\caption{\so2 model (blue line) and observations (black histograms) of R~Dor. In the case of overlapping lines, the top line listed is always the line centred at $\upsilon_\mathrm{LSR}=5.7~\kms$.}
\label{rdorso2results}
\end{center}
\end{figure*}
\addtocounter{figure}{-1}
\begin{figure*}[t]
\begin{center}
\includegraphics[width=\textwidth]{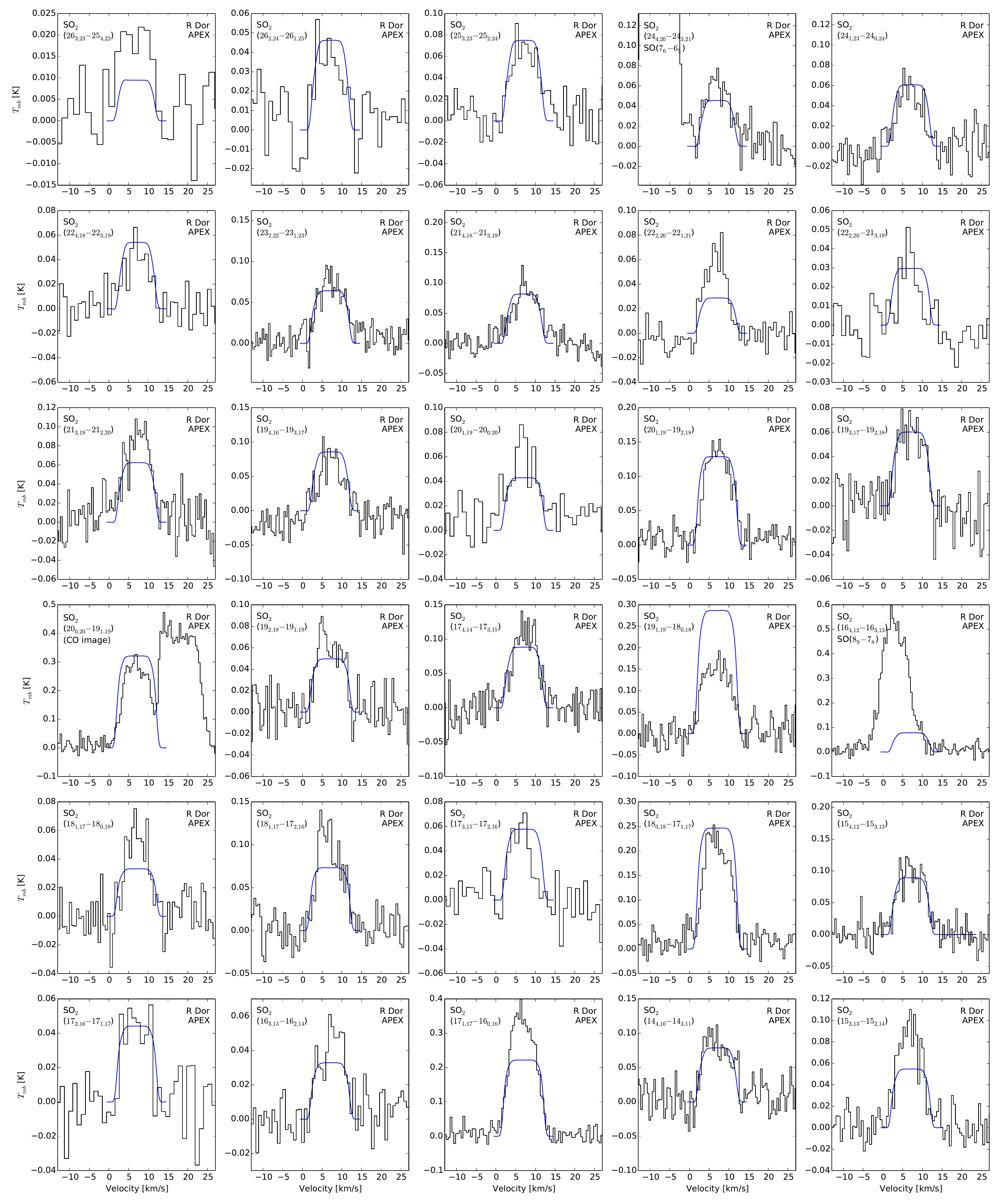}
\caption{continued.}
\end{center}
\end{figure*}
\addtocounter{figure}{-1}
\begin{figure*}[t]
\begin{center}
\includegraphics[width=\textwidth]{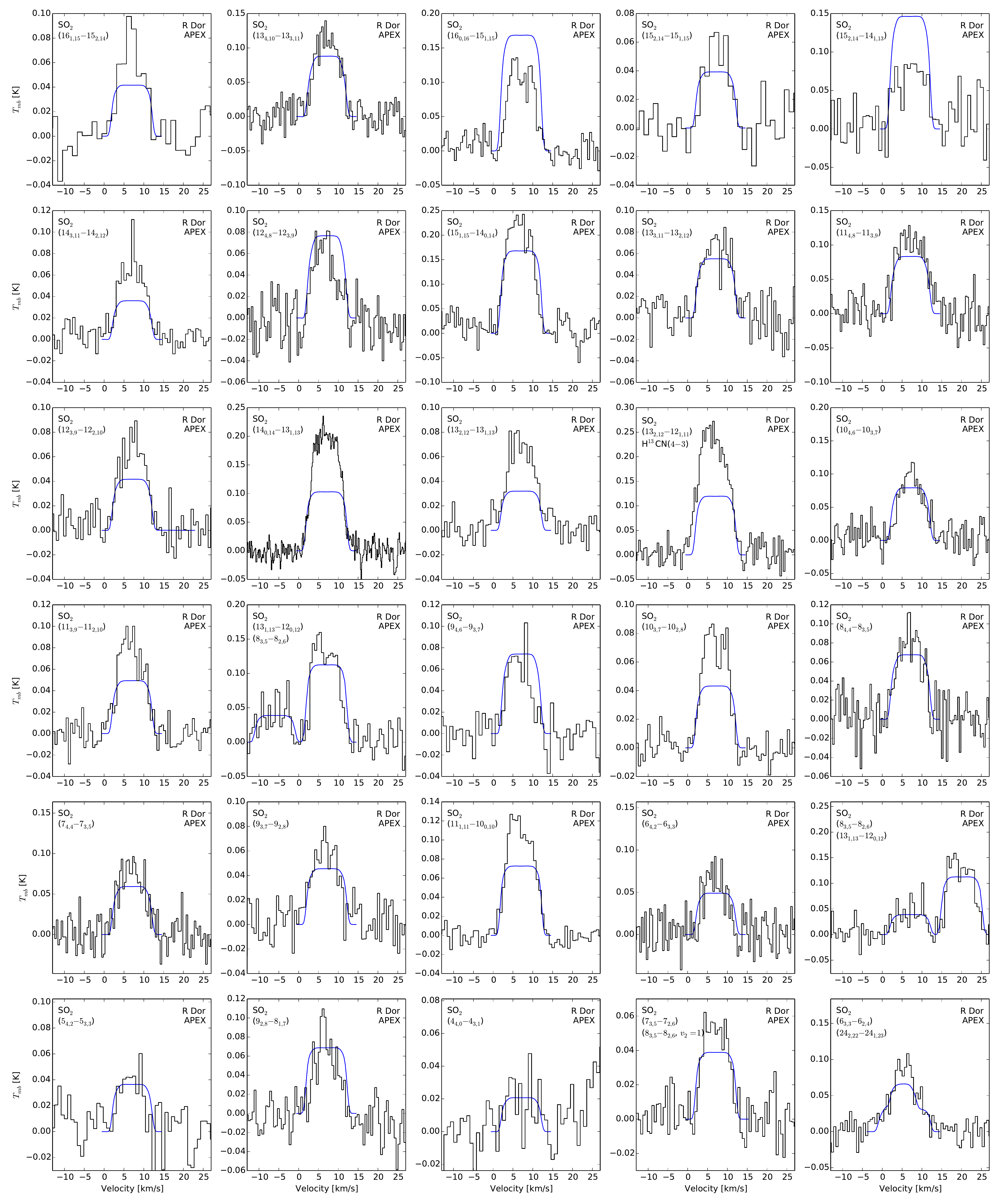}
\caption{continued.}
\end{center}
\end{figure*}
\addtocounter{figure}{-1}
\begin{figure*}[t]
\begin{center}
\includegraphics[width=\textwidth]{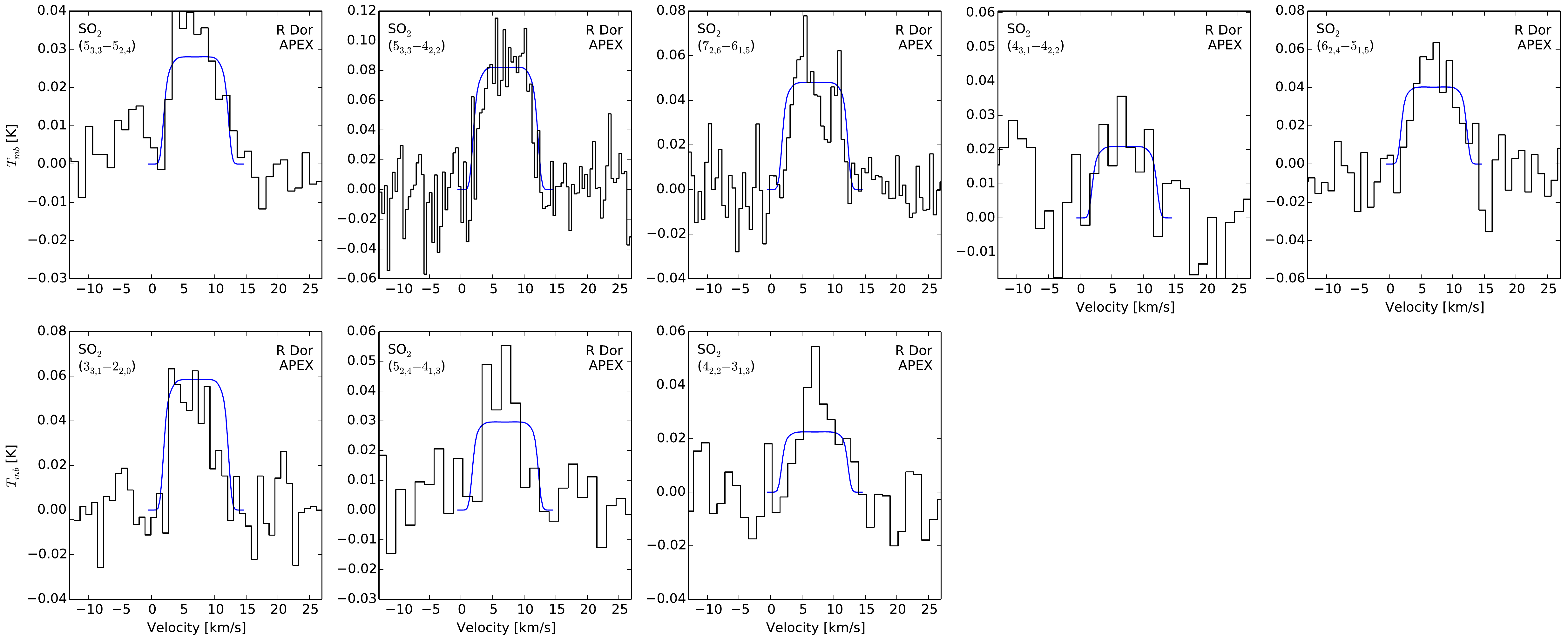}
\caption{continued.}
\end{center}
\end{figure*}

\begin{figure}[t]
\center
\includegraphics[width=0.5\textwidth]{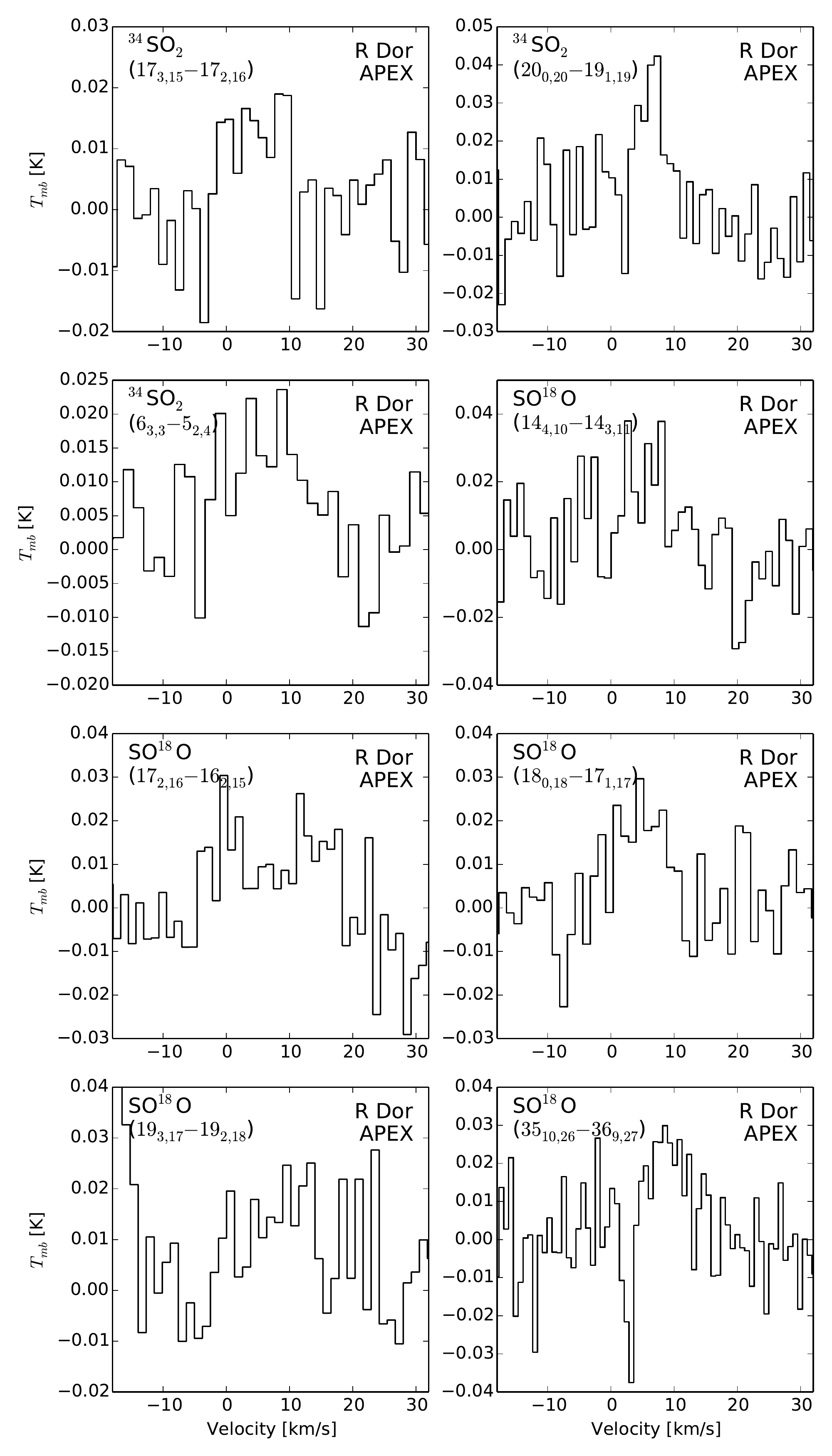}
\caption{Detections of the isotopologues \up{34}\so2 and SO\up{18}O towards R~Dor.}
\label{leftoverisotopologues}
\end{figure}

%
%

\section{HIFI OBSIDs}

The observation IDs for the HIFI observations used in this work are given in Table \ref{obsids}, including the non-detections used to constrain our TX Cam SO model.

\begin{table}[t]
\caption{HIFI OBSIDs for observations used in this work.}\label{obsids}
\begin{center}
\begin{tabular}{c c}
\hline\hline
Star & OBSID\\
\hline
IK Tau & 1342190198\\
		& 1342191594\\
R Dor 	& 1342198355\\
		& 1342197982\\
		& 1342200969\\
		& 1342200906\\
TX Cam & \phantom{* }1342205330 *\\
		& \phantom{* }1342205309 *\\
W Hya & 1342200951\\
		& 1342200981\\
		& 1342200929\\
R Cas & 1342200974\\
		& 1342198335\\
\hline
\end{tabular}
\tablefoot{* indicates only upper limits were derived for the SO lines.}
\end{center}
\end{table}

\end{document}